\documentclass[a4paper,11pt]{article}
\pdfoutput=1 

\usepackage{setspace}
\usepackage[bottom]{footmisc}
\usepackage{jheppub} 
\usepackage{amssymb}
\usepackage{amsmath}
\usepackage{amsthm}
\usepackage[usenames,dvipsnames]{xcolor}
\usepackage{epsfig}
\usepackage{dcolumn}
\usepackage{tikz}
\usetikzlibrary{shapes.geometric,arrows,positioning,decorations.markings}
\definecolor{DDFblue}{RGB}{45,92,143}
\definecolor{DDFred}{RGB}{181,36,53}
\tikzset{ddf axis/.style={draw=black!30,line width=.4pt},
  ddf seam/.style={draw=DDFblue,line width=.8pt},
  ddf contour/.style={draw=DDFred,line width=.9pt,>=stealth},
  ddf point/.style={circle,fill=black,inner sep=1.5pt},
  ddf zero/.style={diamond,draw=DDFblue,fill=white,inner sep=1.8pt},
  ddf leg/.style={draw=black!75,line width=.65pt},
  ddf edge/.style={draw=DDFblue,line width=.85pt,>=stealth},
  ddf forward/.style={postaction={decorate},decoration={markings,
    mark=at position .55 with {\arrow{stealth}}}},
  ddf backward/.style={postaction={decorate},decoration={markings,
    mark=at position .45 with {\arrowreversed{stealth}}}},
  ddf left/.style={circle,draw=DDFblue,fill=DDFblue,inner sep=2pt},
  ddf right/.style={circle,draw=DDFblue,fill=white,line width=.8pt,inner sep=2pt}}
\usepackage{upgreek}
\usepackage{array,multirow,bigdelim,arydshln}
\usepackage{appendix}
\usepackage{xparse}
\usepackage{stmaryrd}
\usepackage[T1]{fontenc} 
\usepackage{mathtools}
\usepackage{bm}
\usepackage{physics}
\usepackage{adjustbox}
\usepackage{graphicx} 
\graphicspath{{./images/}}
\usepackage[nottoc]{tocbibind}
\usepackage{hyperref}
\usepackage[utf8]{inputenc}
\usepackage{CJK}
\hypersetup{
	colorlinks,
	urlcolor=Maroon,
	linkcolor=Maroon,
	citecolor=Maroon
	}

\allowdisplaybreaks[2]
\numberwithin{equation}{section}
\let\DDFtableofcontents\tableofcontents
\renewcommand{\tableofcontents}{%
  \begingroup\setstretch{1}\small\DDFtableofcontents\endgroup}

\renewcommand{\dd}{\mathrm d}
\newcommand{\Id}{\mathbb I}

\newcommand{\diag}{\operatorname{diag}}
\newcommand{\col}{\operatorname{col}}
\renewcommand{\Res}{\operatorname{Res}}
\newcommand{\lhaf}{\operatorname{lhaf}}
\newcommand{\FDDF}{\mathrm{FDDF}}
\newcommand{\phys}{\mathrm{phys}}

\title{Multiparticle residues of open-string tachyon amplitudes from DDF sewing}
\author[]{Dripto Biswas,}
\emailAdd{dripto.biswas@unito.it}
\author[a]{Qu Cao}
\affiliation[a]{Department of Physics, School of Science, Westlake University, Hangzhou 310030, China}
\emailAdd{caoqu@westlake.edu.cn}

\date{}

\abstract{We compute the maximal simultaneous residues of color-ordered tachyon
disk amplitudes in the critical open bosonic string by summing all
physical intermediate states. Sewing cubic Del Giudice--Di Vecchia--Fubini
(DDF) vertices reduces this sum to Gaussian contractions for any planar
cubic tree, with the internal levels specified independently. Each
internal level fixes an exact finite mode cutoff. The residue then
follows by coefficient extraction in the sewing variables from a
determinant prefactor times the exponential of a quadratic form in
the sources. With consistent sewing conventions, the local-coordinate
scales cancel between the vertices and the sewing metric. At four
points, cubic local maps and a one-boson stress-tensor Ward identity
give an all-order resummation that recovers the Veneziano pole
coefficients. The same construction gives explicit matrices for the
five-point double residue and the six-point comb and snowflake, with
low-level checks against independent Koba--Nielsen expansions. The
determinant is the vacuum sewing factor of the transverse bosons.
We formulate the operator, coordinate-map and measure identities
needed to relate this construction to Grunsky operators and the full
higher-point Koba--Nielsen top forms.}


\begin{document}
\maketitle

\section{Introduction}
\label{sec:introduction}

String amplitudes must factorize through the same cubic couplings in
every channel. At four points, a pole residue sums
products of two three-point couplings. At higher multiplicity,
simultaneous poles in compatible channels probe whole trees of cubic
vertices, including couplings among three excited strings. Requiring
these residues to arise from a common set of couplings constrains
deformations of string amplitudes~\cite{ArkaniHamedEtAl2024}.
Evaluating them directly entails a sum over all physical states at
each exchanged level. We perform this sum for tachyon disk amplitudes
in the critical open bosonic string, with the internal levels
specified independently.

At four points, the string S-matrix bootstrap combines crossing
symmetry with assumptions about the spectrum, maximal exchanged spin
and high-energy behavior
\cite{CheungRemmen2023Bootstrap,HaringZhiboedov2024,CheungHillmanRemmen2024}.
String-like asymptotics also follow from massive higher-spin
scattering under suitable assumptions~\cite{CaronHuotEtAl2017}.
Residue zeros give further constraints when supplemented by Regge
and minimal-zero assumptions~\cite{CheungRemmenSciottiTarquini2026},
while positivity of partial-wave coefficients tests tree-level
unitarity~\cite{ArkaniHamedUnitarity2022}.\footnote{For further recent
developments, see refs.~\cite{ElvangHerderscheeMorales2026,WanZhou2026Analytic,
BermanEtAl2026Parity,Sheng2026Monodromy,
sung2026sixpointconsistencyuniquenessveneziano,
huang2026infraredconsistencyuniquenessstring,
WanWangZhou2026Extremality,CaoADR2026}.}

Higher-point factorization asks whether the couplings entering
four-point residues can consistently reproduce residues in other
channels and tree topologies
\cite{ArkaniHamedEtAl2024,BasileRemmenStaudt2026}.
For scalar external states, half-ladder partial waves resolve the
exchanges by spin and distinguish contributions from degenerate
intermediate states~\cite{backus2026halfladderpartialwavesnd}.
The degeneracy at each string level must therefore be retained when
comparing these multiparticle residues.

The DDF construction keeps these states explicit in a transverse
oscillator basis. The Del Giudice--Di Vecchia--Fubini operators
create positive-norm physical states
\cite{DelGiudiceDiVecchiaFubini1972}, whose completeness in the
critical theory follows from the no-ghost theorem
\cite{GoddardThorn1972,Brower1972}. The corresponding physical-state
projector~\cite{BrinkOlive1973} supplies the intermediate-state sums
in on-shell sewing. This description in terms of vertices and
projectors is closely related to repeated factorization and Reggeon
vertices~\cite{Yu1970}, as well as to the interacting vertices of
light-cone string theory and string field theory
\cite{Mandelstam1973,KakuKikkawa1974}.
The no-ghost theorem can also be used in on-shell recursion to replace
the physical-state sum by a covariant Fock-space
sum~\cite{ChangEtAl2013}.

The cubic couplings can be organized by oscillator generating
functions. At four points, an ansatz for the coupling of two tachyons
to a string state at arbitrary level reproduces the Veneziano pole
coefficients after summing oscillator configurations in a BCFW
reconstruction~\cite{FotopoulosVeneziano}.\footnote{The
large-complex-momentum behavior underlying BCFW recursion for disk
amplitudes in the presence of D-branes was studied in
ref.~\cite{FotopoulosPrezasPomerons}.}
DDF coherent-state amplitudes extend this organization to higher-point
gluing~\cite{BianchiFirrotta2020}. In the framed-DDF formulation, the
Reggeon vertex gives explicit Gaussian generating functions for the
cubic couplings~\cite{BiswasPesando2024,BiswasMarottaPesando2025}.
These functions supply the local blocks used in our sewing formula.

Residues of string amplitudes can be reconstructed from monodromy
relations and on-shell recursion~\cite{BoelsHansen2014}, or obtained
from the Koba--Nielsen (KN) representation in terms of Lauricella
functions~\cite{LaiLeeYang2023}. The worldsheet description gives
access to the cubic couplings of arbitrarily excited strings and their
role in four-point factorization~\cite{Firrotta2024}. It also admits
a binary-tree organization of amplitudes based on a weakly associative,
point-split product of on-shell
vertices~\cite{gomez2026fieldinspiredderivationopenstring}.\footnote{Related
source-functional methods have been developed for excited closed-string
amplitudes at fixed loop momenta and arbitrary
genus~\cite{SklirosCopelandSaffinGenerating}.}
Here we start from the complete sum over physical intermediate states,
with the level on each internal edge specified independently. The
framed-DDF basis reduces this sum to finite matrices, from which the
simultaneous residue follows by multivariate coefficient extraction.

Planar factorization on the disk has a geometric description in which
compatible channels correspond to noncrossing diagonals and a
triangulation is dual to a cubic tree~\cite{ArkaniHamedBaiHeYan2018}.
Positive coordinates make the corresponding factorization boundaries
manifest in stringy canonical forms and binary
geometries~\cite{ArkaniHamedHeLam2019,ArkaniHamedHeLamThomas2023},
while the curve-integral formalism, or surfaceology, organizes
amplitudes through curves on
surfaces~\cite{ArkaniHamedCounting2025,ArkaniHamedScaffolded2025}.
For general ray-like diagrams and the six-point Mercedes-Benz
configuration, residue generating functions are available directly in
positive coordinates~\cite{ArkaniHamedFigueiredoRemmen2025}.
These covariant expressions provide independent checks of the DDF
residues. A comparison of the full top forms further requires a map
between the sewing variables and coordinates on disk moduli space.

We take $M$ color-ordered external tachyons in the critical open bosonic
string, with all momenta outgoing, a mostly-plus metric,
$k_r^2=1/\alpha'$ and $\sum_r k_r=0$.
A planar cubic tree $\mathcal T$ selects $M-3$ compatible channels.
Each internal edge $e$ carries a level $N_e\geq0$ and a momentum
satisfying $\alpha'P_e^2=1-N_e$.
We work at $D=26$, with $d_\perp=24$ transverse directions, and choose
a common light-cone frame in which all momenta entering the DDF
operators have nonzero plus components. The momentum-conserving delta
function is suppressed. Our coupling and pole-residue normalizations
are specified in section~\ref{sec:direct-sewing}.

Every tree $\mathcal T$ is bipartite. Grouping its vertices into the
two color classes assembles the cubic quadratic couplings into
matrices $G_L,G_R$ and the linear sources into vectors $u_i,w_i$.
In a common transverse basis, the matrices carry internal-edge and
oscillator-mode indices, while $i=1,\ldots,24$ labels the transverse
direction. With a mode cutoff $L_e$ on each edge, define
\begin{equation}
 Z=\bigoplus_{e\in E}\operatorname{diag}_{1\leq n\leq L_e}
       \left(\frac{z_e^n}{n}\right),\qquad
 K_L=\Id-ZG_RZG_L,\qquad K_R=\Id-ZG_LZG_R .
 \label{eq:intro-sewing-matrices}
\end{equation}
\begin{samepage}
The inverse DDF norms are encoded by $Z$. Gaussian contraction
\mbox{\cite{BulmerEtAl2026,Tarasov2025}} gives
\begin{equation}
 \mathcal R_{\{N_e\}}^{(M;\mathcal T)}
 =\left[\prod_{e\in E}z_e^{N_e}\right]
       (\det K_L)^{-12}\exp\!\left(\mathcal E_{\mathcal T}\right),
 \label{eq:intro-main-result}
\end{equation}
where
\begin{equation}
 \mathcal E_{\mathcal T}=\sum_{i=1}^{24}\biggl[
 u_i^TK_L^{-1}Zw_i
 +\frac12u_i^TK_L^{-1}ZG_RZu_i
 +\frac12w_i^TK_R^{-1}ZG_LZw_i\biggr].
 \label{eq:intro-source-exponent}
\end{equation}
\end{samepage}
The sewing variable $z_e$ includes the coordinate scales at both ends
of edge $e$. The matrices have size $\sum_e L_e$ and are evaluated on
the chosen joint mass shell. Changing an internal level therefore
changes both the current contractions and the coefficient to be
extracted. The determinant branch is fixed by its value one at
$\bm z=0$.

Since no oscillator with mode number $n>N_e$ can occur at level $N_e$,
the cutoff $L_e=N_e$ is exact. Equation~\eqref{eq:intro-main-result}
includes all occupations and transverse polarizations at the prescribed
levels. We derive it in section~\ref{subsec:generality-proof} from
on-shell factorization, DDF completeness in the critical theory and
Gaussian contraction. It gives the maximal simultaneous pole
coefficient for $\mathcal T$; determining the regular part of the
amplitude requires further input.

The local-coordinate factors cancel between the cubic vertices and
the sewing metric when all blocks of a vertex use consistent
coordinate and Belavin--Polyakov--Zamolodchikov (BPZ) conventions.
This includes vertices with three DDF legs.

At four points, the oscillator sum admits an all-order resummation.
Two formal reparametrization identities remove the auxiliary
light-cone ratios and give the pole coefficient $(\alpha't+2)_N/N!$,
where $(x)_N=x(x+1)\cdots(x+N-1)$ and $(x)_0=1$.
Appendix~\ref{app:four-point-identities-proof} derives these identities
from the cubic local maps, free-boson sewing and a stress-tensor Ward
identity. They hold as formal series whose coefficients stabilize
as the mode cutoff increases.

At higher points, mixed blocks couple oscillator sectors on different
internal legs. We display the matrices for the five-point double
residue and the six-point comb and snowflake at arbitrary internal
levels, and compare explicit low-level coefficients with KN
expansions. Keeping $d_\perp$ symbolic isolates contributions
proportional to $d_\perp-24$; the oscillator sum represents the
complete physical spectrum only in the critical theory.

The determinant has a direct interpretation in oscillator sewing.
With canonical normalization, the contraction is a Berezin product
of Gaussian normal symbols, and its source-free part is the vacuum
sewing factor of the transverse bosons. A geometric realization in
terms of Grunsky operators~\cite{AlekseevShatashviliTakhtajan2024}
would relate these matrices to Mandelstam
maps~\cite{BiswasPesandoMandelstam} and to gluing and
resmoothing~\cite{LeClairPeskinPreitschopf1989,AsakawaKugoTakahashi1998}.
Section~\ref{sec:berezin-grunsky} formulates the additional operator,
coordinate-map and measure identities required to match the
higher-point KN top forms. These geometric identities remain open;
the physical-state derivation of the residues does not rely on them.

Section~\ref{sec:ddf-essentials} fixes the framed-DDF conventions and
cubic blocks. Sections~\ref{sec:direct-sewing} and
\ref{sec:four-point-matrix} derive the physical-state sum and evaluate
it at four points. Section~\ref{sec:arbitrary-tree-sewing} extends the
matrix formula to arbitrary planar cubic trees and proves the
cancellation of coordinate scales. Five- and six-point examples follow
in sections~\ref{sec:five-point-matrix} and \ref{sec:six-point}.
Section~\ref{sec:berezin-grunsky} discusses the Berezin--Grunsky
interpretation and the comparison with moduli space.
Appendix~\ref{app:four-point-identities-proof} proves the four-point
reparametrization identities; appendices~\ref{app:transverse-basis} and
\ref{app:cubic-coordinate-factors} treat transverse-basis independence,
local phase conventions and fixed-level coordinate cancellation.

\section{Framed DDF states and cubic vertices}
\label{sec:ddf-essentials}

DDF operators provide a transverse oscillator basis of physical states.
The intermediate-state sum can then be performed without constructing
covariant polarization tensors at each level. We fix the operator
normalization and the reduced cubic coefficients used below.

\subsection{DDF oscillators and physical states}
\label{subsec:ddf-states}

Choose a target-space orthonormal frame \(E^\mu{}_{\nu}\), with
light-cone components $k^\pm=(k^0\pm k^{D-1})/\sqrt2$ in that frame.
We write
\begin{equation}
 k\cdot\ell=-k^+\ell^- -k^-\ell^+ +\bm k\cdot\bm\ell,
 \label{eq:lightcone-product}
\end{equation}
where bold symbols denote the \(d_\perp=D-2\) transverse components.
The dimensionless chiral free bosons are normalized by
\[
L^\mu(z)L^\nu(w)\sim-\eta^{\mu\nu}\log(z-w).
\]
The framed-DDF operators \cite{BiswasPesando2024} are
\begin{equation}
 A_n^i(E)=i\oint\frac{\dd z}{2\pi i}\,
 \partial L^i(z)\exp\!\left[
 \frac{i n L^+(z)}{\sqrt{2\alpha'}\,\widehat p^+}\right],
 \qquad i=1,\ldots,d_\perp .
 \label{eq:framed-ddf-operator}
\end{equation}
The contour surrounds the state insertion in the local worldsheet
coordinate $z$. For $n>0$, $A_{-n}^i$ creates an oscillator and $A_n^i$
annihilates one. The light-cone zero-mode operator $\widehat p^+$ acts
on $\ket p$ with eigenvalue $p^+$. DDF excitations leave this component
unchanged, so the seed tachyon and the excited state have the same
$p^+$. The factor $1/\widehat p^+$ fixes the normalization in each
momentum sector.

The operators are defined for nonzero \(p^+\). We choose a common
light-cone frame in which all external and internal plus momenta are
nonzero,\footnote{In the light-cone description, $|p^+|$ sets the width
of the interacting string.} and suppress the frame label $E$ below.
The oscillator algebra is
\begin{equation}
 [A_m^i,A_n^j]=m\,\delta^{ij}\delta_{m+n,0}.
 \label{eq:ddf-algebra}
\end{equation}
For the matter sector, the on-shell physical-state conditions are
\begin{equation}
 L_m\ket{\psi}=0\quad(m>0),\qquad
 (L_0-1)\ket{\psi}=0,
 \label{eq:physical-state-conditions}
\end{equation}
where $L_m$ are the Virasoro generators. The framed-DDF operators
commute with them and hence preserve the physical-state
conditions~\cite{BiswasPesando2024}.
Let $\lvert k_T\rangle$ be the oscillator vacuum with tachyon momentum
$k_T^2=1/\alpha'$, annihilated by $A_n^i$ for $n>0$.
Acting with $r_{n,i}$ oscillators of mode $n$ and transverse component
$i$ gives total level $N=\sum_{n>0,i}n r_{n,i}$.
The longitudinal exponentials in the DDF operators shift the momentum
to
\begin{equation}
 K^+=k_T^+,\qquad \bm K=\bm k_T,\qquad
 K^-=k_T^-+\frac{N}{2\alpha' k_T^+}.
 \label{eq:framed-momentum}
\end{equation}
It follows that
\begin{equation}
 K^2=-2K^+K^-+\bm K^2
 =-2k_T^+k_T^-+\bm k_T^2-\frac{N}{\alpha'}
 =\frac{1-N}{\alpha'}.
 \label{eq:ddf-mass-shell}
\end{equation}
Thus, the seed and excited state share their plus and transverse
momenta, while their minus momenta differ. The physical mass is
$m_N^2=-K^2=(N-1)/\alpha'$: $N=0$ is the tachyon, $N=1$ is massless,
and $N\geq2$ is massive. The associated on-shell matter vertex has
conformal weight one; multiplying it by the weight-$(-1)$ ghost $c$ gives an
unintegrated vertex of weight zero.

Writing $P$ for a fixed physical momentum on this mass shell, the
unnormalized occupation states and their oscillator norms are
\begin{equation}
 \lvert\{r\};P\rangle
 =\prod_{n,i}(A_{-n}^i)^{r_{n,i}}\lvert k_T\rangle,\qquad
 \sum_{n,i}nr_{n,i}=N,\qquad
 \langle\{r\};P\vert\{r\};P\rangle
 =\prod_{n,i}n^{r_{n,i}}r_{n,i}! .
 \label{eq:ddf-basis-and-norm}
\end{equation}
The momentum delta function is suppressed. At $D=26$, these states
span the physical spectrum modulo null states. Their inverse norms
therefore define the physical-state projector used in
section~\ref{sec:direct-sewing}, with the full degeneracy of each level
included.

We refer to each end of an internal edge as a half-edge and allow it
an independent transverse basis. The light-cone directions are common
to all legs. Orthogonal rotations of the transverse bases leave the
sewing invariant, as shown in appendix~\ref{app:transverse-basis}.
The physical residue is also independent of the auxiliary light-cone
direction by DDF completeness at $D=26$, although the individual
matrices can depend on it. We retain symbolic $d_\perp$ only to track
the dimension dependence of the oscillator sum.

\subsection{Reduced three-point blocks}
\label{subsec:reduced-three-point}

Consider one level-$N$ DDF state of momentum $K$ and two tachyons of
momenta $k,\ell$, all outgoing:
\begin{equation}
 K+k+\ell=0.
 \label{eq:three-point-conservation}
\end{equation}
Define the light-cone ratio and relative momentum
\begin{equation}
 \rho=\frac{k^+}{K^+},\qquad Q=k-\rho K.
 \label{eq:rho-Q}
\end{equation}
Then \(Q^+=0\).  Moreover,
\begin{equation}
 \frac{\ell^+}{K^+}=-1-\rho,
 \qquad
 \bm\ell-\frac{\ell^+}{K^+}\bm K
 =-\bm K-\bm k+(1+\rho)\bm K=-\bm Q,
 \label{eq:other-three-point-leg}
\end{equation}
so the two tachyon momenta determine a single transverse vector.
More generally, $Q^+=Q'^+=0$ implies
$Q\cdot Q'=\bm Q\cdot\bm Q'$. We will use this identity to express
transverse current contractions in terms of Lorentz invariants.

After removing the local-coordinate factors, the three-point framed-DDF
coefficients derived in ref.~\cite{BiswasMarottaPesando2025} are
\begin{equation}
 \widehat A_n^i=\sqrt{2\alpha'}\,Q^i f_n(\rho),
 \qquad
 \widehat B_{nm}^{ij}=\delta^{ij}g_{nm}(\rho),
 \label{eq:reduced-three-point-blocks}
\end{equation}
where
\begin{align}
 f_n(\rho)
 &=(-1)^{n-1}\binom{n\rho+n-1}{n-1},
 \label{eq:f-def}\\
 g_{nm}(\rho)
 &=(-1)^{n+m}
 \binom{n(1+\rho)}{n}\binom{m(1+\rho)}{m}
 \frac{nm}{n+m}\frac{\rho}{1+\rho}
 \nonumber\\
 &=\rho(1+\rho)\frac{nm}{n+m}f_n(\rho)f_m(\rho).
 \label{eq:g-def}
\end{align}
Here the binomial coefficients are understood through their finite-product
definitions, and $f_1=1$. The second expression for $g_{nm}$ makes its
polynomial dependence on $\rho$ explicit and avoids an apparent singularity
at $\rho=-1$.

The coefficient $\widehat A_n^i$ couples one oscillator to the tachyon
momenta, while $\widehat B_{nm}^{ij}$ contracts two oscillators on the
DDF leg. With the cubic coupling suppressed, the generating function
in commuting sources $\lambda_{ni}$ is
\begin{equation}
 \begin{aligned}
 \widehat{\mathcal M}(\lambda)
 &=\exp\!\left[\sum_{n,i}\lambda_{ni}\widehat A_n^i
 +\frac12\sum_{n,m,i,j}\lambda_{ni}\lambda_{mj}
                         \widehat B_{nm}^{ij}\right],\\
 \left.\partial_{\lambda_{ni}}\partial_{\lambda_{mj}}
 \widehat{\mathcal M}\right|_{\lambda=0}
 &=\widehat A_n^i\widehat A_m^j+\widehat B_{nm}^{ij}.
 \end{aligned}
 \label{eq:cubic-block-source-meaning}
\end{equation}
Differentiation generates the unpaired current terms and pairwise
contractions of Wick's theorem~\cite{BiswasMarottaPesando2025}.
For a level-$N$ state, the momenta and currents are evaluated on its
mass shell, and derivatives of weighted degree $N$ give the on-shell
cubic couplings. The generating function does not prescribe an
off-shell continuation.
This vertex suffices at the ends of a four-point diagram. Vertices
with two or three DDF legs are introduced in
section~\ref{sec:arbitrary-tree-sewing}.

An on-shell three-point disk amplitude has no modulus. At fixed levels,
the seed-tachyon KN factor, oscillator coordinate powers and
three-ghost correlator combine into a puncture-independent
amplitude~\cite{BiswasMarottaPesando2025}. We use this reduced cubic
amplitude below. The coherent Reggeon vertex retains its seed factor
until the levels are selected.
Section~\ref{subsec:punc_coord_mats} derives the fixed-level
cancellation. Section~\ref{subsec:reduced_mat_res} shows the same
cancellation in the matrices by keeping the oscillator coordinate
factors and the compensating rescalings of the sewing metric explicit.

\section{On-shell sewing}
\label{sec:direct-sewing}
\suppressfloats[t]

Repeated on-shell factorization expresses a simultaneous pole
coefficient as cubic amplitudes joined by one physical-state projector
on each internal edge. We first write this sum at four and five points,
then evaluate it for an arbitrary planar cubic tree in
section~\ref{sec:arbitrary-tree-sewing}.

The internal pairing is the BPZ bilinear form induced by conformal
inversion of a local coordinate
\cite{LeClairPeskinPreitschopfKinematics1989}. With all cubic momenta
outgoing, the two ends of an internal line carry opposite momenta.
We choose BPZ-dual oscillator bases so that sewing uses the inverse
norms in Eq.~\eqref{eq:ddf-basis-and-norm}. Bras below denote these
dual states. Analytically continued momenta and sources are paired
algebraically, without complex conjugation; the associated oscillator
phases are displayed in section~\ref{sec:arbitrary-tree-sewing}.

\subsection{Four-point sewing}
\label{subsec:four-point-sewing}

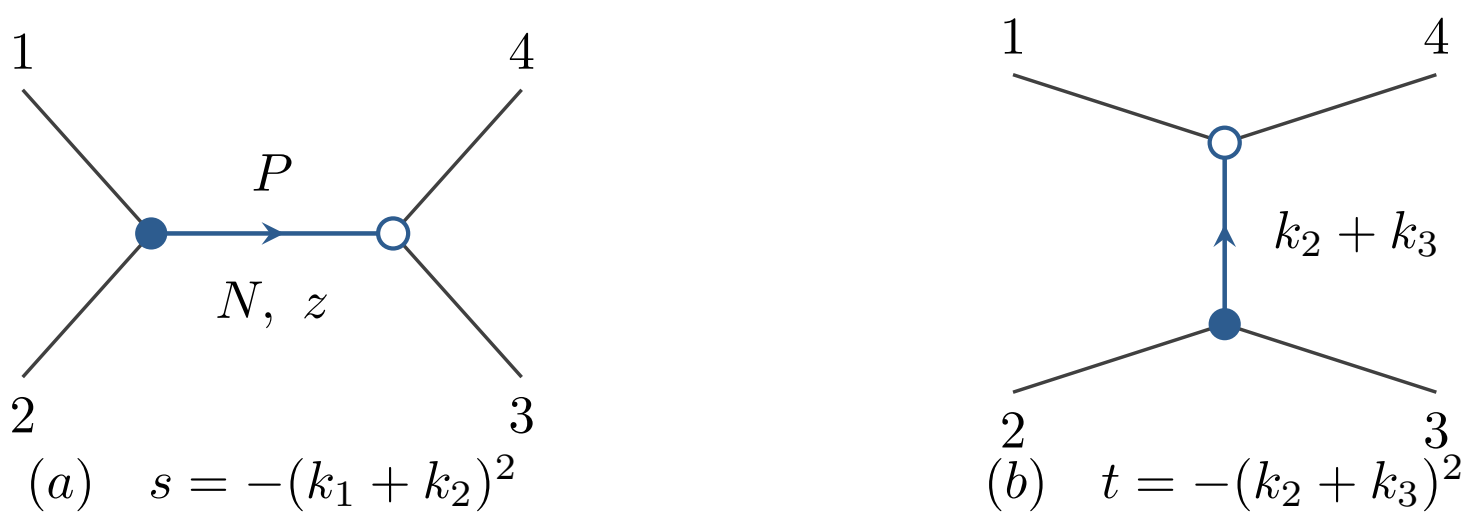
\begin{figure}[tbp]
\centering
\begin{tikzpicture}[font=\small]
  \begin{scope}[xshift=-3.15cm]
    \coordinate (a) at (-.8,0); \coordinate (b) at (.8,0);
    \draw[ddf leg] (-1.65,.95) node[above] {$1$} -- (a)
      -- (-1.65,-.95) node[below] {$2$};
    \draw[ddf leg] (1.65,-.95) node[below] {$3$} -- (b)
      -- (1.65,.95) node[above] {$4$};
    \draw[ddf edge,ddf forward] (a) -- (b)
      node[midway,above=4pt] {$P$} node[midway,below=5pt] {$N,\ z$};
    \node[ddf left] at (a) {}; \node[ddf right] at (b) {};
    \node at (0,-1.65) {$(a)\quad s=-(k_1+k_2)^2$};
  \end{scope}
  \begin{scope}[xshift=3.15cm]
    \coordinate (a) at (0,-.6); \coordinate (b) at (0,.6);
    \draw[ddf leg] (-1.4,1.05) node[above] {$1$} -- (b)
      -- (1.4,1.05) node[above] {$4$};
    \draw[ddf leg] (-1.4,-1.05) node[below] {$2$} -- (a)
      -- (1.4,-1.05) node[below] {$3$};
    \draw[ddf edge,ddf forward] (a) -- (b)
      node[midway,right=5pt] {$k_2+k_3$};
    \node[ddf left] at (a) {}; \node[ddf right] at (b) {};
    \node at (0,-1.65) {$(b)\quad t=-(k_2+k_3)^2$};
  \end{scope}
\end{tikzpicture}
\caption{Alternative pole expansions of the ordered amplitude
$\mathcal V_{1234}$. External labels denote outgoing momenta $k_r$ in
cyclic order. In the $s$ channel, $P=k_1+k_2$, and extracting $[z^N]$
selects the complete level-$N$ physical-state sum on the internal edge,
with $P^2=(1-N)/\alpha'$. The $t$-channel construction follows by cyclic
relabeling. Each panel is used in its own pole expansion. Filled and open
vertices denote the $L$ and $R$ BPZ classes; blue arrows indicate
$L\to R$, with momentum signs fixed separately by outgoing momentum
conservation at each vertex.}
\label{fig:four-point-channels}
\end{figure}

Take ordered four-tachyon kinematics
\begin{equation}
 k_1+k_2+k_3+k_4=0,\qquad k_r^2=\frac1{\alpha'},
 \label{eq:four-point-kinematics}
\end{equation}
and define
\begin{equation}
 P=k_1+k_2=-(k_3+k_4),\qquad
 s=-P^2,\qquad t=-(k_2+k_3)^2.
 \label{eq:four-point-channels}
\end{equation}
The ordered four-tachyon amplitude~\cite{Veneziano1968} fixes our
normalization:
\begin{equation}
 \mathcal V(s,t)
 =g_o^2\int_0^1\dd x\,x^{-\alpha's-2}(1-x)^{-\alpha't-2}
 =g_o^2\mathrm B(-\alpha's-1,-\alpha't-1),
 \label{eq:Veneziano-benchmark}
\end{equation}
where $\mathrm B$ is Euler's beta function and the integral is understood
by meromorphic continuation. We will recover its pole coefficients
from the physical-state sum.
The two planar four-point pole factorizations are displayed in
figure~\ref{fig:four-point-channels}; we work with the $s$ channel below.

The channel momenta entering the left and right three-point amplitudes are

\begin{equation}
 K_L=-P,\qquad K_R=P.
 \label{eq:four-point-massive-momenta}
\end{equation}
Introduce the nonzero ratios
\begin{equation}
 a=\frac{k_2^+}{K_L^+}=-\frac{k_2^+}{P^+},
 \qquad
 b=\frac{k_3^+}{K_R^+}=\frac{k_3^+}{P^+},
 \label{eq:four-point-ratios}
\end{equation}
with \(a,b,1+a,1+b\neq0\), and the transverse momenta
\begin{equation}
 \bm Q_L=\bm k_2+a\bm P,\qquad
 \bm Q_R=\bm k_3-b\bm P.
 \label{eq:four-point-Q}
\end{equation}
Both corresponding spacetime vectors have vanishing plus component.  With
the right amplitude BPZ-oriented as a bra, define
\begin{equation}
 J_L^i=\sqrt{2\alpha'}\,Q_L^i,\qquad
 J_R^i=-\sqrt{2\alpha'}\,Q_R^i.
 \label{eq:four-point-currents}
\end{equation}
Using the reduced coefficients of section~\ref{subsec:reduced-three-point},
the BPZ-oriented cubic generating functions are
\begin{align}
 \mathcal M_L(\lambda)
 &=g_o\exp\!\left[
 \sum_{n,i}\lambda_{ni}f_n(a)J_L^i
 +\frac12\sum_{n,m,i}\lambda_{ni}\lambda_{mi}g_{nm}(a)
 \right],
 \label{eq:left-cubic-generator}\\
 \mathcal M_R(\widetilde\lambda)
 &=g_o\exp\!\left[
 \sum_{n,i}\widetilde\lambda_{ni}f_n(b)J_R^i
 +\frac12\sum_{n,m,i}\widetilde\lambda_{ni}
              \widetilde\lambda_{mi}g_{nm}(b)
 \right].
 \label{eq:right-cubic-generator}
\end{align}
The coupling to an unnormalized occupation state $\{r_{n,i}\}$ is
obtained by differentiating with respect to its sources and then
setting all sources to zero.

At the level-\(N\) pole,
\begin{equation}
 \alpha's=N-1,\qquad P^2=\frac{1-N}{\alpha'},
 \label{eq:four-point-pole}
\end{equation}
and the two legs being sewn are on shell. Using the occupation basis
and norms in Eq.~\eqref{eq:ddf-basis-and-norm}, the transverse
framed DDF (FDDF) projector is
\begin{equation}
 \Pi_N^{\FDDF}(P)=
 \sum_{\sum nr_{n,i}=N}
 \frac{\lvert\{r\};P\rangle\langle\{r\};P\rvert}
      {\prod_{n,i}n^{r_{n,i}}r_{n,i}!}.
 \label{eq:ddf-projector}
\end{equation}
At $D=26$, this projector represents the complete physical
Becchi--Rouet--Stora--Tyutin (BRST) cohomology in on-shell matrix
elements:
\begin{equation}
 \Pi_N^{\phys}(P)\doteq\Pi_N^{\FDDF}(P),\qquad D=26,
 \label{eq:physical-ddf-projector}
\end{equation}
Here null and BRST-exact states decouple from the pole
residue~\cite{GoddardThorn1972,Brower1972}.

On a level-\(N\) state,
\begin{equation}
 \frac{\Pi_N^{\FDDF}(P)}{L_0-1}
 =\frac{\Pi_N^{\FDDF}(P)}{\alpha'P^2+N-1}.
 \label{eq:propagator-projector}
\end{equation}
The level contribution obtained by sewing the two cubic amplitudes therefore
has the pole
\begin{equation}
 \mathcal V(s,t)=-\frac{g_o^2 R_N}{\alpha's-(N-1)}+O(1),
 \label{eq:four-point-pole-form}
\end{equation}
where
\begin{equation}
 R_N=\frac1{g_o^2}
 \sum_{\sum nr_{n,i}=N}
 \frac{\mathcal M_L(\{r\})\mathcal M_R(\{r\})}
      {\prod_{n,i}n^{r_{n,i}}r_{n,i}!}.
 \label{eq:four-point-state-sum}
\end{equation}
Equivalently,
\begin{equation}
 R_N=[z^N]
 \left.
 \exp\!\left(\sum_{n,i}\frac{z^n}{n}
       \partial_{\lambda_{ni}}\partial_{\widetilde\lambda_{ni}}\right)
 \frac{\mathcal M_L(\lambda)\mathcal M_R(\widetilde\lambda)}{g_o^2}
 \right|_{\lambda=\widetilde\lambda=0},
 \label{eq:four-point-derivative-sum}
\end{equation}
Here $[z^n]F(z)$ denotes the coefficient of $z^n$ in $F(z)$; the same
notation will be used for multivariate coefficient extraction.
The quantity $R_N$ is the dimensionless pole coefficient in the
convention used above. In particular,
\begin{equation}
  \operatorname*{Res}_{s=(N-1)/\alpha'} \mathcal V(s,t)
    = -\frac{g_o^2}{\alpha'}R_N .
  \label{eq:residue-normalization}
\end{equation}
The minus sign follows from the on-shell condition:
$\alpha'P^2+N-1=-[\alpha's-(N-1)]$.

We resum Eq.~\eqref{eq:four-point-derivative-sum} in
section~\ref{sec:four-point-matrix}.
\subsection{Five-point sewing}
\label{subsec:five-point-pattern}

For the planar comb in which legs 1 and 2 fuse first and leg 3 is then
attached, let
\begin{equation}
 P=k_1+k_2,\qquad Q=P+k_3=-(k_4+k_5),\qquad
 s=-P^2,\qquad \widetilde s=-Q^2.
 \label{eq:five-point-direct-channels}
\end{equation}
In the Koba--Nielsen representation~\cite{KobaNielsen1969}, fix
\(z_1=0\), \(z_4=1\), \(z_5=\infty\), and put
\(z_2=xy\), \(z_3=y\).  With
\begin{equation}
 X=-2\alpha'k_2\cdot k_3,\qquad
 Y=-2\alpha'k_3\cdot k_4,\qquad
 W=-2\alpha'k_2\cdot k_4,
 \label{eq:five-point-direct-invariants}
\end{equation}
the reduced ordered amplitude becomes
\begin{equation}
 \mathcal V_{12345}=g_o^3\int_0^1\dd x\int_0^1\dd y\,
 x^{-\alpha's-2}y^{-\alpha'\widetilde s-2}
 (1-x)^{-X}(1-y)^{-Y}(1-xy)^{-W}.
 \label{eq:five-point-ray-integral}
\end{equation}
The direct double expansion defines
\begin{equation}
 R_{N,M}^{(5)}(X,Y,W)
 =\sum_{c=0}^{\min(N,M)}
 \frac{(X)_{N-c}(Y)_{M-c}(W)_c}
 {(N-c)!(M-c)!c!}.
 \label{eq:five-point-direct-residue}
\end{equation}
In particular,
\begin{equation}
 R_{1,1}^{(5)}=XY+W,
 \label{eq:five-point-first-mixed-residue}
\end{equation}
The additional term $W$ comes from the middle
DDF--tachyon--DDF vertex. It is absent from the product of the two
four-point residue polynomials.

At the simultaneous pole
\begin{equation}
 \alpha's=N-1,\qquad \alpha'\widetilde s=M-1,
 \label{eq:five-point-double-pole}
\end{equation}
the two covariant projectors may independently be replaced by framed-DDF
projectors. Let $\mathcal M_L(r)$, $\mathcal M_C(s,r)$, and
$\mathcal M_R(s)$ be the three on-shell cubic amplitudes, and define
\begin{equation}
 \mathcal N_r=\prod_{n,i}n^{r_{n,i}}r_{n,i}!,\qquad
 \mathcal N_s=\prod_{m,j}m^{s_{m,j}}s_{m,j}!,
 \label{eq:five-point-norms}
\end{equation}
Here $|r|=\sum_{n,i}nr_{n,i}$ and $|s|=\sum_{m,j}ms_{m,j}$ denote the
oscillator levels, with each mode weighted by its mode number.
Summing over the two intermediate states gives
\begin{equation}
 \left.
 \frac1{g_o^3}
 \sum_{\substack{|r|=N\\|s|=M}}
 \frac{\mathcal M_R(s)\mathcal M_C(s,r)\mathcal M_L(r)}
      {\mathcal N_s\mathcal N_r}
 \right|_{\alpha's=N-1,\;\alpha'\widetilde s=M-1}
 =R_{N,M}^{(5)}(X,Y,W).
 \label{eq:five-point-state-sum-identity}
\end{equation}
The two propagator signs cancel, and near the simultaneous pole the amplitude
contains
\begin{equation}
 \mathcal V_{12345}
 =\frac{g_o^3R_{N,M}^{(5)}(X,Y,W)}
 {\bigl[\alpha's-(N-1)\bigr]
  \bigl[\alpha'\widetilde s-(M-1)\bigr]}
 + (\text{terms with at most one of the two poles}).
 \label{eq:five-point-double-pole-form}
\end{equation}
Thus the five-point residue contracts three Gaussian vertices through
two inverse-norm sewing metrics. Which vertices share an internal
edge determines the blocks in the matrix formula.

\section{Four-point sewing in matrix form}
\label{sec:four-point-matrix}

We evaluate the occupation-number sum in
Eq.~\eqref{eq:four-point-derivative-sum} by Gaussian contraction.
The local-coordinate factors are absorbed into the invariant sewing
variable \(z\). Section~\ref{sec:arbitrary-tree-sewing} treats these
factors explicitly and proves their cancellation.

\subsection{Finite mode matrices}

Let $L$ denote an oscillator-mode cutoff, independently of the
physical pole level $N$. Every occurrence of mode $n$ in a sewn
term carries at least $z^n$, so $L\geq N$ suffices to extract
$[z^N]$. Define
\begin{equation}
 \begin{gathered}
  f^{[L]}(\rho)=
    \bigl(f_1(\rho),\ldots,f_L(\rho)\bigr)^T,\qquad
  g^{[L]}(\rho)=
    \bigl(g_{nm}(\rho)\bigr)_{1\leq n,m\leq L},\\
  Z^{[L]}=\operatorname{diag}
    \left(z,\frac{z^2}{2},\ldots,\frac{z^L}{L}\right).
 \end{gathered}
\end{equation}
We omit the cutoff superscript when it is unambiguous. The ratios
$a$ and $b$ belong to different cubic vertices, so plus-momentum
conservation does not relate them by $1+a+b=0$.

For every transverse direction we set
\begin{equation}
 \lambda_i=\col(\lambda_{1i},\ldots,\lambda_{Li}),\qquad
 \widetilde\lambda_i=\col(\widetilde\lambda_{1i},\ldots,
 \widetilde\lambda_{Li}),\qquad
 \Lambda_i=\binom{\lambda_i}{\widetilde\lambda_i}.
 \label{eq:four-point-source-columns}
\end{equation}
The quadratic coupling matrix, sewing matrix and source vector are
\begin{equation}
 G=\begin{pmatrix}g(a)&0\\0&g(b)\end{pmatrix},\qquad
 \mathbf Z=\begin{pmatrix}0&Z\\Z&0\end{pmatrix},\qquad
 v_i=\binom{J_L^i f(a)}{J_R^i f(b)}.
 \label{eq:four-point-big-matrices}
\end{equation}
Both \(G\) and \(\mathbf Z\) are symmetric.  The two cubic vertices and the
sewing operator are respectively
\begin{align}
 \frac{\mathcal M_L\mathcal M_R}{g_o^2}
 &=\prod_{i=1}^{d_\perp}
 \exp\!\left(v_i^T\Lambda_i+\frac12\Lambda_i^TG\Lambda_i\right),
 \label{eq:four-point-gaussian-product}\\
 \exp\!\left(\sum_{n,i}\frac{z^n}{n}
 \partial_{\lambda_{ni}}\partial_{\widetilde\lambda_{ni}}\right)
 &=\prod_{i=1}^{d_\perp}
 \exp\!\left(\frac12\partial_{\Lambda_i}^T
                   \mathbf Z\partial_{\Lambda_i}\right).
 \label{eq:four-point-matrix-sewing}
\end{align}

\subsection{Loop-hafnian generating function}

The loop-hafnian generating function was derived in quantum
optics~\cite{BulmerEtAl2026} and extended to arbitrary complex symmetric
matrices in ref.~\cite{Tarasov2025}. We use it as a formal Gaussian
contraction identity.

For a symmetric complex \(2m\times2m\) matrix \(S\), a vector
\(v\in\mathbb C^{2m}\), and
\begin{equation}
 \mathbf Z(\bm t)=
 \begin{pmatrix}0&D(\bm t)\\D(\bm t)&0\end{pmatrix},\qquad
 D(\bm t)=\diag(t_1,\ldots,t_m),
 \label{eq:Tarasov-Z}
\end{equation}
the identity reads
\begin{equation}
 \frac{\exp\!\left[
 \frac12v^T(\Id-\mathbf Z(\bm t)S)^{-1}\mathbf Z(\bm t)v\right]}
 {\sqrt{\det(\Id-\mathbf Z(\bm t)S)}}
 =\sum_{\bm r\in\mathbb N_0^m}
 \lhaf\!\left(\widetilde S_{\bm r,\bm r};
                 \widetilde v_{\bm r,\bm r}\right)
 \prod_{n=1}^{m}\frac{t_n^{r_n}}{r_n!}.
 \label{eq:Tarasov-theorem}
\end{equation}
The tilded matrix repeats the $n$-th row and column of each
$m$-component block $r_n$ times, with the same repetition of vector
entries. The loop hafnian sums partitions into pairs and singletons:
a pair contributes a matrix element, and a singleton a vector entry.
We take the square-root branch with value one at $\bm t=0$ and use
the identity as a formal power series. At fixed finite levels, no
convergence assumption is needed. The transposes implement the
algebraic BPZ pairing of section~\ref{sec:direct-sewing}.

The corresponding differential-operator identity also follows directly
from a finite-dimensional heat equation. Define
\begin{equation}
 F_\tau(\Lambda)=
 \exp\!\left(\frac\tau2\partial_\Lambda^T\mathbf Z
                         \partial_\Lambda\right)
 \exp\!\left(v^T\Lambda+\frac12\Lambda^TS\Lambda\right).
 \label{eq:gaussian-heat-definition}
\end{equation}
Writing $F_\tau=\exp(c_\tau+r_\tau^T\Lambda+
\tfrac12\Lambda^TM_\tau\Lambda)$ gives
\begin{equation}
 \dot M_\tau=M_\tau\mathbf ZM_\tau,\qquad
 \dot r_\tau=M_\tau\mathbf Zr_\tau,\qquad
 \dot c_\tau=\frac12\operatorname{Tr}(\mathbf ZM_\tau)
             +\frac12r_\tau^T\mathbf Zr_\tau,
 \label{eq:gaussian-heat-system}
\end{equation}
with initial data $(M_0,r_0,c_0)=(S,v,0)$. Its solution is
\begin{align}
 M_\tau&=S(\Id-\tau\mathbf ZS)^{-1},&
 r_\tau&=(\Id-\tau S\mathbf Z)^{-1}v,\nonumber\\
 c_\tau&=-\frac12\log\det(\Id-\tau\mathbf ZS)
 +\frac\tau2v^T(\Id-\tau\mathbf ZS)^{-1}\mathbf Zv.
 \label{eq:gaussian-heat-solution}
\end{align}
Evaluating $F_1(0)$ gives the left-hand side of
Eq.~\eqref{eq:Tarasov-theorem}. The cubic blocks and sewing metric
supply all the string-theory input to this Gaussian identity.

For DDF sewing, setting \(t_n=z^n/n\) gives
\begin{equation}
 \prod_{n=1}^{L}\frac{t_n^{r_{n,i}}}{r_{n,i}!}
 =z^{\sum_nnr_{n,i}}
  \prod_{n=1}^{L}\frac1{n^{r_{n,i}}r_{n,i}!},
 \label{eq:Tarasov-ddf-norm}
\end{equation}
which is exactly the inverse DDF norm.  Multiplication over all
transverse directions therefore performs the complete occupation-number
sum.  Defining
\begin{equation}
 R_N=[z^N]\mathcal G_N^{(d_\perp)}(z),
 \label{eq:four-point-residue-generator-def}
\end{equation}
Eq.~\eqref{eq:Tarasov-theorem} gives
\begin{equation}
 \mathcal G_N^{(d_\perp)}(z)=
 \frac{\exp\!\left[
 \frac12\sum_{i=1}^{d_\perp}
 v_i^T(\Id-\mathbf ZG)^{-1}\mathbf Zv_i\right]}
 {\det(\Id-\mathbf ZG)^{d_\perp/2}}.
 \label{eq:four-point-Tarasov-result}
\end{equation}
\subsection{Reduction to finite matrices}

The matrix entering Eq.~\eqref{eq:four-point-Tarasov-result} has block form
\begin{equation}
 \Id-\mathbf ZG=
 \begin{pmatrix}\Id_L&-Zg(b)\\-Zg(a)&\Id_L\end{pmatrix}.
 \label{eq:four-point-block-matrix}
\end{equation}
Introducing
\begin{equation}
 K=\Id_L-Zg(b)Zg(a),\qquad
 \widetilde K=\Id_L-Zg(a)Zg(b),
 \label{eq:four-point-Ks}
\end{equation}
block elimination and \(\det(\Id-AB)=\det(\Id-BA)\) give
\begin{equation}
 \det(\Id-\mathbf ZG)=\det K=\det\widetilde K,
 \label{eq:four-point-det-reduction}
\end{equation}
and
\begin{equation}
 (\Id-\mathbf ZG)^{-1}\mathbf Z=
 \begin{pmatrix}
 K^{-1}Zg(b)Z&K^{-1}Z\\
 \widetilde K^{-1}Z&\widetilde K^{-1}Zg(a)Z
 \end{pmatrix}.
 \label{eq:four-point-block-inverse}
\end{equation}
The four-point calculation reduces to three scalar matrix functions:
\begin{align}
 \mathcal A(z)&=f(a)^TK^{-1}Zf(b),
 \label{eq:A-def}\\
 \mathcal S(z)&=
 \frac{f(a)^TK^{-1}Zg(b)Zf(a)}{b(1+b)}
 =\frac{f(b)^T\widetilde K^{-1}Zg(a)Zf(b)}{a(1+a)},
 \label{eq:S-def}\\
 \mathcal C(z)&=-\log\det K.
 \label{eq:C-def}
\end{align}
The expansion $K^{-1}=\sum_{r\geq0}(Zg(b)Zg(a))^r$ sums chains of
alternating cubic pairings and sewing contractions. The chains
contributing to $\mathcal A$ and $\mathcal S$ end on sources.
Closed chains are summed by
$-\log\det K=\sum_{r\geq1}\operatorname{Tr}[(Zg(b)Zg(a))^r]/r$.
These are Wick contractions on the disk and do not change the
worldsheet topology.
Appendix~\ref{app:four-point-identities-proof} resums the open and
closed chains to all orders.

The source exponent in the numerator of Eq.~\eqref{eq:four-point-Tarasov-result} reduces to
\begin{equation}
 \frac12\sum_{i=1}^{d_\perp}
 v_i^T(\Id-\mathbf ZG)^{-1}\mathbf Zv_i
 =C_N\mathcal A+\frac12
 \bigl[b(1+b)U_N+a(1+a)V_N\bigr]\mathcal S.
 \label{eq:four-point-source-reduction}
\end{equation}
Writing $X=-2\alpha'k_2\cdot k_3=\alpha't+2$, the on-shell
contractions follow from
\begin{equation}
  k_2^2=k_3^2=\frac{1}{\alpha'},\qquad
  P\cdot k_2=\frac{P^2}{2},\qquad
  P\cdot k_3=-\frac{P^2}{2},\qquad
  2\alpha'k_2\cdot k_3=-X .
\end{equation}
Since $k_2+aP$ and $k_3-bP$ have vanishing plus components,
their Lorentzian scalar products equal their transverse scalar
products. Thus,
\begin{align}
  J_L\cdot J_R
    &=-2\alpha'(k_2+aP)\cdot(k_3-bP)
     =X+\alpha'P^2(a+b+2ab),\\
  J_L^2
    &=2\alpha'(k_2+aP)^2
     =2\bigl[1+\alpha'P^2a(1+a)\bigr],\\
  J_R^2
    &=2\alpha'(k_3-bP)^2
     =2\bigl[1+\alpha'P^2b(1+b)\bigr].
\end{align}
Substitution of $\alpha'P^2=1-N$ gives
\begin{align}
 C_N&=\sum_iJ_L^iJ_R^i
 =X-(N-1)(a+b+2ab),
 \nonumber\\
 U_N&=\sum_iJ_L^iJ_L^i
 =2\bigl[1-(N-1)a(1+a)\bigr],
 \nonumber\\
 V_N&=\sum_iJ_R^iJ_R^i
 =2\bigl[1-(N-1)b(1+b)\bigr],
 \qquad X=\alpha't+2.
 \label{eq:four-point-current-contractions}
\end{align}
For this four-point calculation, we define
\begin{equation}
 \alpha=a+b+2ab,\qquad q=a(1+a)b(1+b).
 \label{eq:alpha-q}
\end{equation}
Then
\begin{equation}
 \frac12\bigl[b(1+b)U_N+a(1+a)V_N\bigr]
 =a(1+a)+b(1+b)-2(N-1)q.
 \label{eq:weighted-U-V}
\end{equation}
Substitution in Eq.~\eqref{eq:four-point-Tarasov-result} gives the exact
finite-matrix result
\begin{align}
 R_N&=[z^N]\mathcal G_N^{(d_\perp)}(z),
 \nonumber\\
 \mathcal G_N^{(d_\perp)}(z)
 &=(\det K)^{-d_\perp/2}
 \exp\!\left\{
 X\mathcal A+[a(1+a)+b(1+b)]\mathcal S
 -(N-1)[\alpha\mathcal A+2q\mathcal S]\right\}.
 \label{eq:four-point-final-matrix}
\end{align}
Equivalently,
\begin{equation}
 \log\mathcal G_N^{(d_\perp)}
 =X\mathcal A+[a(1+a)+b(1+b)]\mathcal S
 +\frac{d_\perp}{2}\mathcal C
 -(N-1)[\alpha\mathcal A+2q\mathcal S].
 \label{eq:four-point-log-generator}
\end{equation}
The inverses and determinants of these finite matrices can be evaluated
as rational functions of $z$, or directly in the ring
$\mathbb Q(a,b,X)[z]/(z^{N+1})$. This ring keeps precisely the terms needed
for $R_N$. At the critical dimension, the change of variable below
removes the auxiliary light-cone dependence.

For the all-order analysis, write the finite-cutoff functions as
$\mathcal A_{[L]},\mathcal S_{[L]},\mathcal C_{[L]}$.
Their coefficients of $z^k$ are independent of $L$ once $L\geq k$.
The symbols $\mathcal A,\mathcal S,\mathcal C$ below denote the formal
series defined by these stable coefficients. An identity through
$z^k$ involving $\log(\chi/z)$ or $\chi'$ requires $\chi$ through
$z^{k+1}$, for which $L\geq k+1$ suffices.

\subsection{Low-level residues}

The first matrix coefficients are
\begin{align}
 \mathcal A(z)
 &=z+\left(\frac12+\alpha\right)z^2
 +\left[\frac13+\frac32\alpha(1+\alpha)+q\right]z^3+O(z^4),
 \label{eq:A-expansion}\\
 \mathcal S(z)
 &=\frac12z^2+\frac23(1+2\alpha)z^3+O(z^4),
 \label{eq:S-expansion}\\
 \mathcal C(z)
 &=\frac q4z^2+\frac{4q}{9}(1+2\alpha)z^3+O(z^4).
 \label{eq:C-expansion}
\end{align}
Equation~\eqref{eq:four-point-log-generator} then gives
\begin{align}
 R_0&=1,\qquad R_1=X,
 \nonumber\\
 R_2&=\frac{X(X+1)}2+\frac{d_\perp-24}{8}q,
 \nonumber\\
 R_3&=\frac{X(X+1)(X+2)}6
 +\frac{d_\perp-24}{72}q(9X+14\alpha+16).
 \label{eq:four-point-low-residues}
\end{align}
At $d_\perp=24$, the displayed coefficients are $(X)_N/N!$ and no
longer depend on the auxiliary ratios. Away from the critical
dimension, the formal corrections depend on $a,b$ through $q$ and
$\alpha$. We now show how the cancellation extends to every level.

\subsection{Critical reparametrization and all-level residue}
\label{subsec:chi-reparametrization}

At \(d_\perp=24\), define the formal series
\begin{equation}
 \chi(z)=1-e^{-\mathcal A(z)}.
 \label{eq:chi-def}
\end{equation}
Since $\mathcal A(z)=z+O(z^2)$, we have $\chi(z)=z+O(z^2)$, so this
change of variable has a formal inverse. The two identities needed
for the residue are
\begin{align}
 \alpha\mathcal A+2q\mathcal S&=\log\frac{\chi}{z},
 \label{eq:chi-identity-one}\\
 [a(1+a)+b(1+b)]\mathcal S+12\mathcal C
 &=\log\chi'-2\log\frac{\chi}{z}.
 \label{eq:chi-identity-two}
\end{align}
Appendix~\ref{app:four-point-identities-proof} proves both identities by
evaluating the open chains with the cubic local maps and the closed chains
with the one-boson Ward identity. Substitution into
Eq.~\eqref{eq:four-point-log-generator} gives
\begin{equation}
 \log\mathcal G_N^{(24)}
 =-X\log(1-\chi)+\log\chi'-(N+1)\log\frac\chi z,
 \label{eq:critical-log-G}
\end{equation}
or
\begin{equation}
 \mathcal G_N^{(24)}(z)
 =\chi'(z)\left(\frac z{\chi(z)}\right)^{N+1}
  (1-\chi(z))^{-X}.
 \label{eq:critical-G}
\end{equation}
Formal residue invariance under \(\chi=z+O(z^2)\) gives
\begin{align}
 \left.R_N\right|_{d_\perp=24}
 &=\Res_{z=0}\frac{\mathcal G_N^{(24)}(z)}{z^{N+1}}\dd z
 =\Res_{z=0}\frac{\chi'(z)}{\chi(z)^{N+1}}
 (1-\chi(z))^{-X}\dd z
 \nonumber\\
 &=\Res_{\chi=0}\frac{(1-\chi)^{-X}}{\chi^{N+1}}\dd\chi
 =[\chi^N](1-\chi)^{-X}
 =\frac{(X)_N}{N!}.
 \label{eq:critical-residue}
\end{align}
The change to $\chi$ absorbs the light-cone ratios into the residue
measure. The determinant and source exponent both enter this
cancellation: the $24$ transverse bosons supply the closed-chain term
in the Jacobian identity \eqref{eq:chi-identity-two}, and the
level-dependent factor supplies the pole measure.

The fixed-level oscillator sum of ref.~\cite{FotopoulosVeneziano}
also gives the numerator in Eq.~\eqref{eq:critical-residue}.
In that work, a level-$N$ pole in the $t$ channel has coefficient
$(\alpha's+2)_N/N!$. Exchanging $s$ and $t$ yields our convention,
$R_N=(X)_N/N!$ with $X=\alpha't+2$.
\section{Finite-matrix sewing for arbitrary planar trees}
\label{sec:arbitrary-tree-sewing}

Let $\mathcal T=(V,E)$ be a planar cubic tree with $M$ external
tachyons, cubic vertices $V$ and $|E|=M-3$ internal edges. Assign a
physical level $N_e\geq0$ to each edge. We suppress the overall coupling
and momentum-conserving delta function and use the reduced cubic
normalization described above. Unless stated otherwise, matrices act
on both mode and transverse indices. The construction includes
vertices with one, two or three internal DDF legs.

For a residue at prescribed levels, the mode cutoffs can be set to
$L_e=N_e$. To study formal coordinate maps, however, the cutoffs must
be varied independently of the physical levels until each coefficient
stabilizes. At $N_e=0$, the minimal cutoff removes the oscillator
block but keeps the tachyonic channel momentum and its mass-shell
condition. Empty determinants and products are one.

\subsection{Tree data, coloring, and BPZ orientation}

At every vertex \(v \in V\), choose the cyclic order induced by the planar embedding
and take all three momenta to be outgoing.  If \(e\) joins \(v\) and \(v'\),
then
\begin{equation}
 p_{v,e}=-p_{v',e},\qquad
 \sum_{a\,\text{incident on }v}p_{v,a}=0.
 \label{eq:tree-momenta}
\end{equation}
Color the vertices of the tree alternately, giving a bipartition
$V=V_L\sqcup V_R$. Orient each internal edge as $L(e)\to R(e)$ and
choose the oscillator basis at $R(e)$ to be BPZ dual to that at
$L(e)$:
\begin{equation}
 \bigl\langle A^i_{e,n;L},A^j_{e,m;R}\bigr\rangle_{\rm BPZ}
 =n\delta_{nm}\delta^{ij}.
 \label{eq:bpz-dual-basis}
\end{equation}
We distinguish the cyclic local-vertex convention from the
choice of BPZ-dual half-edge bases. In the local oscillator
coordinate, the standard weight-one reflector has phase
\begin{equation}
  \kappa_n=(-1)^{n+1}.
\end{equation}
Before choosing BPZ-dual bases, its inverse-norm sewing block is
\begin{equation}
  Z^{\mathrm{cyc}}_e
    =\operatorname{diag}_{1\leq n\leq L_e}
       \left(\kappa_n\frac{\zeta_e^n}{n}\right)
       \otimes I_\perp .
\end{equation}

Choosing half-edge phases $\theta_{v,e}(n)\in\{+1,-1\}$ satisfying
\begin{equation}
  \theta_{L(e),e}(n)\theta_{R(e),e}(n)=\kappa_n ,
  \label{eq:halfedge-phase-condition}
\end{equation}
and putting
\begin{equation}
  \Theta_{v,e}
    =\operatorname{diag}_{1\leq n\leq L_e}
       \bigl(\theta_{v,e}(n)\bigr)\otimes I_\perp,
  \qquad
  \Theta_v=\bigoplus_{e\in I_v}\Theta_{v,e},
\end{equation}
the stripped local data in these bases are
\begin{equation}
  \widehat j_v=\Theta_v j_v^{\mathrm{cyc}},
  \qquad
  \widehat G_v=\Theta_v G_v^{\mathrm{cyc}}\Theta_v .
  \label{eq:halfedge-rephasing}
\end{equation}
Consequently,
\begin{equation}
  \Theta_{L(e),e}^{-1}Z^{\mathrm{cyc}}_e
    \Theta_{R(e),e}^{-1}
    =
    \operatorname{diag}_{1\leq n\leq L_e}
      \left(\frac{\zeta_e^n}{n}\right)\otimes I_\perp .
\end{equation}
The BPZ-dual sewing block therefore contains the positive inverse
oscillator norms. The coloring fixes the matrix assembly, while the
half-edge phases specify the local oscillator bases.

\subsection{Local cubic vertex}

Let $I_v\subset E$ be the internal edges incident on $v$; for $M\geq4$,
$|I_v|$ is one, two or three. In the cyclic ordering fixed by the planar
embedding, let $s_v(e)$ be the successor of an internal half-edge $e$
among all three outgoing legs, including external tachyons.
Before rephasing the oscillator bases, the cyclic vertex coefficients
are
\begin{equation}
\begin{split}
\rho_{v,e}^{\mathrm{cyc}}
&=
\frac{p^+_{v,s_v(e)}}{p^+_{v,e}},
\\
J_{v,e;n}^{\mathrm{cyc},i}
&=
\sqrt{2\alpha'}\,\varepsilon^i_{v,e;(n)}
\cdot
\left(
p_{v,s_v(e)}
-\rho_{v,e}^{\mathrm{cyc}}p_{v,e}
\right).
\end{split}
 \label{eq:local-ratio-current}
\end{equation}
The vector contracted with $\varepsilon^i_{v,e;(n)}$ has vanishing plus
component. In the scalar-mode formulas below, the transverse bases
are independent of $n$, so
$J_{v,e;n}^{\mathrm{cyc},i}=J_{v,e}^{\mathrm{cyc},i}$.
Appendix~\ref{app:transverse-basis} treats orthogonal changes of these
bases.

The linear source and same-puncture coefficients of the DDF Reggeon,
with local-coordinate factors removed, are
\begin{equation}
 f_n(\rho)=(-1)^{n-1}\prod_{q=1}^{n-1}\frac{n\rho+q}{q},\qquad
 g_{nm}(\rho)=\rho(1+\rho)\frac{nm}{n+m}f_n(\rho)f_m(\rho),
 \label{eq:fg-general-tree}
\end{equation}
and the ordered distinct-puncture coefficient is
\begin{equation}
 b_{nm}(\rho)=(-1)^{m-1}
 \frac{nm(1+\rho)}{n\rho+m}f_n(\rho)f_m(1/\rho),
 \label{eq:b-general-tree}
\end{equation}
where an apparent pole at $n\rho+m=0$ is evaluated after cancellation.
At $\rho=-m/n$, the denominator cancels against the $q=m$ factor of
$f_n(\rho)$ for $m<n$, the $q=n$ factor of $f_m(1/\rho)$ for $n<m$,
or $1+\rho$ for $n=m$. It therefore imposes no additional restriction
on the modes or on the allowed nonzero plus-momentum ratios.

We write
\begin{equation}
R_{v;ef}^{ij}=\varepsilon_{v,e}^{i}\cdot\varepsilon_{v,f}^{j}.
 \label{eq:frame-overlap}
\end{equation}
The BPZ-oriented local source and symmetric quadratic coupling matrix are
\begin{align}
(\widehat j_{v,e})_{ni}
    &=
    \theta_{v,e}(n)
    f_n(\rho_{v,e}^{\mathrm{cyc}})
    J_{v,e}^{\mathrm{cyc},i},\\
  (\widehat G_{v;ee})_{ni,mj}
    &=
    \theta_{v,e}(n)\theta_{v,e}(m)
    g_{nm}(\rho_{v,e}^{\mathrm{cyc}})\delta^{ij},\\
  (\widehat G_{v;ef})_{ni,mj}
    &=
    \theta_{v,e}(n)\theta_{v,f}(m)
    b_{nm}\!\left(\frac{p_{v,f}^+}{p_{v,e}^+}\right)
    R_{v;ef}^{ij},
    \qquad
    \widehat G_{v;fe}=\widehat G_{v;ef}^{\,T}.
 \label{eq:local-G}
\end{align}
For each pair $\{e,f\}$ of internal legs, choose the cyclic orientation
$f=s_v(e)$. The reverse block is its transpose.

For the comb vertex $(P,k,-Q)$, where $Q=P+k$, use
\[
 c=\frac{k^+}{P^+},\qquad e=-\frac{c}{1+c},\qquad
 \theta_P(n)=-1,\qquad \theta_{-Q}(m)=(-1)^m.
\]
The reflection identity $f_n(-1-\rho)=(-1)^{n-1}f_n(\rho)$ converts
the cyclic successor data into the comb sources and same-leg blocks.
The mixed block becomes
\begin{equation}
 (\widehat G_{P,-Q})_{nm}=h_{mn}(c)
 =(-1)^{m+1}b_{mn}\!\left(-\frac{1}{1+c}\right)
 =(-1)^{n+1}b_{nm}(-1-c).
 \label{eq:h-from-b}
\end{equation}
Appendix~\ref{app:cubic-coordinate-factors} gives the source conversion
explicitly. The same basis phases must be used in every block of the
vertex.
The coordinate-stripped cubic vertex is
\begin{equation}
 \widehat V_v(\Lambda_v)=
 \exp\!\left(\Lambda_v^T\widehat j_v
 +\frac12\Lambda_v^T\widehat G_v\Lambda_v\right).
 \label{eq:local-gaussian}
\end{equation}
Wick's theorem fixes this Gaussian for any oscillator occupation on
the three DDF legs. In a common light-cone frame, the longitudinal
exponentials contract neither with one another nor with transverse
derivatives. Only linear and quadratic source terms remain, with
coefficients $f$, $g$ and $b$. They are the single- and double-contour
coefficients of the three-DDF Reggeon derived in sections 4 and 5 of
ref.~\cite{BiswasMarottaPesando2025}.
At fixed physical levels, including the seed factor and ghosts gives
the complete on-shell cubic amplitude.

\subsection{Retaining the puncture-coordinate matrices}
\label{subsec:punc_coord_mats}

For an ordered local vertex \((a,b,c)\), put
\(\beta_{v,a}=\sqrt{2\alpha'}p^+_{v,a}\),
\(x_{v,ab}=x_{v,a}-x_{v,b}\), and
\begin{equation}
 \Xi_v=x_{v,ac}^{\beta_{v,b}}x_{v,ab}^{\beta_{v,c}}
 x_{v,bc}^{\beta_{v,a}},\qquad
 \sigma_{v,a}=\epsilon_{v,a}\Xi_v^{1/\beta_{v,a}}.
 \label{eq:Xi-sigma-tree}
\end{equation}
All roots at a given vertex use the same puncture coordinates and
branch convention. The phase $\epsilon_{v,a}$ fixes the local-coordinate
orientation and contributes its $n$-th power at mode $n$. The
oscillator-basis phases $\theta_{v,e}(n)$ have already been included in
the hatted data. The coordinate matrices on an internal half-edge are
\begin{equation}
 S_{v,e}=\diag_{1\leq n\leq L_e}(\sigma_{v,e}^{n})\otimes I_\perp,
 \qquad S_v=\bigoplus_{e\in I_v}S_{v,e}.
 \label{eq:half-edge-S}
\end{equation}
Restoring the local-coordinate factors gives the congruence transformations
\begin{equation}
 j_v^{(x)}=S_v\widehat j_v,\qquad
 G_v^{(x)}=S_v\widehat G_vS_v,\qquad
 V_v^{(x)}=\exp\!\left(
 \Lambda_v^Tj_v^{(x)}+\frac12\Lambda_v^TG_v^{(x)}\Lambda_v\right).
 \label{eq:local-S-factorization}
\end{equation}
At a three-DDF vertex, the three diagonal blocks and three
distinct-puncture blocks must all use the puncture gauge of
Eq.~\eqref{eq:Xi-sigma-tree} for this factorization to hold.

\paragraph{Puncture independence at fixed levels.}
Let $N_{v,r}$ be the physical level on outgoing leg $r\in\{a,b,c\}$,
with $N_{v,r}=0$ for an external tachyon. Denote the seed-tachyon
Koba--Nielsen factor by $\mathcal K_{T,v}$. Its coordinate powers combine
with the level-dependent oscillator factors and the three ghosts as
\begin{align}
 \mathcal K_{T,v}
 &=\frac{\Xi_v^{-\sum_rN_{v,r}/\beta_{v,r}}}
 {x_{v,ab}x_{v,ac}x_{v,bc}},\nonumber\\
 \mathcal K_{T,v}\left(\prod_r\sigma_{v,r}^{N_{v,r}}\right)
 \langle c(x_{v,a})c(x_{v,b})c(x_{v,c})\rangle
 &=\prod_r\epsilon_{v,r}^{N_{v,r}}.
 \label{eq:fixed-level-coordinate-cancellation}
\end{align}
Appendix~\ref{app:cubic-coordinate-factors} derives the cancellation
from the relation between physical and seed momenta. Only constant
orientation phases remain in the fixed-level cubic amplitude.
The matrices $S_{v,e}$ retain the oscillator coordinate factors during
sewing; section~\ref{subsec:reduced_mat_res} shows their cancellation
at fixed invariant plumbing variables.

\subsection{Assembly on a cubic tree}

We order the internal edges once and define
\begin{equation}
 \mathcal H_L=\bigoplus_{e\in E}\mathcal H_{L(e),e},\qquad
 \mathcal H_R=\bigoplus_{e\in E}\mathcal H_{R(e),e},\qquad
 \dim\mathcal H_L=\dim\mathcal H_R
 =d_\perp\sum_{e\in E}L_e.
 \label{eq:global-half-edge-spaces}
\end{equation}
The coherent variables are stacked in this order as
\(\Lambda_L=\col_e\lambda_{L(e),e}\) and
\(\Lambda_R=\col_e\lambda_{R(e),e}\).  Let \(\iota_v^{L,R}\) be the
injection matrix placing the local blocks labelled by \(I_v\) into the
common edge ordering. The sources and quadratic coupling matrices are
\begin{align}
 u^{(x)}&=\sum_{v\in V_L}\iota_v^Lj_v^{(x)},&
 G_L^{(x)}&=\sum_{v\in V_L}\iota_v^LG_v^{(x)}(\iota_v^L)^T,
 \nonumber\\
 w^{(x)}&=\sum_{v\in V_R}\iota_v^Rj_v^{(x)},&
 G_R^{(x)}&=\sum_{v\in V_R}\iota_v^RG_v^{(x)}(\iota_v^R)^T.
 \label{eq:incidence-assembly}
\end{align}
Each color contains one half-edge for every internal edge. Defining
\begin{equation}
 S_L=\bigoplus_{e\in E}S_{L(e),e},\qquad
 S_R=\bigoplus_{e\in E}S_{R(e),e},
 \label{eq:global-Ss}
\end{equation}
Eq.~\eqref{eq:incidence-assembly} implies
\begin{equation}
 u^{(x)}=S_L\widehat u,\qquad w^{(x)}=S_R\widehat w,\qquad
 G_L^{(x)}=S_L\widehat G_LS_L,\qquad
 G_R^{(x)}=S_R\widehat G_RS_R.
 \label{eq:global-S-factorization}
\end{equation}
Accordingly,
\begin{equation}
 \prod_{v\in V}V_v^{(x)}
 =\exp\!\left[
 \Lambda_L^Tu^{(x)}+\frac12\Lambda_L^TG_L^{(x)}\Lambda_L
 +\Lambda_R^Tw^{(x)}+\frac12\Lambda_R^TG_R^{(x)}\Lambda_R\right].
 \label{eq:global-vertex-product}
\end{equation}
The incidence maps and ordered distinct-puncture blocks contain all
dependence on the tree topology.

\subsection{BPZ sewing and loop-hafnian resummation}

At fixed invariant plumbing coordinates \(\zeta_e\), define
\begin{equation}
 \widehat Z=\bigoplus_{e\in E}\left[
 \diag_{1\leq n\leq L_e}\!\left(\frac{\zeta_e^n}{n}\right)
 \otimes I_\perp\right],
 \qquad Z_x=S_L^{-1}\widehat ZS_R^{-1}.
 \label{eq:coordinate-BPZ-sewing}
\end{equation}
Equivalently, if \(z_e\) is the bare local sewing parameter, then
\begin{equation}
 \zeta_e=\sigma_{L(e),e}\sigma_{R(e),e}z_e.
 \label{eq:invariant-plumbing}
\end{equation}
The product of the half-edge scales is absorbed into $\zeta_e$. The
simultaneous sewing operator is
\begin{equation}
 \mathcal S_{\mathcal T}
 =\exp\!\left((\partial_{\Lambda_L})^TZ_x
 \partial_{\Lambda_R}\right)
 =\exp\!\left[\frac12\partial_\Lambda^T
 \begin{pmatrix}0&Z_x\\Z_x^T&0\end{pmatrix}
 \partial_\Lambda\right].
 \label{eq:tree-sewing-operator}
\end{equation}
Setting
\begin{equation}
 \mathbf j_x=\binom{u^{(x)}}{w^{(x)}},\qquad
 \mathbf G_x=\begin{pmatrix}G_L^{(x)}&0\\0&G_R^{(x)}\end{pmatrix},
 \qquad
 \mathbf Z_x=\begin{pmatrix}0&Z_x\\Z_x^T&0\end{pmatrix},
 \label{eq:bold-global-data}
\end{equation}
the Gaussian identity gives the formal series
\begin{equation}
 \mathcal G_{\mathcal T}^{(x)}(\bm\zeta)=
 \frac{\exp\!\left[
 \frac12\mathbf j_x^T(\Id-\mathbf Z_x\mathbf G_x)^{-1}
 \mathbf Z_x\mathbf j_x\right]}
 {\sqrt{\det(\Id-\mathbf Z_x\mathbf G_x)}}.
 \label{eq:Tarasov-full-tree}
\end{equation}

\subsection{Reduced matrices and cancellation of coordinate factors}
\label{subsec:reduced_mat_res}
We introduce
\begin{equation}
 K_L^{(x)}=\Id-Z_xG_R^{(x)}Z_x^TG_L^{(x)},\qquad
 K_R^{(x)}=\Id-Z_x^TG_L^{(x)}Z_xG_R^{(x)}.
 \label{eq:KLR-retained}
\end{equation}
Block elimination gives
\begin{equation}
 \det(\Id-\mathbf Z_x\mathbf G_x)
 =\det K_L^{(x)}=\det K_R^{(x)},
 \label{eq:tree-determinant-reduction}
\end{equation}
and
\begin{equation}
 (\Id-\mathbf Z_x\mathbf G_x)^{-1}\mathbf Z_x=
 \begin{pmatrix}
 (K_L^{(x)})^{-1}Z_xG_R^{(x)}Z_x^T&(K_L^{(x)})^{-1}Z_x\\
 (K_R^{(x)})^{-1}Z_x^T&(K_R^{(x)})^{-1}Z_x^TG_L^{(x)}Z_x
 \end{pmatrix}.
 \label{eq:tree-block-inverse}
\end{equation}
The source exponent is therefore
\begin{align}
 \mathcal E_x={}&(u^{(x)})^T(K_L^{(x)})^{-1}Z_xw^{(x)}
 \nonumber\\
 &+\frac12(u^{(x)})^T(K_L^{(x)})^{-1}
 Z_xG_R^{(x)}Z_x^Tu^{(x)}
 \nonumber\\
 &+\frac12(w^{(x)})^T(K_R^{(x)})^{-1}
 Z_x^TG_L^{(x)}Z_xw^{(x)}.
 \label{eq:tree-source-exponent}
\end{align}

Substituting Eqs.~\eqref{eq:global-S-factorization} and
\eqref{eq:coordinate-BPZ-sewing}, without commuting \(S_L\) or \(S_R\)
through either Gaussian matrix, gives
\begin{align}
 K_L^{(x)}
 &=\Id-S_L^{-1}\widehat ZS_R^{-1}
 (S_R\widehat G_RS_R)S_R^{-1}\widehat ZS_L^{-1}
 (S_L\widehat G_LS_L)
 \nonumber\\
 &=S_L^{-1}(\Id-\widehat Z\widehat G_R\widehat Z\widehat G_L)S_L
 \equiv S_L^{-1}\widehat K_LS_L,
 \nonumber\\
 K_R^{(x)}&=S_R^{-1}\widehat K_RS_R.
 \label{eq:K-similarity}
\end{align}
Hence,
\begin{equation}
 \det K_{L,R}^{(x)}=\det\widehat K_{L,R},\qquad
 (K_L^{(x)})^{-1}=S_L^{-1}\widehat K_L^{-1}S_L,\qquad
 (K_R^{(x)})^{-1}=S_R^{-1}\widehat K_R^{-1}S_R.
 \label{eq:K-inverse-determinant}
\end{equation}
The coordinate factors cancel separately in each of the three source contractions:
\begin{align}
 (u^{(x)})^T(K_L^{(x)})^{-1}Z_xw^{(x)}
 &=\widehat u^T\widehat K_L^{-1}\widehat Z\widehat w,
 \nonumber\\
 (u^{(x)})^T(K_L^{(x)})^{-1}Z_xG_R^{(x)}Z_x^Tu^{(x)}
 &=\widehat u^T\widehat K_L^{-1}
 \widehat Z\widehat G_R\widehat Z\widehat u,
 \nonumber\\
 (w^{(x)})^T(K_R^{(x)})^{-1}Z_x^TG_L^{(x)}Z_xw^{(x)}
 &=\widehat w^T\widehat K_R^{-1}
 \widehat Z\widehat G_L\widehat Z\widehat w.
 \label{eq:source-S-cancellation}
\end{align}
Thus, \(\mathcal E_x=\widehat {\mathcal E}\) and
\begin{equation}
 \boxed{\mathcal G_{\mathcal T}^{(x)}(\bm\zeta)
 =\widehat{\mathcal G}_{\mathcal T}(\bm\zeta)},\qquad
 \boxed{\mathcal R_{\{N_e\}}^{(M;\mathcal T)}
 =\left[\prod_{e\in E}\zeta_e^{N_e}\right]
 \frac{e^{\widehat {\mathcal E}}}{\sqrt{\det\widehat K_L}}}.
 \label{eq:final-arbitrary-tree-residue}
\end{equation}
The sources in Eq.~\eqref{eq:final-arbitrary-tree-residue} must be
evaluated on the joint mass shell $P_e^2=(1-N_e)/\alpha'$ before
extracting the coefficient. Thus changing the physical levels changes
$\widehat{\mathcal G}_{\mathcal T}$ through its current contractions.
In section~\ref{subsec:generality-proof} we show that the extracted
coefficient equals the complete physical-state sum for any planar
cubic tree and finite level assignment.

Restoring the cubic coupling in the convention of
Eq.~\eqref{eq:four-point-pole-form}, and writing $s_e=-P_e^2$, the
simultaneous pole contribution is
\begin{equation}
 \mathcal V_M=
 \frac{(-1)^{M-3}g_o^{M-2}\mathcal R_{\{N_e\}}^{(M;\mathcal T)}}
 {\displaystyle\prod_{e\in E}[\alpha's_e-(N_e-1)]}
 +\text{terms with fewer of the selected poles}.
 \label{eq:tree-pole-normalization}
\end{equation}
The matrix formula computes this dimensionless pole coefficient.
Taking the iterated residue in the dimensionful variables $s_e$
introduces a further factor $(\alpha')^{-(M-3)}$.

In practice, one assembles the local blocks using the incidence maps
and evaluates the inverse, trace logarithm and source exponential
with $\zeta_e^{N_e+1}=0$ on each edge. We next prove that this finite
calculation gives the full physical-state sum.

For a comb, an end vertex has \(|I_v|=1\), an interior vertex has
\(|I_v|=2\), and the incidence sums produce alternating sparse matrices.
For the snowflake, we use $\theta_{v,-E}(n)=(-1)^n$ at each
outer vertex $(k_{2r-1},k_{2r},-E)$, and
$\theta_{\star,P}(n)=\theta_{\star,Q}(n)
   =\theta_{\star,R}(n)=-1$
at the ordered central vertex $(P,Q,R)$.
These choices satisfy the edge condition and give
\begin{equation}
 \widehat w_\star=-\col\bigl(f(r)C_P,f(s)C_Q,f(t)C_R\bigr),
 \label{eq:snowflake-BPZ-source}
\end{equation}
whereas its dense three-leg matrix is unchanged and contains the ordered
blocks \(B^{PQ}\), \(B^{QR}\), and \(B^{RP}\).

Each central half-edge of the snowflake occurs in two
distinct-puncture blocks. Both occurrences must carry the same phase
and coordinate scale, as prescribed in
section~\ref{subsec:punc_coord_mats}.

\subsection{Generality, exact truncation, and domain of validity}
\label{subsec:generality-proof}

We now prove Eq.~\eqref{eq:final-arbitrary-tree-residue} for the
color-ordered tachyon disk amplitude at $D=26$, for any finite
$M\geq4$, planar cubic tree $\mathcal T$ and levels $N_e\geq0$.
Work first at a generic point of the joint mass shell
$\alpha'P_e^2=1-N_e$, away from other poles, where the selected channel
denominators are independent local coordinates. Meromorphic
continuation then extends the result.

\paragraph{From factorization to the state sum.}
For each internal edge define the finite set of DDF occupations
\begin{equation}
 \begin{split}
 \mathfrak B_e(N_e)&=\left\{\mathbf r_e=(r_{e,ni}):\quad
 r_{e,ni}\in\mathbb N_0,\quad
 \sum_{n\geq1}\sum_{i=1}^{24}n r_{e,ni}=N_e\right\},
 \\
 \nu(\mathbf r_e)&=\prod_{n,i}n^{r_{e,ni}}r_{e,ni}! .
 \end{split}
 \label{eq:tree-occupation-set}
\end{equation}
Let $C_v(\{\mathbf r_e\}_{e\in I_v})$ be the color-ordered on-shell
cubic amplitude with its coupling suppressed, in the fixed BPZ-dual
bases and local conventions above. It is obtained from
Eq.~\eqref{eq:local-gaussian} as
\begin{equation}
 C_v(\{\mathbf r_e\})=
 \left.
 \left(\prod_{e\in I_v}\prod_{n,i}
 \partial_{\lambda_{v,e,ni}}^{r_{e,ni}}\right)
 \widehat V_v(\Lambda_v)\right|_{\Lambda_v=0}.
 \label{eq:tree-cubic-component}
\end{equation}
The physical levels and momenta are fixed before differentiation.
The seed factor and ghosts are included as in
section~\ref{subsec:punc_coord_mats}; constant orientation phases are
already absorbed into the vertex bases and dual sewing metric.
They must therefore be used consistently in the local blocks, without
an additional phase factor. Channel momentum assignments enforce the
momentum delta functions.

Factorization on an internal edge inserts the physical-state
projector at its prescribed level. Applying the same step to the two
subtrees reduces the residue to cubic amplitudes. In these on-shell
matrix elements, Eq.~\eqref{eq:physical-ddf-projector} replaces each
projector by its DDF representation, since null and BRST-exact states
decouple. The dimensionless simultaneous pole coefficient is
therefore
\begin{equation}
 \mathcal R_{\{N_e\}}^{(M;\mathcal T)}
 =\sum_{\substack{\mathbf r_e\in\mathfrak B_e(N_e)\\ e\in E}}
 \frac{\displaystyle\prod_{v\in V}
 C_v(\{\mathbf r_e\}_{e\in I_v})}
 {\displaystyle\prod_{e\in E}\nu(\mathbf r_e)}.
 \label{eq:general-tree-physical-state-sum}
\end{equation}
The induction ends at cubic amplitudes with up to three DDF legs.
Every internal edge joins two vertices and contributes one inverse
norm, summed over all oscillator partitions and polarizations.
Terms missing any selected pole have zero simultaneous residue.
The argument also applies when two adjacent vertices each have three
internal DDF legs.

\paragraph{From the state sum to the finite matrices.}
In the stripped BPZ-dual bases, the right-hand side of
Eq.~\eqref{eq:general-tree-physical-state-sum} is equivalently
\begin{equation}
 \left[\prod_{e\in E}\zeta_e^{N_e}\right]
 \left.
 \exp\!\left(\sum_{e\in E}\sum_{n,i}
 \frac{\zeta_e^n}{n}
 \partial_{\lambda_{L(e),e,ni}}
 \partial_{\lambda_{R(e),e,ni}}\right)
 \prod_{v\in V}\widehat V_v(\Lambda_v)
 \right|_{\Lambda=0}.
 \label{eq:general-tree-differential-state-sum}
\end{equation}
Indeed, choosing $r_{e,ni}$ copies of one derivative pair contributes
$\zeta_e^{n r_{e,ni}}/(n^{r_{e,ni}}r_{e,ni}!)$.
The power of $\zeta_e$ imposes the level constraint in
Eq.~\eqref{eq:tree-occupation-set}, and the derivatives give exactly the
two component amplitudes joined by that edge. There is therefore one
factorial per occupation and no additional symmetry factor for a fixed
color-ordered tree.

The bipartite coloring assembles the local Gaussians into
Eq.~\eqref{eq:global-vertex-product}, with one half-edge of each color
per internal edge. Gaussian contraction, block reduction and
cancellation of the coordinate factors give
Eq.~\eqref{eq:final-arbitrary-tree-residue}.
The phase condition \eqref{eq:halfedge-phase-condition} is solved
edge by edge. Rephasing a half-edge transforms its source, every
incident block and the inverse sewing metric together, so BPZ duality
imposes no further restriction on the tree.

For prescribed levels $\{N_e\}$, an oscillator with $n>N_e$ cannot occur
in $\mathfrak B_e(N_e)$. Such modes can therefore be omitted before
resummation, proving the exact bound $L_e=N_e$. More explicitly, one can work
in the quotient ring
\begin{equation}
 \mathcal A_{\mathbf N}
 =\mathbb F[\zeta_e:e\in E]/(\zeta_e^{N_e+1}:e\in E),
 \qquad
 (\mathbf Z\mathbf G)^{1+\sum_eN_e}=0.
 \label{eq:tree-nilpotent-truncation}
\end{equation}
The coefficient field $\mathbb F$ contains the fixed on-shell
kinematics, and $\mathbf Z,\mathbf G$ are the stripped matrices of
Eq.~\eqref{eq:bold-global-data}. Every entry of $\mathbf Z\mathbf G$
has positive total sewing degree, giving the stated nilpotence bound.
The inverse, trace logarithm and source exponential therefore
terminate in this ring. The determinant has constant term one, which
fixes the square-root branch without a convergence or positivity
assumption.
For $N_e=0$, only the empty occupation array contributes: the edge
carries a tachyon with unit inverse norm, while its momentum and
mass shell still enter the cubic vertices. The proof places no upper
bound on $N_e$ or $M$, although the matrices and coefficient
extraction become more costly as they increase.

\paragraph{Kinematic domain and scope.}
The common light-cone direction must give nonzero plus components
for all momenta in the DDF blocks. For finitely many nonzero leg and
channel momenta, this holds generically: the forbidden choices are a
finite union of hyperplane sections of the null cone.
Special loci can be treated in another frame or by analytic
continuation. The physical projector identity guarantees that the
residue is independent of this auxiliary direction; it does not
require the unprojected matrices in different frames to be similar.
Apparent poles in Eq.~\eqref{eq:b-general-tree} are removed before
evaluation.

Each level assignment has its own joint mass shell, which can be
built recursively from complex on-shell three-point kinematics.
The remaining invariants satisfy momentum conservation and the
$D=26$ Gram relations. A common real Lorentzian scattering region
for all assignments is unnecessary, but the currents must always be
evaluated on the mass shell of the chosen residue.

This proves the formula for the maximal simultaneous residue in all
$M-3$ channels of $\mathcal T$. A residue in fewer channels retains
higher-point subamplitudes, and the full amplitude requires further
input. DDF completeness is used at $D=26$; symbolic $d_\perp$ gives
only the formal oscillator sum. The three-tachyon amplitude provides
the empty-edge base case, with reduced coefficient one.
The higher-point coordinate-map and measure identities considered in
section~\ref{sec:berezin-grunsky} are separate from this proof.

\section{Five-point double residues}
\label{sec:five-point-matrix}
\suppressfloats[t]

Let $N,M$ be the two channel levels. Here and in the six-point examples,
the sewing matrices use invariant DDF sewing variables. When the same
symbols appear in a KN expression, they denote coordinates in the KN
chart. We compare extracted residues; matching the full top forms
also requires the coordinate map discussed in
section~\ref{sec:berezin-grunsky}.

\begin{figure}[tbp]
\centering
\begin{tikzpicture}[font=\small]
  \coordinate (a) at (-2.6,0); \coordinate (b) at (0,0);
  \coordinate (c) at (2.6,0);
  \draw[ddf leg] (-3.65,1) node[above] {$1$} -- (a)
    -- (-3.65,-1) node[below] {$2$};
  \draw[ddf leg] (b) -- (0,-1.3) node[below] {$3$};
  \draw[ddf leg] (3.65,-1) node[below] {$4$} -- (c)
    -- (3.65,1) node[above] {$5$};
  \draw[ddf edge,ddf forward] (a) -- (b)
    node[midway,above=5pt] {$P$} node[midway,below=5pt] {$N,\ x$};
  \draw[ddf edge,ddf backward] (b) -- (c)
    node[midway,above=5pt] {$Q$} node[midway,below=5pt] {$M,\ y$};
  \node[ddf left,label=above:{$v_1$}] at (a) {};
  \node[ddf right,label=above:{$v_2$}] at (b) {};
  \node[ddf left,label=above:{$v_3$}] at (c) {};
  \node[text=DDFblue] at (0,1.15) {$h(c):\ P\leftrightarrow Q$};
  \node at (0,-2.15) {$[x^Ny^M]\,\mathcal F_5$};
\end{tikzpicture}
\caption{The five-point comb with $P=k_1+k_2$ and
$Q=P+k_3=-(k_4+k_5)$. Labels below each internal edge give its physical
level and sewing variable. The three outgoing cubic momentum triples are
$(k_1,k_2,-P)$, $(P,k_3,-Q)$, and $(Q,k_4,k_5)$.
The middle vertex couples the two oscillator sectors through $h(c)$;
the coefficient of $x^Ny^M$ in the sewn Gaussian
$\mathcal F_5=e^{\mathcal E_5}(\det K_5)^{-d_\perp/2}$
gives the double residue in Eq.~\eqref{eq:five-point-final-matrix}.
The $L$ vertices are $v_1,v_3$, and the $R$ vertex is $v_2$;
arrows denote BPZ orientation as in figure~\ref{fig:four-point-channels}.}
\label{fig:five-point-comb}
\end{figure}
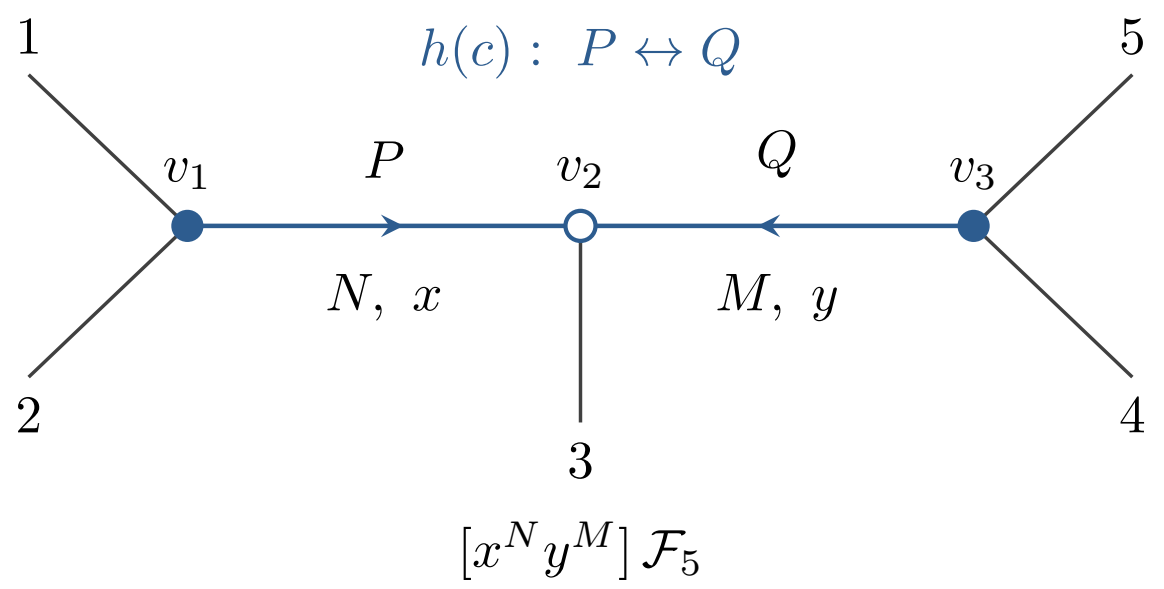

For the five-point comb,
\begin{equation}
 P=k_1+k_2,\qquad Q=P+k_3=-(k_4+k_5),\qquad
 P^2=\frac{1-N}{\alpha'},\qquad Q^2=\frac{1-M}{\alpha'}.
 \label{eq:five-point-matrix-channels}
\end{equation}
Figure~\ref{fig:five-point-comb} shows these two compatible channels and
the associated matrix blocks.

Choose \(V_L=\{v_1,v_3\}\), \(V_R=\{v_2\}\), and order the internal
edges as \((P,Q)\).  The required half-edge ratios are
\begin{equation}
 a=-\frac{k_2^+}{P^+},\qquad
 c=\frac{k_3^+}{P^+},\qquad
 e=-\frac{k_3^+}{Q^+}=-\frac{c}{1+c},\qquad
 b=\frac{k_4^+}{Q^+}.
 \label{eq:five-point-matrix-ratios}
\end{equation}
The BPZ-oriented currents obtained from the local prescription are
\begin{align}
 J_L&=\sqrt{2\alpha'}(k_2+aP)_\perp,
 &J_P&=-\sqrt{2\alpha'}(k_3-cP)_\perp,
 \nonumber\\
 J_Q&=\sqrt{2\alpha'}(k_3+eQ)_\perp=-\frac{J_P}{1+c},
 &J_R&=-\sqrt{2\alpha'}(k_4-bQ)_\perp.
 \label{eq:five-point-matrix-currents}
\end{align}
The signs in this equation already include the BPZ orientation.

Writing
\begin{equation}
 f^{[r]}(\rho)=(f_1(\rho),\ldots,f_r(\rho))^T,\qquad
 g^{[r]}(\rho)=(g_{nm}(\rho))_{n,m\leq r},
 \label{eq:finite-fg-blocks}
\end{equation}
the block coupling the two central DDF legs is the \(M\times N\) matrix
\begin{equation}
 h(c)=(h_{mn}(c)),\qquad
 h_{mn}(c)=f_m(e)f_n(c)\frac{mn c}{n(1+c)-m},
 \label{eq:five-point-h}
\end{equation}
where removable denominator zeros are evaluated after cancelling
common factors. The entries are generally rational in $c$; for example,
\[
  h_{11}=1,\qquad
  h_{21}=\frac{2c}{1+c},\qquad
  h_{12}=-2c .
\]

For each transverse direction, the incidence sums of
Eq.~\eqref{eq:incidence-assembly} give
\begin{equation}
 u_i=\binom{f^{[N]}(a)J_L^i}{f^{[M]}(b)J_R^i},\qquad
 w_i=\binom{f^{[N]}(c)J_P^i}{f^{[M]}(e)J_Q^i}.
 \label{eq:five-point-u-w}
\end{equation}
In a common transverse frame,
\begin{equation}
 G_L^{(0)}=
 \begin{pmatrix}g^{[N]}(a)&0\\0&g^{[M]}(b)\end{pmatrix},\qquad
 G_R^{(0)}=
 \begin{pmatrix}g^{[N]}(c)&h(c)^T\\h(c)&g^{[M]}(e)\end{pmatrix},
 \label{eq:five-point-GL-GR}
\end{equation}
and the full matrices are \(G_{L,R}=G_{L,R}^{(0)}\otimes I_{d_\perp}\).
The sewing matrix is assembled edge by edge:
\begin{equation}
 Z_0=\diag(Z_x,Z_y),\qquad
 Z_x=\diag\!\left(x,\frac{x^2}{2},\ldots,\frac{x^N}{N}\right),\qquad
 Z_y=\diag\!\left(y,\frac{y^2}{2},\ldots,\frac{y^M}{M}\right).
 \label{eq:five-point-Z0}
\end{equation}
The reduced matrices are
\begin{equation}
 K_5=\Id_{N+M}-Z_0G_R^{(0)}Z_0G_L^{(0)},\qquad
 \widetilde K_5=\Id_{N+M}-Z_0G_L^{(0)}Z_0G_R^{(0)}.
 \label{eq:five-point-Ks}
\end{equation}
In blocks,
\begin{equation}
 K_5=
 \begin{pmatrix}
 \Id_N-Z_xg(c)Z_xg(a)&-Z_xh(c)^TZ_yg(b)\\
 -Z_yh(c)Z_xg(a)&\Id_M-Z_yg(e)Z_yg(b)
 \end{pmatrix}.
 \label{eq:five-point-K-explicit}
\end{equation}
The source exponent is
\begin{align}
 \mathcal E_5=\sum_{i=1}^{d_\perp}\biggl[
 &u_i^TK_5^{-1}Z_0w_i
 +\frac12u_i^TK_5^{-1}Z_0G_R^{(0)}Z_0u_i
 \nonumber\\[-1mm]
 &+\frac12w_i^T\widetilde K_5^{-1}Z_0G_L^{(0)}Z_0w_i
 \biggr].
 \label{eq:five-point-E}
\end{align}
The resulting oscillator sum is
\begin{equation}
 \boxed{
 R_{N,M}^{(5,d_\perp)}
 =[x^Ny^M]\frac{\exp(\mathcal E_5)}{(\det K_5)^{d_\perp/2}}.}
 \label{eq:five-point-final-matrix}
\end{equation}
At $d_\perp=24$, this gives the physical double-residue coefficient.
The mode matrices $G_L^{(0)}$, $G_R^{(0)}$, $Z_0$, $K_5$ and $\widetilde K_5$ are
$(N+M)\times(N+M)$, with diagonal blocks of size $N\times N$ and
$M\times M$ and mixed block $h(c)$ of size $M\times N$.

For comparison with section~\ref{subsec:five-point-pattern}, use
$X,Y,W$ from Eq.~\eqref{eq:five-point-direct-invariants} and set
\begin{equation}
 q_P=a(1+a)c(1+c),\qquad
 q_Q=e(1+e)b(1+b).
 \label{eq:five-point-qs}
\end{equation}
Direct coefficient extraction using~\eqref{eq:five-point-final-matrix} gives
\begin{align}
 R_{1,0}^{(5,d_\perp)}&=X,\qquad
 R_{0,1}^{(5,d_\perp)}=Y,\qquad
 R_{1,1}^{(5,d_\perp)}=XY+W,
 \nonumber\\
 R_{2,0}^{(5,d_\perp)}&=\frac{(X)_2}{2}
 +\frac{d_\perp-24}{8}q_P,
 \qquad
 R_{0,2}^{(5,d_\perp)}=\frac{(Y)_2}{2}
 +\frac{d_\perp-24}{8}q_Q.
 \label{eq:five-point-low-checks}
\end{align}
These coefficients agree with Eq.~\eqref{eq:five-point-direct-residue}
at $d_\perp=24$. At the next mixed levels, the KN expansion gives
\begin{equation}
 R_{2,1}^{(5)}=\frac{(X)_2Y}{2}+XW,\qquad
 R_{2,2}^{(5)}=\frac{(X)_2(Y)_2}{4}+XYW+\frac{(W)_2}{2}.
 \label{eq:five-point-higher-benchmarks}
\end{equation}
These follow directly from the three-factor KN expansion
\eqref{eq:five-point-ray-integral} and can be used to test the DDF
formula at higher mixed levels. Their $W$-dependent terms involve
the middle cubic vertex, which is not probed by the single-channel
coefficients in Eq.~\eqref{eq:five-point-low-checks}.

\section{Six-point residues}
\label{sec:six-point}

\begin{figure}[tbp]
\centering
\begin{tikzpicture}[font=\small]
  \begin{scope}[yshift=2.05cm]
    \coordinate (a) at (-3,0); \coordinate (b) at (-1,0);
    \coordinate (c) at (1,0); \coordinate (d) at (3,0);
    \draw[ddf leg] (-4,.85) node[above] {$1$} -- (a)
      -- (-4,-.85) node[below] {$2$};
    \draw[ddf leg] (b) -- (-1,-1.1) node[below] {$3$};
    \draw[ddf leg] (c) -- (1,-1.1) node[below] {$4$};
    \draw[ddf leg] (4,-.85) node[below] {$5$} -- (d)
      -- (4,.85) node[above] {$6$};
    \draw[ddf edge,ddf forward] (a) -- (b)
      node[midway,above=4pt] {$P$} node[midway,below=5pt] {$N_P,\ x$};
    \draw[ddf edge,ddf backward] (b) -- (c)
      node[midway,above=4pt] {$Q$} node[midway,below=5pt] {$N_Q,\ y$};
    \draw[ddf edge,ddf forward] (c) -- (d)
      node[midway,above=4pt] {$R$} node[midway,below=5pt] {$N_R,\ z$};
    \node[ddf left,label=above:{$v_1$}] at (a) {};
    \node[ddf right,label=above:{$v_2$}] at (b) {};
    \node[ddf left,label=above:{$v_3$}] at (c) {};
    \node[ddf right,label=above:{$v_4$}] at (d) {};
    \node at (0,1.35) {$(a)\quad\text{comb}$};
  \end{scope}
  \begin{scope}[yshift=-2.05cm]
    \coordinate (o) at (0,0); \coordinate (p) at (-2.2,.95);
    \coordinate (q) at (0,-1.6); \coordinate (r) at (2.2,.95);
    \draw[ddf leg] (-3.3,1.65) node[above] {$1$} -- (p)
      -- (-3.3,.25) node[left] {$2$};
    \draw[ddf leg] (-.9,-2.35) node[below] {$3$} -- (q)
      -- (.9,-2.35) node[below] {$4$};
    \draw[ddf leg] (3.3,.25) node[right] {$5$} -- (r)
      -- (3.3,1.65) node[above] {$6$};
    \draw[ddf edge,ddf forward] (p) -- (o)
      node[midway,above=5pt] {$P$}
      node[midway,below left=3pt] {$N_P,\ \zeta_P$};
    \draw[ddf edge,ddf forward] (q) -- (o)
      node[midway,right=5pt] {$Q$}
      node[midway,left=5pt] {$N_Q,\ \zeta_Q$};
    \draw[ddf edge,ddf forward] (r) -- (o)
      node[midway,above=5pt] {$R$}
      node[midway,below right=3pt] {$N_R,\ \zeta_R$};
    \node[ddf left] at (p) {}; \node[ddf left] at (q) {};
    \node[ddf left] at (r) {};
    \node[ddf right,label=above:{$v_\star$}] at (o) {};
    \node at (0,2.05) {$(b)\quad\text{snowflake}$};
  \end{scope}
\end{tikzpicture}
\caption{Six-point planar cubic topologies in the cyclic order
$1,\ldots,6$. (a) Nested comb channels:
$P=k_1+k_2$, $Q=k_1+k_2+k_3$, $R=-(k_5+k_6)$.
The distinct-edge blocks connect $P,Q$ at $v_2$ and $Q,R$ at $v_3$.
(b) Snowflake channels: $P=k_1+k_2$, $Q=k_3+k_4$, $R=k_5+k_6$,
with $P+Q+R=0$ at the center. All three pairs of oscillator sectors
meet at $v_\star$, producing the blocks $B^{PQ},B^{QR},B^{RP}$.
Every internal edge is on shell at $e^2=(1-N_e)/\alpha'$.
The labels give the levels and sewing variables used in the corresponding
coefficient extractions, Eqs.~\eqref{eq:six-comb-final} and
\eqref{eq:snowflake-final}. Filled/open vertices and arrows denote
$L/R$ BPZ orientation. The two panels describe different simultaneous
pole loci of the same ordered amplitude.}
\label{fig:six-point-topologies}
\end{figure}
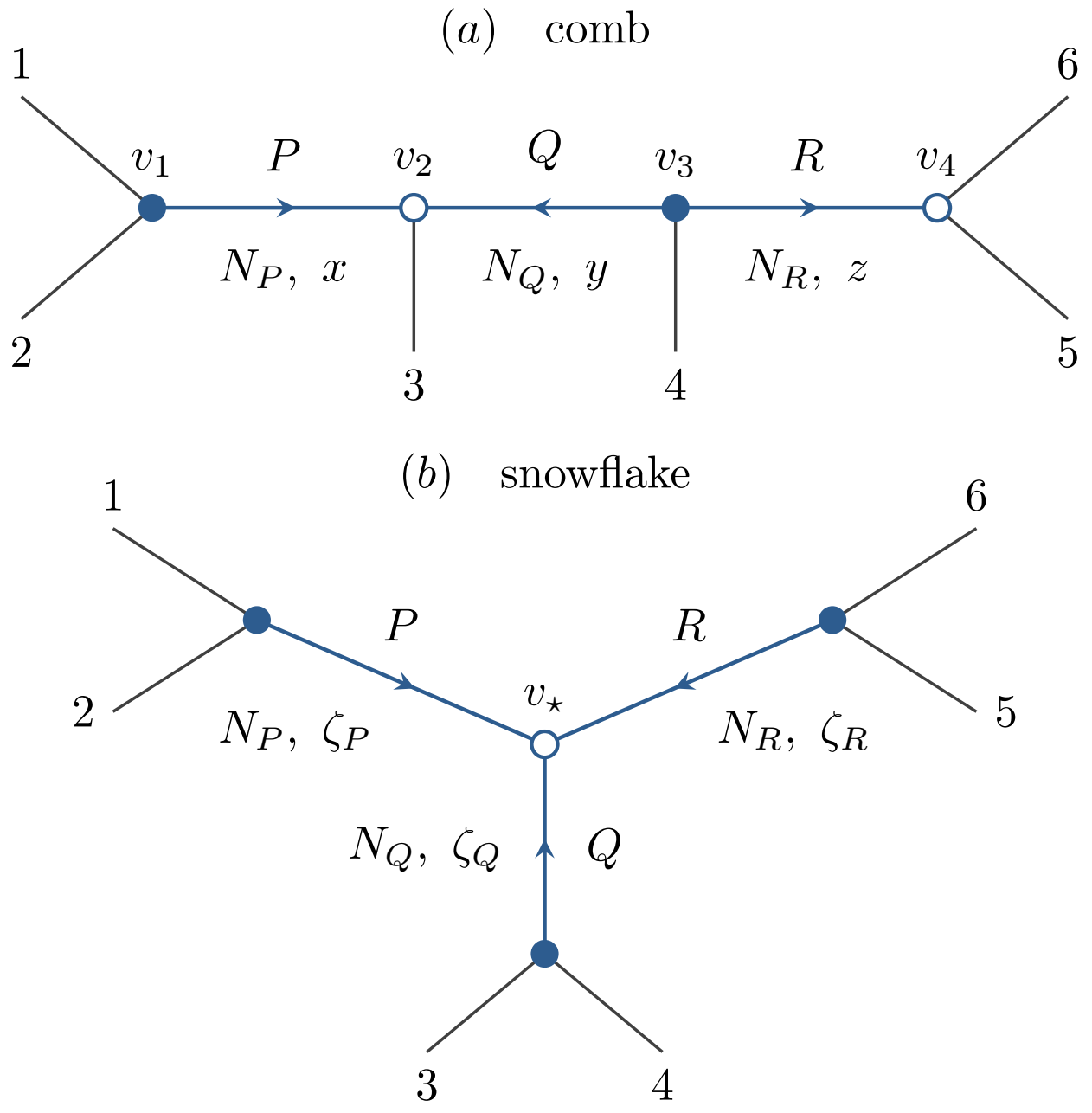

At six points, the two cubic tree topologies are the comb and the
snowflake, shown in figure~\ref{fig:six-point-topologies}.
The comb joins two successive two-DDF vertices, whereas the snowflake
contains a central three-DDF vertex.

All external momenta are outgoing and satisfy
\begin{equation}
 k_r^2=\frac1{\alpha'},\qquad \sum_{r=1}^{6}k_r=0.
 \label{eq:six-point-kinematics}
\end{equation}
We use the same reduced cubic normalization and suppress the overall
coupling. The matrices below have the exact cutoffs $L_e=N_e$.
Increasing a cutoff leaves the chosen residue unchanged; independent
cutoffs are useful when constructing the stabilized series at fixed
physical levels.
\suppressfloats[t]

\subsection{Comb in three nested channels}

We define the internal momenta and pole levels by
\begin{equation}
 \begin{gathered}
 P=k_1+k_2,\qquad Q=P+k_3,\qquad
 R=Q+k_4=-(k_5+k_6),\\
 P^2=\frac{1-N_P}{\alpha'},\qquad
 Q^2=\frac{1-N_Q}{\alpha'},\qquad
 R^2=\frac{1-N_R}{\alpha'}.
 \end{gathered}
 \label{eq:six-comb-channels}
\end{equation}
The four ordered cubic vertices are
\begin{equation}
 (k_1,k_2,-P),\qquad(P,k_3,-Q),\qquad
 (Q,k_4,-R),\qquad(R,k_5,k_6).
 \label{eq:six-comb-vertices}
\end{equation}
The six half-edge ratios are
\begin{equation}
 \begin{gathered}
 a=-\frac{k_2^+}{P^+},\qquad
 c=\frac{k_3^+}{P^+},\qquad
 e=-\frac{k_3^+}{Q^+}=-\frac{c}{1+c},\\
 \gamma=\frac{k_4^+}{Q^+},\qquad
 \eta=-\frac{k_4^+}{R^+}=-\frac{\gamma}{1+\gamma},\qquad
 b=\frac{k_5^+}{R^+}.
 \end{gathered}
 \label{eq:six-comb-ratios}
\end{equation}
With the same BPZ convention as at five points, the oriented currents are
\begin{align}
 J_{1P}&=\sqrt{2\alpha'}(k_2+aP)_\perp,&
 J_{2P}&=-\sqrt{2\alpha'}(k_3-cP)_\perp,
 \nonumber\\
 J_{2Q}&=\sqrt{2\alpha'}(k_3+eQ)_\perp,&
 J_{3Q}&=-\sqrt{2\alpha'}(k_4-\gamma Q)_\perp,
 \nonumber\\
 J_{3R}&=\sqrt{2\alpha'}(k_4+\eta R)_\perp,&
 J_{4R}&=-\sqrt{2\alpha'}(k_5-bR)_\perp,
 \label{eq:six-comb-currents}
\end{align}
with
\begin{equation}
 J_{2Q}=-\frac{J_{2P}}{1+c},\qquad
 J_{3R}=-\frac{J_{3Q}}{1+\gamma}.
 \label{eq:six-comb-current-relations}
\end{equation}

The distinct-leg blocks have dimensions
\(h_{PQ}\in\mathbb C^{N_Q\times N_P}\) and
\(h_{QR}\in\mathbb C^{N_R\times N_Q}\), with
\begin{equation}
 (h_{PQ})_{mn}=f_m(e)f_n(c)\frac{mn c}{n(1+c)-m},\qquad
 (h_{QR})_{\ell m}=f_\ell(\eta)f_m(\gamma)
 \frac{\ell m\gamma}{m(1+\gamma)-\ell}.
 \label{eq:six-comb-h-blocks}
\end{equation}
Removable singularities are evaluated by analytic continuation, so
\((h_{PQ})_{11}=(h_{QR})_{11}=1\).

We choose \(V_L=\{v_1,v_3\}\), \(V_R=\{v_2,v_4\}\), and the edge order
\((P,Q,R)\).  The source columns are
\begin{equation}
 u_i=\begin{pmatrix}
 f^{[N_P]}(a)J_{1P}^i\\
 f^{[N_Q]}(\gamma)J_{3Q}^i\\
 f^{[N_R]}(\eta)J_{3R}^i
 \end{pmatrix},\qquad
 w_i=\begin{pmatrix}
 f^{[N_P]}(c)J_{2P}^i\\
 f^{[N_Q]}(e)J_{2Q}^i\\
 f^{[N_R]}(b)J_{4R}^i
 \end{pmatrix}.
 \label{eq:six-comb-sources}
\end{equation}
The scalar mode matrices obtained from the vertex incidence sums are
\begin{equation}
 G_L^c=\begin{pmatrix}
 g(a)&0&0\\
 0&g(\gamma)&h_{QR}^T\\
 0&h_{QR}&g(\eta)
 \end{pmatrix},\qquad
 G_R^c=\begin{pmatrix}
 g(c)&h_{PQ}^T&0\\
 h_{PQ}&g(e)&0\\
 0&0&g(b)
 \end{pmatrix}.
 \label{eq:six-comb-GL-GR}
\end{equation}
Each \(f\) and \(g\) block is cut off at the level carried by its edge.  Let
\begin{equation}
 \begin{gathered}
 Z_P=\diag\!\left(x,\frac{x^2}{2},\ldots,
                         \frac{x^{N_P}}{N_P}\right),\qquad
 Z_Q=\diag\!\left(y,\frac{y^2}{2},\ldots,
                         \frac{y^{N_Q}}{N_Q}\right),\\
 Z_R=\diag\!\left(z,\frac{z^2}{2},\ldots,
                         \frac{z^{N_R}}{N_R}\right),\qquad
 Z_0=\diag(Z_P,Z_Q,Z_R).
 \end{gathered}
 \label{eq:six-comb-Z}
\end{equation}
and
\begin{equation}
 K_L^c=\Id-Z_0G_R^cZ_0G_L^c,\qquad
 K_R^c=\Id-Z_0G_L^cZ_0G_R^c.
 \label{eq:six-comb-Ks}
\end{equation}
The first matrix has the explicit block form
\begin{equation}
 K_L^c=\begin{pmatrix}
 \Id-Z_Pg(c)Z_Pg(a)&-Z_Ph_{PQ}^TZ_Qg(\gamma)&
                    -Z_Ph_{PQ}^TZ_Qh_{QR}^T\\
 -Z_Qh_{PQ}Z_Pg(a)&\Id-Z_Qg(e)Z_Qg(\gamma)&
                    -Z_Qg(e)Z_Qh_{QR}^T\\
 0&-Z_Rg(b)Z_Rh_{QR}&\Id-Z_Rg(b)Z_Rg(\eta)
 \end{pmatrix}.
 \label{eq:six-comb-K-explicit}
\end{equation}
The general result now yields
\begin{align}
 R_{N_P,N_Q,N_R}^{(6,c;d_\perp)}
 &=[x^{N_P}y^{N_Q}z^{N_R}]
 (\det K_L^c)^{-d_\perp/2}\exp(\mathcal E_c),
 \label{eq:six-comb-final}\\
 \mathcal E_c&=\sum_{i=1}^{d_\perp}\biggl[
 u_i^T(K_L^c)^{-1}Z_0w_i
 +\frac12u_i^T(K_L^c)^{-1}Z_0G_R^cZ_0u_i
 \nonumber\\[-1mm]
 &\hspace{37mm}
 +\frac12w_i^T(K_R^c)^{-1}Z_0G_L^cZ_0w_i
 \biggr].
 \label{eq:six-comb-E}
\end{align}

\paragraph{Covariant comparison at level $(1,1,1)$.}
For the first coefficient involving all three channels, define
\begin{equation}
 A_P=J_{1P},\quad A_Q=J_{3Q},\quad A_R=J_{3R},\qquad
 B_P=J_{2P},\quad B_Q=J_{2Q},\quad B_R=J_{4R}.
 \label{eq:six-comb-AB}
\end{equation}
Since \(f_1=1\) and both mixed mode-one entries equal one,
\begin{align}
 R_{1,1,1}^{(6,c;d_\perp)}={}&
 (A_P\!\cdot B_P)(A_Q\!\cdot B_Q)(A_R\!\cdot B_R)
 +(A_P\!\cdot A_Q)(A_R\!\cdot B_R)
 \nonumber\\
 &+(B_Q\!\cdot B_R)(A_P\!\cdot B_P)+A_P\!\cdot B_R.
 \label{eq:six-comb-111}
\end{align}
The four terms correspond to no internal-vertex pairing, a $P$--$Q$
pairing, a $Q$--$R$ pairing, and both pairings, respectively.
There is no $d_\perp$ dependence at this order. A closed contraction
chain on a tree crosses some edge at least twice, while each sewing
variable enters this coefficient only once.
To compare with a covariant expression, set
$T_{ij}=-2\alpha'k_i\cdot k_j$. The KN factor in the comb chart is
\begin{equation}
 K_6^{\rm comb}=
 (1-x)^{-T_{23}}(1-y)^{-T_{34}}(1-z)^{-T_{45}}
 (1-xy)^{-T_{24}}(1-yz)^{-T_{35}}(1-xyz)^{-T_{25}}.
 \label{eq:six-comb-KN-factor}
\end{equation}
Its $xyz$ coefficient is
\begin{equation}
 [xyz]K_6^{\rm comb}
 =T_{23}T_{34}T_{45}+T_{24}T_{45}+T_{35}T_{23}+T_{25}.
 \label{eq:six-comb-KN-111}
\end{equation}

On the pole locus $N_P=N_Q=N_R=1$, the internal momenta obey
$P^2=Q^2=R^2=0$. The currents in
Eq.~\eqref{eq:six-comb-currents} then give
\[
\begin{aligned}
A_P\cdot B_P
&=T_{23}+a,
\\
A_Q\cdot B_Q
&=T_{34}+e+\gamma,
\\
A_R\cdot B_R
&=T_{45}+b,
\\
A_P\cdot A_Q
&=T_{24}-\gamma T_{23}-aT_{34}+a-a\gamma,
\\
B_Q\cdot B_R
&=T_{35}+b(1-T_{34})-eT_{45}-eb,
\\
A_P\cdot B_R
&=T_{25}-b(T_{23}+T_{24})
-a(T_{35}+T_{45})+ab(T_{34}-2).
\end{aligned}
\]
Substituting these contractions into the four mode-one sewing
terms gives
\begin{equation}
\begin{aligned}
\mathcal R_{1,1,1}^{(6,c;d_\perp)}
={}&
(T_{23}+a)(T_{34}+e+\gamma)(T_{45}+b)
\\
&+
(T_{24}-\gamma T_{23}-aT_{34}+a-a\gamma)
(T_{45}+b)
\\
&+
\bigl[T_{35}+b(1-T_{34})-eT_{45}-eb\bigr]
(T_{23}+a)
\\
&+
T_{25}-b(T_{23}+T_{24})
-a(T_{35}+T_{45})+ab(T_{34}-2)
\\
={}&
T_{23}T_{34}T_{45}
+T_{24}T_{45}
+T_{35}T_{23}
+T_{25}.
\end{aligned}
\end{equation}
All dependence on $a,b,e,\gamma$ cancels, leaving the KN coefficient
\eqref{eq:six-comb-KN-111}.

\subsection{The six-point snowflake}
\label{subsec:snowflake}

For the snowflake in figure~\ref{fig:six-point-topologies}b, define
\begin{equation}
 P=k_1+k_2,\qquad Q=k_3+k_4,\qquad R=k_5+k_6,\qquad
 P+Q+R=0,
 \label{eq:snowflake-channels}
\end{equation}
with pole levels \(N_P,N_Q,N_R\).  At the three outer vertices we define
\begin{align}
 a_P&=-\frac{k_2^+}{P^+},&
 A_P&=\sqrt{2\alpha'}(k_2+a_PP)_\perp,
 \nonumber\\
 a_Q&=-\frac{k_4^+}{Q^+},&
 A_Q&=\sqrt{2\alpha'}(k_4+a_QQ)_\perp,
 \nonumber\\
 a_R&=-\frac{k_6^+}{R^+},&
 A_R&=\sqrt{2\alpha'}(k_6+a_RR)_\perp.
 \label{eq:snowflake-outer-data}
\end{align}
We also choose the cyclic ordering \((P,Q,R)\) at the central vertex and set
\begin{equation}
 r=\frac{Q^+}{P^+},\qquad
 s=\frac{R^+}{Q^+}=-\frac{1+r}{r},\qquad
 t=\frac{P^+}{R^+}=-\frac1{1+r}.
 \label{eq:snowflake-central-ratios}
\end{equation}
The corresponding transverse currents are
\begin{equation}
 \begin{gathered}
 C_P=\sqrt{2\alpha'}(Q-rP)_\perp,\qquad
 C_Q=\sqrt{2\alpha'}(R-sQ)_\perp=\frac{C_P}{r},\\
 C_R=\sqrt{2\alpha'}(P-tR)_\perp=-\frac{C_P}{1+r}.
 \end{gathered}
 \label{eq:snowflake-central-currents}
\end{equation}

The three central legs must be evaluated in one \emph{common} puncture gauge.  With
\begin{equation}
 \beta_e=\sqrt{2\alpha'}e^+,\qquad
 \Xi_\star=x_{PQ}^{\beta_R}x_{PR}^{\beta_Q}x_{QR}^{\beta_P},
 \label{eq:snowflake-Xi}
\end{equation}
the exact local-coordinate factors are
\begin{equation}
 \sigma_e=\Xi_\star^{1/\beta_e},\qquad
 S_e=\diag(\sigma_e,\sigma_e^2,\ldots,\sigma_e^{N_e}),\qquad
 e=P,Q,R,
 \label{eq:snowflake-local-scales}
\end{equation}
with a common branch convention. These matrices contain only the
coordinate factors; oscillator phases are already included in the
stripped data through the half-edge rephasings.

The distinct-puncture blocks with local-coordinate factors removed are
\begin{equation}
 (B^{PQ})_{nm}=b_{nm}(r),\qquad
 (B^{QR})_{m\ell}=b_{m\ell}(s),\qquad
 (B^{RP})_{\ell n}=b_{\ell n}(t),
 \label{eq:snowflake-B-blocks}
\end{equation}
where \(b_{nm}\) is given in Eq.~\eqref{eq:b-general-tree}.  Coloring the
three outer vertices \emph{left} and the center \emph{right} gives the stripped sources
\begin{equation}
 \widehat u_i=\begin{pmatrix}
 f(a_P)A_P^i\\f(a_Q)A_Q^i\\f(a_R)A_R^i
 \end{pmatrix},\qquad
 \widehat w_i=-\begin{pmatrix}
 f(r)C_P^i\\f(s)C_Q^i\\f(t)C_R^i
 \end{pmatrix}.
 \label{eq:snowflake-sources}
\end{equation}
The central source carries three minus signs because
$\theta_{\star,P}=\theta_{\star,Q}=\theta_{\star,R}=-1$.
Together with the outer phases $(-1)^n$, these satisfy BPZ duality
on every internal edge. The scalar mode matrices are
\begin{equation}
 \widehat G_o=\begin{pmatrix}
 g(a_P)&0&0\\0&g(a_Q)&0\\0&0&g(a_R)
 \end{pmatrix},\qquad
 \widehat G_c=\begin{pmatrix}
 g(r)&B^{PQ}&(B^{RP})^T\\
 (B^{PQ})^T&g(s)&B^{QR}\\
 B^{RP}&(B^{QR})^T&g(t)
 \end{pmatrix}.
 \label{eq:snowflake-Go-Gc}
\end{equation}
Thus, \(\widehat G_c^T=\widehat G_c\), as required by the loop-hafnian master theorem.

With \(S_c=\diag(S_P,S_Q,S_R)\), restoring the central coordinate factors gives
\begin{equation}
 w_i^{(x)}=S_c\widehat w_i,\qquad
 G_c^{(x)}=S_c\widehat G_cS_c.
 \label{eq:snowflake-retained-data}
\end{equation}
With the outer half-edge scales set to one, the central factors are
absorbed into the invariant plumbing variables:
\begin{equation}
 \zeta_e=\sigma_e z_e,\qquad
 \widehat Z_e=\diag\!\left(\zeta_e,\frac{\zeta_e^2}{2},\ldots,
 \frac{\zeta_e^{N_e}}{N_e}\right),\qquad
 \widehat Z_0=\diag(\widehat Z_P,\widehat Z_Q,\widehat Z_R).
 \label{eq:snowflake-Z}
\end{equation}
For general outer scales, $\zeta_e$ includes both the outer and central
factors, as in Eq.~\eqref{eq:invariant-plumbing}. Define
\begin{equation}
 K_o=\Id-\widehat Z_0\widehat G_c\widehat Z_0\widehat G_o,\qquad
 K_c=\Id-\widehat Z_0\widehat G_o\widehat Z_0\widehat G_c,
 \label{eq:snowflake-Ks}
\end{equation}
the three-edge loop-hafnian resummation gives
\begin{align}
 R_{N_P,N_Q,N_R}^{(6,\star;d_\perp)}
 &=[\zeta_P^{N_P}\zeta_Q^{N_Q}\zeta_R^{N_R}]
 (\det K_o)^{-d_\perp/2}\exp(E_\star),
 \label{eq:snowflake-final}\\
 E_\star&=\sum_{i=1}^{d_\perp}\biggl[
 \widehat u_i^TK_o^{-1}\widehat Z_0\widehat w_i
 +\frac12\widehat u_i^TK_o^{-1}\widehat Z_0
 \widehat G_c\widehat Z_0\widehat u_i
 \nonumber\\[-1mm]
 &\hspace{39mm}
 +\frac12\widehat w_i^TK_c^{-1}\widehat Z_0
 \widehat G_o\widehat Z_0\widehat w_i
 \biggr].
 \label{eq:snowflake-E}
\end{align}
At mode one, \(f_1=b_{11}=1\), and direct expansion gives
\begin{align}
 R_{1,1,1}^{(6,\star)}=-\bigl[&
 (A_P\!\cdot C_P)(A_Q\!\cdot C_Q)(A_R\!\cdot C_R)
 +(A_P\!\cdot A_Q)(A_R\!\cdot C_R)
 \nonumber\\
 &+(A_Q\!\cdot A_R)(A_P\!\cdot C_P)
 +(A_R\!\cdot A_P)(A_Q\!\cdot C_Q)\bigr].
 \label{eq:snowflake-111}
\end{align}
Both central blocks incident on edge $e$ contain the same $S_e$.
This common scale is needed for the similarity transformation
\eqref{eq:K-similarity} and the cancellation of coordinate factors.
Choosing a separate gauge for each pairwise block would give
inconsistent local coordinates for the central three-DDF vertex.

\paragraph{The snowflake residue in an adapted Koba--Nielsen chart.}
An independent KN calculation uses the degeneration-adapted chart
\begin{equation}
(x_1,x_2,x_3,x_4,x_5,x_6)
=
\left(0,\xi,1,1+\eta,\frac1\omega,\infty\right),
\end{equation}
where $\xi,\eta,\omega\to0$ select the $(P,Q,R)$ channels.
The change $\dd x_5=-\dd\omega/\omega^2$, together with reversal of the
$\omega$-integration limits, gives the positive measure
$\dd\xi\,\dd\eta\,\dd\omega/\omega^2$.

In the same stripped-residue normalization, the top form at the
simultaneous pole is
\[
\xi^{-N_P-1}\eta^{-N_Q-1}\omega^{-N_R-1}
\mathcal K_6^\star(\xi,\eta,\omega)\,\dd\xi\,\dd\eta\,\dd\omega,
\]
where
\begin{equation}
\begin{aligned}
\mathcal K_6^\star(\xi,\eta,\omega)
={}&
(1+\eta)^{-T_{14}}
(1-\xi)^{-T_{23}}
(1+\eta-\xi)^{-T_{24}}
\\
&\times
(1-\xi\omega)^{-T_{25}}
(1-\omega)^{-T_{35}}
(1-\omega-\eta\omega)^{-T_{45}}.
\end{aligned}
\end{equation}
All $T_{ij}$ are evaluated on the selected pole locus.
Consequently, an independent covariant benchmark is
\[
\mathcal R_{N_P,N_Q,N_R}^{(6,\star),\mathrm{KN}}
=
[\xi^{N_P}\eta^{N_Q}\omega^{N_R}]
\mathcal K_6^\star(\xi,\eta,\omega).
\]

For the mode-one coefficient, define
\begin{equation}
\mathsf A=T_{23}+T_{24},
\qquad
\mathsf B=-T_{14}-T_{24},
\qquad
\mathsf C=T_{35}+T_{45}.
\end{equation}
The required coefficients of the logarithm are
\[
\begin{aligned}
[\xi]\log\mathcal K_6^\star&=\mathsf A,
&
[\eta]\log\mathcal K_6^\star&=\mathsf B,
&
[\omega]\log\mathcal K_6^\star&=\mathsf C,
\\
[\xi\eta]\log\mathcal K_6^\star&=-T_{24},
&
[\xi\omega]\log\mathcal K_6^\star&=T_{25},
&
[\eta\omega]\log\mathcal K_6^\star&=T_{45},
\end{aligned}
\]
and $[\xi\eta\omega]\log\mathcal K_6^\star=0$. Thus,
\[
[\xi\eta\omega]\mathcal K_6^\star
=
\mathsf A\mathsf B\mathsf C
-T_{24}\mathsf C
+T_{25}\mathsf B
+T_{45}\mathsf A.
\]

To reduce the DDF expression to this polynomial, impose
$P^2=Q^2=R^2=0$ and $P+Q+R=0$ at
$N_P=N_Q=N_R=1$. The outer and central currents then satisfy
\[
A_P\cdot C_P=-\mathsf A,
\qquad
A_Q\cdot C_Q=-\mathsf B,
\qquad
A_R\cdot C_R=-\mathsf C,
\]
and
\begin{equation}
\begin{aligned}
A_P\cdot A_Q
&=-T_{24}-a_Q\mathsf A+a_P\mathsf B,
\\
A_Q\cdot A_R
&=-T_{46}-a_R\mathsf B+a_Q\mathsf C,
\\
A_R\cdot A_P
&=-T_{26}-a_P\mathsf C+a_R\mathsf A.
\end{aligned}
\end{equation}
Therefore, the mode-one sewing result is
\begin{equation}
\begin{aligned}
\mathcal R_{1,1,1}^{(6,\star)}
={}&
\mathsf A\mathsf B\mathsf C
+\mathsf C(A_P\cdot A_Q)
+\mathsf A(A_Q\cdot A_R)
+\mathsf B(A_R\cdot A_P)
\\
={}&
\mathsf A\mathsf B\mathsf C
-T_{24}\mathsf C
-T_{46}\mathsf A
-T_{26}\mathsf B.
\end{aligned}
\end{equation}
The terms involving $a_P,a_Q,a_R$ cancel identically.
Finally, momentum conservation implies
\[
T_{45}+T_{46}=\mathsf B,
\qquad
T_{25}+T_{26}=-\mathsf A,
\]
and hence
\begin{equation}
\mathcal R_{1,1,1}^{(6,\star)}
=
\mathsf A\mathsf B\mathsf C
-T_{24}\mathsf C
+T_{25}\mathsf B
+T_{45}\mathsf A
=
[\xi\eta\omega]\mathcal K_6^\star.
\end{equation}
Thus the DDF and KN expressions agree algebraically at
$(N_P,N_Q,N_R)=(1,1,1)$, with the central-source sign included.
\section{Berezin--Grunsky interpretation and the moduli-space map}
\label{sec:berezin-grunsky}

We now ask how the residue generator is related to the KN top form.
Throughout this section, $d_\perp=24$ and
$(2\pi)^{26}\delta^{26}(\sum_rk_r)$ is suppressed.
We use the BPZ-oriented cubic blocks of
section~\ref{sec:arbitrary-tree-sewing}, with coordinate factors
removed and half-edge scales absorbed into the invariant sewing
variables.

Gaussian composition identifies the determinant algebraically.
To interpret it geometrically, one must also identify the operators
as in section~\ref{subsec:gaussian-geometric-dictionary}, construct a
resmoothing map and include the $bc$-ghost measure.
We formulate this problem in a ray chart adapted to the nested comb.
Equality of top forms in several variables does not by itself fix a
unique coordinate map. The conditions below concern this geometric
comparison; the physical-state proof in
section~\ref{subsec:generality-proof} is already complete.

\subsection{Gaussian composition and coordinate maps}
\label{subsec:gaussian-geometric-dictionary}

\paragraph{Canonical normalization.}
For one transverse boson, take the mode matrices $G_s$ and sources
$u,w$ of section~\ref{sec:arbitrary-tree-sewing}, with coordinate
factors removed and BPZ phases included. At finite cutoffs $L_e$,
define
\begin{equation}
 D=\bigoplus_e\diag_{1\leq n\leq L_e}(n),\qquad
 P=\bigoplus_e\diag_{1\leq n\leq L_e}(z_e^n),\qquad
 q=D^{1/2}\lambda.
 \label{eq:Berezin-D-P-q}
\end{equation}
Since $Z=D^{-1/2}PD^{-1/2}$, the contraction becomes
\begin{equation}
 \exp(\partial_{\lambda_L}^TZ\partial_{\lambda_R})
 =\exp(\partial_{q_L}^TP\partial_{q_R}).
 \label{eq:Berezin-contraction}
\end{equation}
The sources and quadratic matrices transform as
\begin{equation}
 u^{\mathrm{can}}=D^{-1/2}u,\qquad
 w^{\mathrm{can}}=D^{-1/2}w,\qquad
 G_s^{\mathrm{can}}=D^{-1/2}G_sD^{-1/2},\quad s=L,R.
 \label{eq:canonical-source-normalization}
\end{equation}
Thus $u^T\lambda=(u^{\mathrm{can}})^Tq$ and
$\lambda^TG_s\lambda=q^TG_s^{\mathrm{can}}q$.
We use $\mathrm{can}$ for canonical normalization and reserve hats
for coordinate stripping and BPZ orientation.

\paragraph{Gaussian composition.}
The normal symbol of an operator is obtained by replacing its
normally ordered creation and annihilation operators with commuting
variables. Composition is then represented by a differential product
of symbols. Define the symmetric blocks
\begin{align}
 C_L&:=G_L^{\mathrm{can}},
 \label{eq:Berezin-left-block}\\
 B_R^{(P)}&:=-PG_R^{\mathrm{can}}P.
 \label{eq:Berezin-right-block}
\end{align}
The minus sign is fixed by the normal-symbol convention
$-\tfrac12q^{*T}Bq^*$; the BPZ phases are already included in $G_R$.
Define the source-free Gaussians
\begin{equation}
 N_L(q,0)=\exp\!\left(\frac12q^TC_Lq\right),\qquad
 N_R^{(P)}(0,q^*)=\exp\!\left(-\frac12q^{*T}B_R^{(P)}q^*\right).
 \label{eq:Berezin-DDF-quadratics}
\end{equation}
Here $q,q^*$ are independent oscillator coordinates. Including the linear
sources gives
\begin{equation}
 \begin{aligned}
 F_L(q)&=\exp\!\left((u^{\mathrm{can}})^Tq\right)N_L(q,0),\\
 F_R^{(P)}(q^*)&=\exp\!\left((w^{\mathrm{can}})^TPq^*\right)
                  N_R^{(P)}(0,q^*).
 \end{aligned}
 \label{eq:Berezin-FLR}
\end{equation}
The Berezin normal-symbol product~\cite{AlekseevShatashviliTakhtajan2024}
therefore reproduces the DDF contraction:
\begin{equation}
 \begin{aligned}
 (F_L*F_R^{(P)})(0,0)
 &:=\left.e^{\partial_q^T\partial_{q^*}}F_L(q)F_R^{(P)}(q^*)\right|_0\\
 &=\left.e^{\partial_{\lambda_L}^TZ\partial_{\lambda_R}}
                 V_L(\lambda_L)V_R(\lambda_R)\right|_0.
 \end{aligned}
 \label{eq:Berezin-star-DDF}
\end{equation}
Gaussian composition gives
\begin{equation}
 \begin{aligned}
 N_L*N_R^{(P)}
 &=\frac{N_{\mathrm{sewn}}}
        {\sqrt{\det(\Id+B_R^{(P)}C_L)}},\\
 K_L^{\mathrm{can}}
 &:=\Id+B_R^{(P)}C_L
   =\Id-PG_R^{\mathrm{can}}PG_L^{\mathrm{can}},
 \end{aligned}
 \label{eq:Berezin-composition}
\end{equation}
where $N_{\mathrm{sewn}}(0,0)=1$ and the square root has constant term
one. The one-boson vacuum factor is therefore
\begin{equation}
 (N_L*N_R^{(P)})(0,0)=(\det K_L^{\mathrm{can}})^{-1/2}.
 \label{eq:one-boson-sewing-cocycle}
\end{equation}
Taking the product over the $24$ transverse bosons defines
\begin{equation}
 \mathcal C_{\mathrm{sewn}}^{24}
 :=\bigl[(N_L*N_R^{(P)})(0,0)\bigr]^{24}.
 \label{eq:Csew-definition}
\end{equation}
Since
\[
 K_L=D^{-1/2}K_L^{\mathrm{can}}D^{1/2},\qquad
 \det K_L=\det K_L^{\mathrm{can}},
\]
we obtain
\begin{equation}
 \boxed{\mathcal C_{\mathrm{sewn}}^{24}=(\det K_L)^{-12}}.
 \label{eq:Csew-result}
\end{equation}
Restoring the linear sources in every transverse direction gives
\begin{equation}
 \boxed{\mathcal G_{\mathcal T,\bm N}^{(24)}
 =\mathcal C_{\mathrm{sewn}}^{24}e^{\mathcal E_{\mathcal T}}
 =(\det K_L)^{-12}e^{\mathcal E_{\mathcal T}}}.
 \label{eq:critical-tree-generator}
\end{equation}
This recovers the generator of section~\ref{sec:arbitrary-tree-sewing}
in canonically normalized variables.

\paragraph{Coordinate-map interpretation.}
The same algebra has a symplectic formulation. On the polarized oscillator
space $\mathcal H=\mathcal H_+\oplus\mathcal H_-$, with coordinates
$a\in\mathcal H_+$ and $a^*\in\mathcal H_-$, use
\begin{equation}
 \Omega((a,a^*),(b,b^*))=b^{*T}a-a^{*T}b,\qquad
 J=\begin{pmatrix}0&\Id\\-\Id&0\end{pmatrix}.
 \label{eq:oscillator-symplectic-form}
\end{equation}
At finite cutoff, $C_L,B_R^{(P)}$ admit the algebraic realization
\[
 A_L^{\mathrm{alg}}=\begin{pmatrix}\Id&0\\C_L&\Id\end{pmatrix},
 \qquad
 A_R^{\mathrm{alg}}=\begin{pmatrix}\Id&B_R^{(P)}\\0&\Id\end{pmatrix}.
\]
Their symplecticity follows from the symmetry of the two blocks.
To obtain a geometric realization, a normalized holomorphic map $f$
acts on the full differentiated free field by
$\psi(z)\mapsto\psi(f(z))f'(z)$. Its mode expansion induces
\begin{equation}
 \begin{pmatrix}\widetilde a\\\widetilde a^*\end{pmatrix}
 =A_f\begin{pmatrix}a\\a^*\end{pmatrix},\qquad
 A_f=\begin{pmatrix}\alpha_f&\beta_f\\\gamma_f&\delta_f\end{pmatrix},
 \qquad A_f^TJA_f=J.
 \label{eq:Af-blocks}
\end{equation}
The last identity expresses invariance of the contour pairing on the
full mode space. A finite principal truncation need not be
symplectic: $f$ can mix modes on opposite sides of the cutoff.
For geometric sewing, one must therefore use the full admissible
operator or its coefficientwise stabilized series. Analytic
implementers and determinants also require the restricted-symplectic
and trace-class conditions of section 4.3 in
ref.~\cite{AlekseevShatashviliTakhtajan2024}.

When $A_f$ is admissible, its projective Fock-space implementer satisfies
\begin{equation}
 \widehat A_f\begin{pmatrix}\widehat a\\\widehat a^*\end{pmatrix}
 \widehat A_f^{-1}
 =A_f\begin{pmatrix}\widehat a\\\widehat a^*\end{pmatrix}.
 \label{eq:Af-implementer}
\end{equation}
The normal symbol replaces the normal-ordered operators by commuting
variables:
\[
 \widehat A_f=\sum_{I,J}c_{I,J}(\widehat a^*)^J\widehat a^I
 \quad\longmapsto\quad
 N_f(a,a^*)=\sum_{I,J}c_{I,J}a^I(a^*)^J.
\]
With normalization $N_f(0,0)=1$, its Gaussian expression is
\begin{equation}
 \begin{gathered}
 N_f(a,a^*)=\exp\!\left[
 a^{*T}(\alpha_f^{-1}-\Id)a
 -\frac12a^{*T}B_fa^*+\frac12a^TC_fa\right],\\
 B_f=\alpha_f^{-1}\beta_f,\qquad C_f=\gamma_f\alpha_f^{-1}.
 \end{gathered}
 \label{eq:Berezin-normal-symbol}
\end{equation}
The symmetric blocks $B_f,C_f$ describe pair creation and annihilation;
the first term describes number-preserving mode mixing. In particular,
\begin{equation}
 N_f(q,0)=e^{q^TC_fq/2},\qquad
 N_f(0,q^*)=e^{-q^{*T}B_fq^*/2}.
 \label{eq:Berezin-pure-symbols}
\end{equation}

For each color $s=L,R$, let $f_{s,v}$, $v\in V_s$, be a candidate
system of local-to-global germs on the components of the cubic tree,
after doubling across the open-string boundary.
If these germs define admissible transformations and implementers,
the disconnected components of a given color combine as
\begin{equation}
 A_{f_s}:=\bigoplus_{v\in V_s}A_{f_{s,v}},\qquad
 N_{f_s}:=\prod_{v\in V_s}N_{f_{s,v}}.
 \label{eq:color-class-symbols}
\end{equation}
The relation between DDF vertices and Mandelstam
maps~\cite{BiswasPesandoMandelstam} suggests that the cubic blocks
should be identified with current kernels. For germs $f_{s,e}$ and
$f_{s,e'}$ on the same component, in the cyclic reference convention,
define
\begin{equation}
 \begin{aligned}
 (\mathcal B_s^{\mathrm{cyc}})_{(e,n),(e',m)}
 &=\frac{1}{\sqrt{nm}}[t^{n-1}u^{m-1}]
 \left[\frac{f'_{s,e}(t)f'_{s,e'}(u)}
 {(f_{s,e}(t)-f_{s,e'}(u))^2}-\frac{\delta_{ee'}}{(t-u)^2}\right],\\
 G_s^{\mathrm{can}}&=\Theta_s\mathcal B_s^{\mathrm{cyc}}\Theta_s.
 \end{aligned}
 \label{eq:Grunsky-DDF}
\end{equation}
The first line defines generalized Grunsky coefficients from the
pulled-back current two-point function. At coincident punctures, the
universal double pole is subtracted. These coefficients encode the
pairings induced by the coordinate maps
\cite{LeClairPeskinPreitschopfKinematics1989,AlekseevShatashviliTakhtajan2024}.
The second line states the required identification with the DDF
vertex. The diagonal phase matrix $\Theta_s$ commutes with $D$ and
converts the cyclic kernels to BPZ-oriented blocks. The direct sum
sets kernels between disconnected components of the same color to
zero.

If this identification and the admissibility conditions hold, then
\begin{equation}
 C_{f_L}=C_L=G_L^{\mathrm{can}},\qquad
 B_{f_R}^{(P)}=B_R^{(P)}=-PG_R^{\mathrm{can}}P.
 \label{eq:Berezin-Grunsky-blocks}
\end{equation}
Under these assumptions, $N_L$ and $N_R^{(P)}$ are the pure normal
symbols of the two transformations. Their vacuum contraction
\eqref{eq:one-boson-sewing-cocycle} gives the scalar cocycle in the
geometric composition law. Linear sources implement Heisenberg
translations and recover $\mathcal E_{\mathcal T}$.
To match the moduli-space top form, one must still determine the
resmoothing map and ghost measure.

\subsection{Gluing, the measure, and a candidate coordinate map}

For compatible local coordinates and the complete matter--ghost
system, the generalized gluing and resmoothing theorem
\cite{LeClairPeskinPreitschopf1989,AsakawaKugoTakahashi1998} relates BPZ
sewing to the Reggeon vertex of the sewn disk. Zero modes supply the
suppressed momentum delta function, and the $bc$ system supplies the
gauge-fixed measure. We now specialize to a comb with nested channels
$X_{1,a+2}=-N_a$, $a=1,\ldots,m$, where $m=M-3$, and define the
dimensionless planar variables
\begin{equation}
 X_{ij}=\alpha'(k_i+\cdots+k_{j-1})^2-1,\qquad
 C_{ij}=-2\alpha'k_i\cdot k_j.
 \label{eq:ray-kinematic-conventions}
\end{equation}
The four-point variable $X=\alpha't+2$ is therefore $1-X_{24}$.
With the half-edge coordinate scales absorbed into $\bm z$, we seek a
normalized resmoothing map
$\Phi_{\mathcal T}:\bm z\mapsto\bm x$ that preserves the boundary
divisors and takes the sewing variables to the comb ray chart.
At fixed light-cone ratios, we require this map to be independent of
the remaining planar invariants and internal levels. The desired
top-form identity is
\begin{equation}
 \mathcal G_{\mathcal T,\bm N}^{(24)}(\bm z)
 \bigwedge_{a=1}^{m}\frac{\dd z_a}{z_a^{N_a+1}}
 =\Phi_{\mathcal T}^*\left[
 K_M^{\rm ray}(\bm x)
 \bigwedge_{a=1}^{m}\frac{\dd x_a}{x_a^{N_a+1}}\right].
 \label{eq:top-form-equality}
\end{equation}
This identity requires more than the finite Gaussian contraction.
The chart must be adapted to the chosen degeneration: the six-point
snowflake, for example, has channels $\{X_{13},X_{35},X_{15}\}$,
whereas the comb ray chart uses $\{X_{13},X_{14},X_{15}\}$.
For the snowflake one must use a chart adapted to its channels, or
transform charts together with the residue
cycle~\cite{ArkaniHamedFigueiredoRemmen2025}.
Writing $x_{r:s}=\prod_{a=r}^{s}x_a$, the ray integrand is
\cite{ArkaniHamedHeLam2019}
\begin{equation}
 K_M^{\rm ray}(\bm x)
 =\prod_{a=1}^{m}(1-x_a)^{X_{a+1,a+3}-1}
 \prod_{1\leq r<s\leq m}(1-x_{r:s})^{-C_{r+1,s+2}}.
 \label{eq:ray-integrand}
\end{equation}
For convenience, we define the logarithmic Jacobian
\begin{equation}
 \mathsf M_{ab}(\bm z)=\frac{\partial\log x_a}{\partial\log z_b},
 \qquad
 \Phi_{\mathcal T}^*\bigwedge_{a=1}^{m}\frac{\dd x_a}{x_a}
 =\det\mathsf M(\bm z)
 \bigwedge_{a=1}^{m}\frac{\dd z_a}{z_a}.
 \label{eq:log-Jacobian}
\end{equation}
Equation~\eqref{eq:top-form-equality} then implies
\begin{equation}
 \log\mathcal G_{\mathcal T,\bm N}^{(24)}(\bm z)
 =\log\det\mathsf M(\bm z)
 +\log K_M^{\rm ray}(\bm x(\bm z))
 +\sum_{a=1}^{m}N_a\log\frac{z_a}{x_a(\bm z)}.
 \label{eq:log-top-form}
\end{equation}

In the following derivatives, the light-cone ratios are fixed and
$\{X_{a+1,a+3},C_{r+1,s+2}\}$ are independent formal variables after
the comb pole conditions are imposed. At sufficiently high
multiplicity, physical $D=26$ kinematics satisfy additional Gram
relations. The derivatives must then be taken before restricting to
that locus. If the map has the required kinematic independence,
differentiating Eq.~\eqref{eq:log-top-form} gives
\begin{equation}
 \frac{\partial}{\partial X_{a+1,a+3}}
 \log\mathcal G_{\mathcal T,\bm N}^{(24)}(\bm z)
 =\log(1-x_a(\bm z)),
 \qquad
 x_a(\bm z)=1-\exp\!\left[
 \frac{\partial}{\partial X_{a+1,a+3}}
 \log\mathcal G_{\mathcal T,\bm N}^{(24)}(\bm z)\right].
 \label{eq:elementary-x-from-G}
\end{equation}
The second equation can also be read as a definition of a candidate
map. At fixed light-cone ratios, the on-shell current contractions
make $\log\mathcal G$ affine in the remaining planar invariants and
internal levels. Its derivative with respect to one of these
invariants is therefore independent of the invariants and levels.
We may evaluate it at $\bm N=0$, provided that the mode cutoffs are
kept independent and the stabilized series is retained. Setting the
physical levels to zero does not truncate the mode space. The
proposed top-form identity then reads
\begin{equation}
 \log\mathcal G_{\mathcal T,0}^{(24)}(\bm z)
 =\log\det\mathsf M(\bm z)
 +\log K_M^{\rm ray}(\bm x(\bm z)).
 \label{eq:N0-top-form}
\end{equation}
Similarly,
\begin{align}
 \frac{\partial}{\partial C_{r+1,s+2}}
 \log\mathcal G_{\mathcal T,0}^{(24)}(\bm z)
 &=-\log(1-x_{r:s}(\bm z)),
 \nonumber\\
 1-\exp\!\left[-\frac{\partial}{\partial C_{r+1,s+2}}
 \log\mathcal G_{\mathcal T,0}^{(24)}(\bm z)\right]
 &=x_{r:s}(\bm z)=\prod_{a=r}^{s}x_a(\bm z).
 \label{eq:composite-x-from-G}
\end{align}

These derivatives are evaluated directly from
\begin{equation}
 \log\mathcal G_{\mathcal T,0}^{(24)}
 =-12\log\det K_L+\left. \mathcal E_{\mathcal T}\right|_{\bm N=0}.
 \label{eq:N0-matrix-G}
\end{equation}
At fixed light-cone ratios, $K_L$ is independent of the remaining
planar invariants. Hence
\begin{equation}
 \frac{\partial}{\partial\xi}
 \log\mathcal G_{\mathcal T,0}^{(24)}
 =\frac{\partial}{\partial\xi}
 \left.\mathcal E_{\mathcal T}\right|_{\bm N=0},
 \qquad
 \xi\in\{X_{a+1,a+3},C_{r+1,s+2}\}.
 \label{eq:planar-derivative-E}
\end{equation}
The derivatives act on current contractions reconstructed from the
on-shell momentum data. In particular,
\begin{equation}
 S_{ij}:=\alpha'(k_i+\cdots+k_{j-1})^2,\qquad
 2\alpha'k_i\cdot k_j
 =S_{i,j+1}-S_{i,j}-S_{i+1,j+1}+S_{i+1,j},
 \label{eq:planar-dot-reconstruction}
\end{equation}
where $S_{ij}=X_{ij}+1$ for non-adjacent chords and $S_{1,a+2}=1$
at $\bm N=0$. These relations implement the on-shell reconstruction
at fixed light-cone ratios when taking planar derivatives.

Let $\mathfrak m_{\bm z}=(z_1,\ldots,z_m)$. The linear source term
$u^TZw$, with $f_1=1$, gives the required tangent normalization
\begin{equation}
 \frac{\partial}{\partial X_{a+1,a+3}}
 \log\mathcal G_{\mathcal T,0}^{(24)}(\bm z)
 =-z_a+O(\mathfrak m_{\bm z}^2).
 \label{eq:linear-x-derivative}
\end{equation}
Preservation of the selected boundary divisors further requires the
$a$-th component of the map to be divisible by $z_a$ to all orders.
If this holds,
\begin{equation}
 x_a(\bm z)=z_a[1+O(\mathfrak m_{\bm z})],\qquad
 \mathsf M_{ab}=\delta_{ab}+O(\mathfrak m_{\bm z}),\qquad
 \det\mathsf M=1+O(\mathfrak m_{\bm z}).
 \label{eq:Jacobian-leading}
\end{equation}
The tangent normalization gives a unique formal inverse. The
candidate must also obey the zero-level top-form identity
\begin{equation}
 \mathcal G_{\mathcal T,0}^{(24)}(\bm z)
 =\det\mathsf M(\bm z)K_M^{\rm ray}(\bm x(\bm z)).
 \label{eq:N0-complete-identity}
\end{equation}

\subsection{Criteria for matrix-series matching}

After divisor preservation, three conditions must be checked:
relations among composite coordinates, the invariant-independent
part of the Jacobian, and the internal-level dependence. To compare
the matrix series, use
\begin{equation}
 \begin{gathered}
 K_L^{-1}=\sum_{\ell\geq0}(ZG_RZG_L)^\ell,\qquad
 K_R^{-1}=\sum_{\ell\geq0}(ZG_LZG_R)^\ell,\\
 -\log\det K_L=\sum_{\ell\geq1}\frac1\ell
 \operatorname{Tr}(ZG_RZG_L)^\ell,
 \end{gathered}
 \label{eq:matrix-series-verification}
\end{equation}
and substitute into Eq.~\eqref{eq:N0-matrix-G}.
Equation~\eqref{eq:elementary-x-from-G} determines the elementary
coordinates at $\bm N=0$, with all modes needed at the chosen series
order retained. The resulting candidate must satisfy the composite
relations \eqref{eq:composite-x-from-G} for every
$1\leq r<s\leq m$.

All exponents in the ray integrand vanish at
\begin{equation}
 X_{a+1,a+3}=1,\qquad C_{r+1,s+2}=0.
 \label{eq:ray-unit-point}
\end{equation}
At this point \(K_M^{\rm ray}=1\), and the Jacobian identity is
\begin{equation}
 \left.\log\mathcal G_{\mathcal T,0}^{(24)}(\bm z)
 \right|_{X=1,C=0}
 =\log\det\!\left(
 \frac{\partial(\log x_1,\ldots,\log x_m)}
 {\partial(\log z_1,\ldots,\log z_m)}\right).
 \label{eq:tested-Jacobian-identity}
\end{equation}
For example, this point is
\begin{equation}
 (X_{24},X_{35},X_{25})=(1,1,2)
 \label{eq:five-point-unit-point}
\end{equation}
at five points and
\begin{equation}
 (X_{24},X_{35},X_{46},X_{25},X_{36},X_{26})=(1,1,1,2,2,3)
 \label{eq:six-point-unit-point}
\end{equation}
at six points in the comb ray chart. The complete zero-level condition is
\begin{align}
 \log\mathcal G_{\mathcal T,0}^{(24)}(\bm z)-\log\det\mathsf M(\bm z)
 ={}&\sum_{a=1}^{m}(X_{a+1,a+3}-1)\log(1-x_a(\bm z))
 \nonumber\\
 -\sum_{1\leq r<s\leq m}&C_{r+1,s+2}
 \log\!\left(1-\prod_{a=r}^{s}x_a(\bm z)\right)
 \nonumber\\
 ={}&\log K_M^{\rm ray}(\bm x(\bm z)).
 \label{eq:complete-ray-check}
\end{align}
Because $\log\mathcal G$ is affine in the remaining formal
invariants, the elementary and composite derivative identities,
together with the Jacobian identity at the unit-integrand point,
would establish the zero-level identity for all values of those
invariants.

Matching the top forms at nonzero internal levels requires a further
condition. Let $\bm e_a$ be the
level vector with one unit on edge $a$ and zero on all other edges, and
define
\begin{equation}
 \mathcal L_a(\bm z)
 :=\log\mathcal G_{\mathcal T,\bm e_a}^{(24)}(\bm z)
   -\log\mathcal G_{\mathcal T,0}^{(24)}(\bm z).
 \label{eq:level-matching-definition}
\end{equation}
At fixed light-cone ratios and remaining independent invariants, the
current contractions make the matrix logarithm affine in $\bm N$.
Consequently,
\begin{equation}
 \log\mathcal G_{\mathcal T,\bm N}^{(24)}
 =\log\mathcal G_{\mathcal T,0}^{(24)}
   +\sum_{a=1}^{m}N_a\mathcal L_a.
 \label{eq:affine-level-dependence}
\end{equation}
At every level, the same map must therefore satisfy
\begin{equation}
 \boxed{\mathcal L_a(\bm z)=\log\frac{z_a}{x_a(\bm z)}},
 \qquad a=1,\ldots,m.
 \label{eq:level-matching-condition}
\end{equation}
The divisor normalization in Eq.~\eqref{eq:Jacobian-leading} makes these
logarithms well-defined formal series with zero constant term.
All generators in these comparisons use the same independent oscillator
cutoffs, including the generator at zero physical levels.

The elementary derivatives define the candidate map. To establish
zero-level matching, one must prove divisor preservation, the
composite-coordinate relations and the Jacobian identity.
Equation~\eqref{eq:level-matching-condition} is then needed for
nonzero levels. Gaussian contraction gives the affine dependence on
kinematics, but does not prove these higher-point geometric
identities. They remain open and are not used in the residue formula
\eqref{eq:final-arbitrary-tree-residue}.
For a finite-order comparison, retain every mode contributing to the
chosen multidegree. Whenever division by $z_a$ or differentiation
lowers the series order, compute the map one order further.

\section{Conclusions and outlook}
\label{sec:conclusion}

We have obtained a finite-matrix formula for the maximal simultaneous
residues of color-ordered tachyon disk amplitudes in the critical
open bosonic string. It applies to every planar cubic tree and every
finite assignment of internal levels. In the DDF basis, the
projectors supply the complete sums over oscillator occupations and
transverse polarizations, with their inverse norms. The Gaussian
cubic vertices allow these sums to be performed with the exact mode
cutoffs $L_e=N_e$. The residue is then a coefficient of a determinant
prefactor times a quadratic source exponential. Consistent BPZ
conventions ensure that the local-coordinate scales cancel between
the vertices and the sewing metric.

The four-point calculation shows how the oscillator sum recovers a
covariant residue at every level. Cubic local maps resum the open
contraction chains, and the one-boson stress-tensor Ward identity
fixes the determinant. Together they determine the change of
variable that yields the Veneziano pole coefficients.
At higher points, the mixed blocks couple distinct internal legs;
the five-point double residue and six-point comb and snowflake give
explicit examples. For both six-point topologies, the coefficient at
$(N_P,N_Q,N_R)=(1,1,1)$ agrees algebraically with an independent KN
expansion, including the central-source sign in the snowflake.

The determinant is already identified algebraically as the vacuum
sewing factor of the transverse bosons. Its higher-point geometric
interpretation requires more: the DDF blocks must be identified with
Grunsky operators, the resmoothing map must be constructed in a chart
adapted to the degeneration, and the ghost measure must be included.
This also requires control of the passage from finite cutoffs to
coefficientwise stabilized series. These are open questions about
the full top forms, separate from the physical-state derivation of
the residues.

\begin{samepage}
For the string S-matrix bootstrap, the resulting residues provide
explicit benchmarks for multiparticle factorization constraints
\cite{ArkaniHamedEtAl2024,BasileRemmenStaudt2026}.
In the present calculation the spectrum and cubic couplings are
fixed. A bootstrap would instead treat the couplings as unknowns,
keeping the labels that distinguish degenerate states explicit, and
constrain them by crossing, unitarity and high-energy behavior
\cite{CheungRemmen2023Bootstrap,HaringZhiboedov2024,CheungHillmanRemmen2024}.
Residue-zero conditions provide another set of constraints
\cite{CheungRemmenSciottiTarquini2026}.
A comparison with the residue-pairing bootstrap for massless scalar
disk integrals~\cite{CaoADR2026} first requires a dictionary for the
external states and kinematics.
\par
\end{samepage}

The external states and spectrum also matter for positivity bounds.
The multipositivity bounds of
ref.~\cite{CheungRemmenMultipositivity2025} apply to massless scalar
amplitudes in a tachyon-free spectrum, so their use here requires an
extension of the present setup.\footnote{A spin-resolved framework
for multipositivity has recently been developed for massless planar
amplitudes, with explicit four-, five- and six-point kinematic
charts~\cite{jeong2026partialwavesmultipositivity}.}

Excited external states can be treated within the same sewing
calculation by retaining their DDF sources in the cubic
vertices.\footnote{At fixed external momenta and levels, mixed
external--internal blocks shift the internal sources, while terms
containing only external sources factor out of the Gaussian
contraction~\cite{BiswasMarottaPesando2025,Firrotta2024}.
Differentiation with respect to the external sources selects the
external states; coefficient extraction in the sewing variables
selects the internal levels.}
Applying the same strategy to a tachyon-free superstring would require
the corresponding cubic vertices and physical-state projectors.

\begin{acknowledgments}
D.B. thanks Westlake University for its hospitality during a visit in
which a substantial part of this work was carried out. Q.C. is
supported by the Westlake Fellows Program at Westlake University.
OpenAI's GPT-6 model, accessed through Codex, was used to assist with
derivations, computational checks and manuscript preparation. The
authors are responsible for the scientific content.

\end{acknowledgments}
\appendix
\section{All-order four-point chain resummation}
\label{app:four-point-identities-proof}

We derive the open- and closed-chain sums from the cubic local maps,
the normalized free-boson Green function on the sewn sphere and the
one-boson stress-tensor Ward identity. The derivation uses neither the
Veneziano residues nor critical-dimension completeness.
All identities below are statements about coefficientwise stabilized
formal series.

Write $\gamma_a=a(1+a)$, $\gamma_b=b(1+b)$,
$\alpha=a+b+2ab$, $q=\gamma_a\gamma_b$, and
$c_*=\alpha(1+\alpha)-2q$, where $c_*$ is an algebraic coefficient.
Start with generic positive $a,b$ and a small enough sewing parameter
for the local maps and sewing series to converge.
The analytic inverse germs give exponential Cauchy bounds on the
Taylor and Grunsky coefficients. On sufficiently small annuli, the
weighted matching operator has small norm and its products are trace
class, so the Green-function and determinant manipulations are
well defined. Each formal coefficient is recovered at a sufficiently
large finite cutoff. Polynomial dependence of the chain coefficients
on $a,b$ then extends the identities coefficientwise by analytic
continuation. Square roots are taken with constant term one.
These statements apply to the stabilized series; they need not hold
for the full functions at a fixed cutoff.

Define
\begin{equation}
 \Delta(\chi)=[1-(1+\alpha)\chi]^2-4q\chi^2.
 \label{eq:Delta-def}
\end{equation}
\subsection*{The chain sums}

Because \(Zg(b)Zg(a)=O(z^2)\),
\begin{equation}
 K^{-1}=\sum_{r=0}^{\infty}[Zg(b)Zg(a)]^r.
 \label{eq:K-Neumann}
\end{equation}
Consequently,
\begin{align}
 \mathcal A
 &=f(a)^TZf(b)+f(a)^TZg(b)Zg(a)Zf(b)+\cdots,
 \label{eq:A-chains}\\
 \mathcal S
 &=\frac1{b(1+b)}\bigl[
 f(a)^TZg(b)Zf(a)
 +f(a)^TZg(b)Zg(a)Zg(b)Zf(a)+\cdots\bigr].
 \label{eq:S-chains}
\end{align}
Using \(Z_{nm}=z^n\delta_{nm}/n\) and
Eq.~\eqref{eq:g-def}, these chains become
\begin{align}
 \mathcal A={}&\sum_{n\geq1}\frac{z^n}{n}f_n(a)f_n(b)
 \nonumber\\
 &+\sum_{r\geq1}q^r
 \sum_{n_0,\ldots,n_{2r}\geq1}
 z^{n_0+\cdots+n_{2r}}
 \prod_{s=0}^{2r}f_{n_s}(a)f_{n_s}(b)
 \frac{\prod_{s=1}^{2r-1}n_s}
      {\prod_{s=0}^{2r-1}(n_s+n_{s+1})},
 \label{eq:A-chain-sum}\\
 \mathcal S={}&\sum_{r\geq0}q^r
 \sum_{n_0,\ldots,n_{2r+1}\geq1}
 z^{n_0+\cdots+n_{2r+1}}
 \prod_{s=0}^{2r+1}f_{n_s}(a)f_{n_s}(b)
 \frac{\prod_{s=1}^{2r}n_s}
      {\prod_{s=0}^{2r}(n_s+n_{s+1})}.
 \label{eq:S-chain-sum}
\end{align}
Empty products are one. At each power of $z$, only finitely many
chains contribute, and their weights depend on $a,b$ through
$f_n(a)f_n(b)$ and $q$. The even open-chain sum is therefore symmetric
under $a\leftrightarrow b$. This proves that the two expressions for
$\mathcal S$ in Eq.~\eqref{eq:S-def} agree, also at any common finite
mode cutoff.

\subsection*{Local coordinates and free-boson sewing}

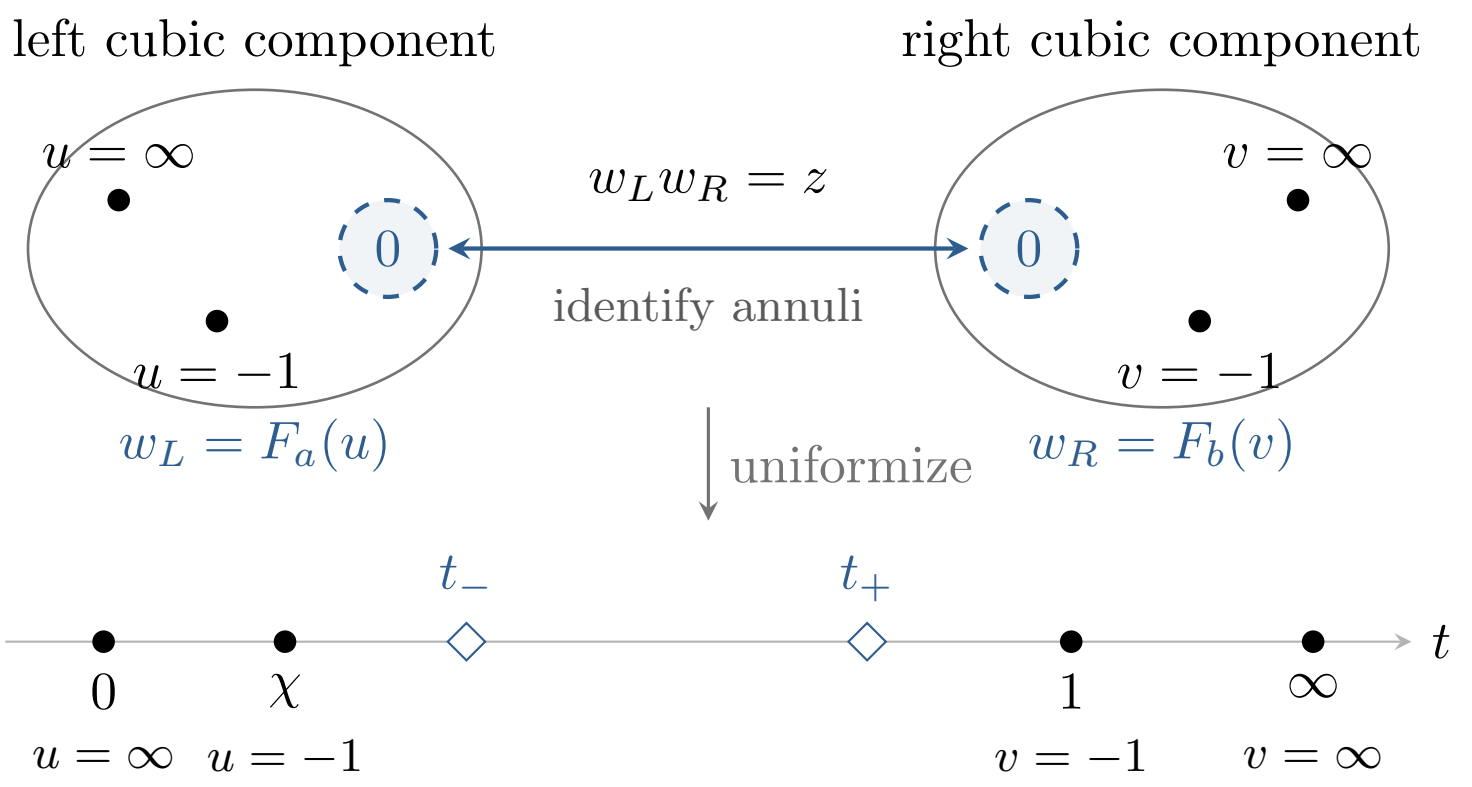
\begin{figure}[tbp]
\centering
\begin{tikzpicture}[font=\small,>=stealth]
  \draw[black!55,line width=.5pt] (-3,0) ellipse (1.5 and 1.05);
  \draw[black!55,line width=.5pt] (3,0) ellipse (1.5 and 1.05);
  \node at (-3,1.36) {left cubic component};
  \node at (3,1.36) {right cubic component};
  \node[ddf point,label=above:{$u=\infty$}] at (-3.9,.32) {};
  \node[ddf point,label=below:{$u=-1$}] at (-3.25,-.48) {};
  \node[ddf point,label=above:{$v=\infty$}] at (3.9,.32) {};
  \node[ddf point,label=below:{$v=-1$}] at (3.25,-.48) {};
  \filldraw[ddf seam,fill=DDFblue!7,dashed] (-2.12,0) circle (.32);
  \filldraw[ddf seam,fill=DDFblue!7,dashed] (2.12,0) circle (.32);
  \node[text=DDFblue] at (-2.12,0) {$0$};
  \node[text=DDFblue] at (2.12,0) {$0$};
  \node[text=DDFblue] at (-3,-1.3) {$w_L=F_a(u)$};
  \node[text=DDFblue] at (3,-1.3) {$w_R=F_b(v)$};
  \draw[ddf seam,<->] (-1.72,0) -- (1.72,0)
    node[midway,above=5pt] {$w_Lw_R=z$};
  \node[font=\footnotesize,text=black!65] at (0,-.4) {identify annuli};
  \draw[->,black!55,line width=.6pt] (0,-1.05) -- (0,-1.8)
    node[right,midway] {uniformize};
  \begin{scope}[yshift=-2.6cm]
    \draw[ddf axis,->] (-4.65,0) -- (4.65,0) node[right,text=black] {$t$};
    \node[ddf point,label=below:{$0$}] at (-4,0) {};
    \node[ddf point,label=below:{$\chi$}] at (-2.8,0) {};
    \node[ddf zero,label={[text=DDFblue]above:{$t_-$}}] at (-1.6,0) {};
    \node[ddf zero,label={[text=DDFblue]above:{$t_+$}}] at (1.05,0) {};
    \node[ddf point,label=below:{$1$}] at (2.4,0) {};
    \node[ddf point,label=below:{$\infty$}] at (4,0) {};
    \node[font=\footnotesize] at (-4,-.77) {$u=\infty$};
    \node[font=\footnotesize] at (-2.8,-.77) {$u=-1$};
    \node[font=\footnotesize] at (2.4,-.77) {$v=-1$};
    \node[font=\footnotesize] at (4,-.77) {$v=\infty$};
  \end{scope}
\end{tikzpicture}
\caption{Cubic plumbing and its uniformization. The dashed circles denote
the node neighborhoods; their annuli are identified by $w_Lw_R=z$.
The four external punctures (black dots) survive on the sewn sphere,
whereas the nodes do not. The open diamonds mark the two interaction
points, which are zeros of $R=\rho'$ (not additional punctures).
The lower line uses the illustrative real ordering for $a,b>0$ and
sufficiently small positive $\chi$; positions are schematic, and
$\infty$ denotes the point at infinity on the sphere.}
\label{fig:chain-plumbing}
\end{figure}

Consider the cubic local coordinate and its inverse germ,
\begin{equation}
 F_a(u)=u(1+u)^a,\qquad U_a(w)=F_a^{-1}(w)=w+O(w^2).
 \label{eq:chain-cubic-map}
\end{equation}
This is the three-puncture Mandelstam coordinate, with punctures
$u=0,-1,\infty$, in the normalization relevant to the DDF
blocks~\cite{BiswasPesandoMandelstam}. Figure~\ref{fig:chain-plumbing}
summarizes the sewing geometry and the puncture assignments used below.
Lagrange inversion gives
\begin{equation}
 [w^n]\log(1+U_a(w))
 =\frac1n[u^{n-1}](1+u)^{-an-1}
 =\frac{f_n(a)}n.
 \label{eq:chain-linear-map-coefficients}
\end{equation}
The quadratic coefficients follow from a generating kernel, without
summing two mode indices explicitly. For $u=U_a(w)$ and
$v=U_a(\eta)$, define
\begin{equation}
 H_a(w,\eta)=\log\frac{U_a(w)-U_a(\eta)}{w-\eta}
             +a\log(1+u)+a\log(1+v).
 \label{eq:chain-Grunsky-kernel}
\end{equation}
It vanishes when either argument is zero. Since
$w\,\partial_wu=u(1+u)/(1+(1+a)u)$, direct differentiation yields
\begin{align}
 (w\partial_w+\eta\partial_\eta)H_a
 &=\gamma_a\frac{u}{1+(1+a)u}\frac{v}{1+(1+a)v},\nonumber\\
 H_a(w,\eta)&=\sum_{n,m\geq1}\frac{g_{nm}(a)}{nm}w^n\eta^m.
 \label{eq:chain-Grunsky-coefficients}
\end{align}
The two factors on the first line have coefficients $f_n(a)$ and
$f_m(a)$. Dividing their product by the eigenvalue $n+m$ of the Euler
derivative gives $\gamma_af_n(a)f_m(a)/(n+m)=g_{nm}(a)/(nm)$,
identifying the quadratic kernel to all orders.

Sew two copies by $w_Lw_R=z$ in the BPZ-dual bases of the main text.
For a chiral boson with
$\langle j(w)j(\eta)\rangle=(w-\eta)^{-2}$, place charges
$+\lambda,-\lambda$ at the left punctures $u=-1,\infty$ and
$-\mu,+\mu$ at the right punctures $v=-1,\infty$.
Each component has zero total charge, hence the sewn channel carries
zero boson momentum. The right-hand sign implements the bra
orientation and gives the oriented source $+\mu f(b)$.
These auxiliary charges generate the scalar Gaussian functions;
they do not alter the external string states.
By Eqs.~\eqref{eq:chain-linear-map-coefficients} and
\eqref{eq:chain-Grunsky-coefficients}, the current and its quadratic
contraction give
\begin{equation}
 \frac{\mathcal Z(\lambda,\mu;z)}{\mathcal Z(0,0;z)}
 =\exp\!\left(\lambda\mu\mathcal A
       +\frac{\lambda^2}{2}T_L+\frac{\mu^2}{2}T_R\right),
 \quad
 \begin{cases}
 T_L=f(a)^TK^{-1}Zg(b)Zf(a),\\
 T_R=f(b)^T\widetilde K^{-1}Zg(a)Zf(b).
 \end{cases}
 \label{eq:chain-charged-Gaussian}
\end{equation}
Here, $\mathcal Z(0,0;z)=(\det K)^{-1/2}=e^{\mathcal C/2}$.

The free-current Green function gives the same result. Expanding its
regular parts on the two sewing circles yields the matching equations
\begin{equation}
 x=Z\bigl(\mu f(b)+g(b)y\bigr),\qquad
 y=Z\bigl(\lambda f(a)+g(a)x\bigr).
 \label{eq:chain-Green-matching}
\end{equation}
Eliminating $y$ produces the inverse $K^{-1}$ in
Eq.~\eqref{eq:chain-charged-Gaussian}. Since $K=\Id+O(z^2)$, the
formal solution is unique. The equations match Laurent modes of the
normalized current Green function, with the oscillator inverse metric
supplying the factors $1/n$ in $Z$.
The matched bidifferential has the standard double pole and no other
poles. On the sewn sphere it is therefore $dt\,ds/(t-s)^2$: the
difference would be a holomorphic bidifferential on a genus-zero
surface and must vanish. Integrating the current with the prescribed
charge residues gives the charged free-boson correlator in the global
coordinate $t$. Dividing by $\mathcal Z(0,0;z)$ removes the
source-independent conformal anomaly. The determinant remains to be
fixed separately.

Normalize the four external punctures to $t=0,\chi,1,\infty$. Let
$t_L(u)$ denote the resmoothing map from the left component and write
\begin{equation}
 d_L=\lim_{u\to\infty}u\,t_L(u),\qquad
 s_L=t_L'(-1),\qquad
 t_L(\infty)=0,\quad t_L(-1)=\chi.
 \label{eq:chain-external-jets}
\end{equation}
The coefficient of $\lambda\mu$ in the global correlator is
$-\log(1-\chi)$. The two left primaries have weight $\lambda^2/2$;
their coordinate factors and separation give the coefficient
$\tfrac12\log(-d_Ls_L/\chi^2)$ of $\lambda^2$. Thus
\begin{equation}
 \mathcal A=-\log(1-\chi),\qquad
 T_L=\log\frac{-d_Ls_L}{\chi^2}.
 \label{eq:chain-Green-identification}
\end{equation}
The sign in the second expression comes from using $1/u$ near the
left infinite puncture. The expansions
$d_L=-\chi+O(\chi^2)$ and $s_L=\chi+O(\chi^2)$ fix the logarithm to
have zero constant term. Thus the geometric cross ratio is the
formal series $\chi=1-e^{-\mathcal A}$ introduced in the main text.

\subsection*{Open chains from the two critical points}

The differential $d\log F_a(u)$ on the left agrees with
$-d\log F_b(v)$ across the seam. It is therefore a global meromorphic
differential with the four prescribed external residues. In the above
normalization it is $d\rho$, where
\begin{equation}
 \begin{aligned}
 \rho(t)&=-(1+a)\log t+a\log(t-\chi)-b\log(1-t),\\
 R(t):=\rho'(t)&=-\frac{1+a}{t}+\frac a{t-\chi}-\frac b{t-1}.
 \end{aligned}
 \label{eq:chain-Mandelstam-differential}
\end{equation}
Each original cubic differential has one simple zero. Their images are
the zeros of $R$, namely
\begin{equation}
 t_\mp=\frac{1+(1+a+b)\chi\mp\sqrt{\Delta(\chi)}}{2(1+b)},\qquad
 t_-=(1+a)\chi+O(\chi^2),\quad
 t_+=\frac1{1+b}+O(\chi).
 \label{eq:chain-interaction-points}
\end{equation}
The discriminant is $\Delta$ because
$(1+\alpha)^2-4q=(1+a+b)^2$.

Let $\omega_a=F_a(-1/(1+a))$ and
$\omega_b=F_b(-1/(1+b))$ be the fixed cubic critical values. The local
maps on the sewn surface obey
\begin{equation}
 \begin{gathered}
 F_a(u)=c_Le^{\rho(t)},\qquad
 F_b(v)=c_Re^{-\rho(t)},\qquad z=c_Lc_R,\\
 c_L=\omega_ae^{-\rho(t_-)},\quad
 c_R=\omega_be^{\rho(t_+)}.
 \end{gathered}
 \label{eq:chain-critical-value-matching}
\end{equation}
Evaluating the local maps at their critical points fixes these
constants, with branches chosen so that $z/\chi\to1$ as $\chi\to0$.

Since $R(t_\mp)=0$, differentiating the critical values gives
\begin{equation}
 \frac{d\log c_L}{d\chi}=\frac a{t_--\chi},\qquad
 \frac{d\log c_R}{d\chi}=-\frac a{t_+-\chi}.
 \label{eq:chain-critical-value-derivatives}
\end{equation}
Using
$(t_--\chi)(t_+-\chi)=a\chi(1-\chi)/(1+b)$, we obtain
\begin{equation}
 \boxed{\frac{d\log z}{d\chi}
 =\frac{\sqrt{\Delta(\chi)}}{\chi(1-\chi)},\qquad
 \frac z\chi=1+O(\chi).}
 \label{eq:chain-map-ODE}
\end{equation}
After subtracting $1/\chi$, the right-hand side is regular at zero.
Together with $z/\chi\to1$, this uniquely fixes the formal series
$z(\chi)$.

The two external jets in Eq.~\eqref{eq:chain-external-jets} follow by
expanding Eq.~\eqref{eq:chain-critical-value-matching} near
$u=\infty,-1$:
\begin{equation}
 d_L^{1+a}=c_L(-\chi)^a,\qquad
 s_L^a=-\frac{\chi^{1+a}(1-\chi)^b}{c_L}.
 \label{eq:chain-jet-equations}
\end{equation}
Only logarithmic derivatives of these relations are needed; their
constant phases have already been fixed above. They give
\begin{align}
 \frac{dT_L}{d\chi}
 &=\frac{1}{\gamma_a}
       \left(\frac1\chi-\frac a{t_--\chi}\right)
       -\frac b{a(1-\chi)}\nonumber\\
 &=\frac{1-(1+\alpha)\chi-\sqrt\Delta}
         {2\gamma_a\chi(1-\chi)}.
 \label{eq:chain-left-self-derivative}
\end{align}
Similarly, if $t_R(v)=d_Rv+O(1)$ at infinity and
$s_R=t_R'(-1)$ at $t=1$, the right self-contraction is
$T_R=\log(s_R/d_R)$. Its derivative is
\begin{equation}
 \frac{dT_R}{d\chi}
 =-\frac a{b(1-\chi)}+\frac a{\gamma_b(t_+-\chi)}
 =\frac{1-(1+\alpha)\chi-\sqrt\Delta}
         {2\gamma_b\chi(1-\chi)}.
 \label{eq:chain-right-self-derivative}
\end{equation}
Both self-contractions vanish at the degeneration, so
$T_L/\gamma_b=T_R/\gamma_a=\mathcal S$, in agreement with the
chain argument for Eq.~\eqref{eq:S-def}. Defining
$D_\chi=1-(1+\alpha)\chi$, we obtain
\begin{equation}
 \boxed{\frac{d\mathcal S}{d\chi}
 =\frac{D_\chi-\sqrt\Delta}{2q\chi(1-\chi)}
 =\frac{2\chi}{(1-\chi)(D_\chi+\sqrt\Delta)},\qquad
 \mathcal S(0)=0.}
 \label{eq:chain-source-ODE}
\end{equation}
The second expression is regular also at $q=0$.

\subsection*{Closed chains from the one-boson Ward identity}

For one boson, the vacuum sewing factor is
$\mathcal Z_0=e^{\mathcal C/2}$. Varying the sewing parameter inserts
the oscillator Hamiltonian $L_0=\sum_{n>0}\alpha_{-n}\alpha_n$ into the
Fock-space contraction. The corresponding Ward identity is
\begin{equation}
 z\partial_z\log\mathcal Z_0
 =\frac{1}{2\pi i}\oint_{w_L}w_L\langle T(w_L)\rangle\,dw_L,
 \qquad
 \langle T(w_L)\rangle=\frac1{12}\{t,w_L\}.
 \label{eq:chain-sewing-Ward-identity}
\end{equation}
For the second equality, point split $T=\tfrac12:j^2:$ and subtract
the local double pole from the pulled-back current bidifferential.
The finite remainder is $\{t,w_L\}/6$, so the factor $1/2$ in $T$
gives $1/12$ for a boson of central charge one.
This uses the normalized bidifferential and does not assume an
expression for $\mathcal Z_0$. The insertion is the plane-coordinate
$L_0$, without a zero-mode term or cylinder vacuum-energy shift.

Since $w_L=c_Le^{\rho(t)}$, the Schwarzian chain rule gives
\begin{equation}
 w_L\{t,w_L\}\,dw_L
 =\left(-\frac{R''}{R^2}+\frac{3(R')^2}{2R^3}+\frac R2\right)dt
 =:Q(t)\,dt.
 \label{eq:chain-Schwarzian-rational-form}
\end{equation}
The positively oriented $w_L$ circle bounds the removed node disk.
Relative to the remaining left component its orientation is negative.
In the $t$ plane that component contains $0,\chi,t_-$, as shown in
figure~\ref{fig:chain-Ward-contour}, so
\begin{equation}
 z\partial_z\mathcal C=-\frac16\mathcal R_-,\qquad
 \mathcal R_-:=\operatorname*{Res}_{t=0,\chi,t_-}Q(t)\,dt.
 \label{eq:chain-closed-Ward-residue}
\end{equation}
The sign and the factor $1/6$ include, respectively, this contour
orientation and $\log\mathcal Z_0=\mathcal C/2$.

\begin{figure}[tbp]
\centering
\begin{tikzpicture}[font=\small,>=stealth]
  \begin{scope}[xshift=-3.35cm]
    \draw[ddf axis,->] (-1.3,0) -- (1.3,0);
    \draw[ddf axis,->] (0,-1.17) -- (0,1.27);
    \fill[black!5] (0,0) circle (.53);
    \draw[black!35,dashed,line width=.5pt] (0,0) circle (.53);
    \node[ddf point,label=below right:{$0$}] at (0,0) {};
    \draw[ddf contour] (0,0) circle (.85);
    \draw[ddf contour,->] (.85,0) arc (0:65:.85);
    \node[text=DDFred] at (-1.1,.8) {$\Gamma_w$};
    \node at (.97,1.22) {$w_L$};
    \node[font=\footnotesize,align=center] at (0,-1.65)
      {removed node disk\\$\Gamma_w$: counterclockwise};
  \end{scope}
  \draw[->,black!55,line width=.6pt] (-1.7,.2) -- (-.65,.2)
    node[midway,above] {$t$};
  \begin{scope}[xshift=1.45cm]
    \draw[ddf axis,->] (-1.7,0) -- (3.1,0) node[right,text=black] {$t$};
    \draw[ddf axis,->] (-1.45,-1.17) -- (-1.45,1.27);
    \node[ddf point,label=below:{$0$}] at (-1,0) {};
    \node[ddf point,label=below:{$\chi$}] at (-.25,0) {};
    \node[ddf zero,label=below:{$t_-$}] at (.5,0) {};
    \node[ddf zero,label=below:{$t_+$}] at (1.75,0) {};
    \node[ddf point,label=below:{$1$}] at (2.6,0) {};
    \draw[ddf contour] (-.25,0) ellipse (1.12 and .78);
    \draw[ddf contour,->] (-.25,.78)
      arc[start angle=90,end angle=25,x radius=1.12,y radius=.78];
    \node[text=DDFred] at (.9,1.08) {$t(\Gamma_w)\sim-\Gamma_-$};
    \node[font=\footnotesize,align=center] at (.5,-1.65)
      {remaining left component\\$t(\Gamma_w)$: clockwise};
  \end{scope}
  \node at (.2,-2.5)
    {$\displaystyle\frac{1}{2\pi i}\oint_{t(\Gamma_w)}Q(t)\,dt
      =-\!\sum_{p\in\{0,\chi,t_-\}}\Res_{t=p}Q(t)\,dt$};
\end{tikzpicture}
\caption{Orientation in the one-boson Ward identity. The positive local
circle surrounds the excised node disk. Its image has negative
orientation around the remaining left component; the red curve on the
right is a homologous clockwise representative. Here $\Gamma_-$ denotes
the counterclockwise contour around $0,\chi,t_-$. Black dots are external
punctures, and open diamonds are zeros of $R$; both types contribute poles
of $Q$. The right interaction point $t_+$ and the punctures $1,\infty$
lie outside the contour. The real ordering is schematic as in
figure~\ref{fig:chain-plumbing}.}
\label{fig:chain-Ward-contour}
\end{figure}

As a normalization check, $g_{11}(a)=\gamma_a/2$ gives
$\mathcal Z_0=1+qz^2/8+O(z^3)$, or
$\mathcal C=qz^2/4+O(z^3)$, directly from the two oscillator vertices.

At a pole of $R$ with residue $r$, the residue of $Q$ is
$(r-r^{-1})/2$. At the simple zero $t_-$ of $R$, it is
$\bigl(R''(t_-)^2-R'(t_-)R'''(t_-)\bigr)/(4R'(t_-)^3)$.
The closed-chain sum thus reduces to the finite rational expression
\begin{equation}
 \mathcal R_-=-\frac12-\frac1{2\gamma_a}
 +\frac{R''(t_-)^2-R'(t_-)R'''(t_-)}{4R'(t_-)^3}.
 \label{eq:chain-residue-local-evaluation}
\end{equation}
Substituting Eq.~\eqref{eq:chain-Mandelstam-differential} and the quadratic
equation for $t_-$ gives
\begin{equation}
 \mathcal R_-=
 \frac{\chi(1-\chi)}{2\sqrt\Delta}
 \left[\frac{1+\alpha}{1-\chi}
       +\frac{\Delta'}{2\Delta}
       +c_*\frac{d\mathcal S}{d\chi}\right].
 \label{eq:chain-residue-reduction}
\end{equation}
Primes on $\Delta$ denote $\chi$ derivatives, whereas primes on $R$
denote $t$ derivatives. The algebra in
this substitution can be verified without a radical: set $\tau=t_-$ and
use
\begin{equation}
 \chi=\frac{\tau((1+b)\tau-1)}{(1+a+b)\tau-(1+a)},\qquad
 \sqrt\Delta=1+(1+a+b)\chi-2(1+b)\tau.
 \label{eq:chain-rational-certificate-parameterization}
\end{equation}
The residue reduction is then a rational identity in $a,b,\tau$.
An explicit polynomial certificate is obtained by defining
\begin{equation}
 \begin{gathered}
 A_0=1+a,\qquad B_0=1+b,\qquad \sigma=1+a+b,\\
 H=\sigma\tau-A_0,\qquad L=B_0\tau-1,\qquad J_0=B_0\tau-A_0,\\
 Y=-A_0+2A_0B_0\tau-\sigma B_0\tau^2,\\
 M_0=(1+3a+b)\tau-(1+2a),\qquad
 V_0=(\partial_\tau Y)H-\sigma Y.
 \end{gathered}
 \label{eq:chain-certificate-polynomials}
\end{equation}
Then $\chi=\tau L/H$, $\sqrt\Delta=Y/H$, and
\begin{align}
 U_0={}&4B_0^2H^2a^2\tau^2(\tau-1)^2
       +2B_0HM_0Ya\tau(\tau-1)\nonumber\\
 &+\bigl[M_0^2-3H^2-3a^2((\tau-1)^2+\tau^2)\bigr]Y^2.
 \label{eq:chain-certificate-U}
\end{align}
After clearing the common denominator $4aA_0Y^3$, the two sides of
Eq.~\eqref{eq:chain-residue-reduction} are equal by the polynomial identity
\begin{align}
 A_0U_0-2(aA_0+1)Y^3
 =2a\bigl[&A_0\tau L(1+\alpha)Y^2
       +A_0\tau L(\tau-1)J_0V_0\nonumber\\
       &-c_*\tau^2Y^2\bigr].
 \label{eq:chain-polynomial-certificate}
\end{align}
Expanding in $\mathbb Z[a,b,\tau]$ verifies this polynomial identity
and proves the residue reduction without truncating the sewing
series.

Combining Eqs.~\eqref{eq:chain-map-ODE},
\eqref{eq:chain-closed-Ward-residue}, and
\eqref{eq:chain-residue-reduction} proves the differential equation
\begin{equation}
 12\frac{d\mathcal C}{d\chi}
 =-\frac{1+\alpha}{1-\chi}-\frac{\Delta'}{2\Delta}
  -c_*\frac{d\mathcal S}{d\chi},\qquad \mathcal C(0)=0.
 \label{eq:chain-determinant-ODE}
\end{equation}
Integrating this equation gives Eq.~\eqref{eq:C-resummation} below.
This completes the closed-chain resummation independently of the
open-chain coefficient formulas.

\subsection*{Coefficient formulas and the two reparametrization identities}

Let \(z(\chi)\) be the compositional inverse of
\(\chi(z)=1-e^{-\mathcal A(z)}\).  Lagrange--B\"urmann inversion,
\begin{equation}
 [\chi^n]F(z(\chi))
 =\frac1n[z^{n-1}]F'(z)\left(\frac z{\chi(z)}\right)^n,
 \label{eq:Lagrange-Burmann}
\end{equation}
relates the chain sums in the two coordinates. To extract explicit
coefficients, expand Eq.~\eqref{eq:chain-source-ODE} using the Catalan
generating function at $u=q\chi^2/D_\chi^2$.
Applying the same expansion to Eq.~\eqref{eq:chain-map-ODE} determines
$\log(z/\chi)$. For $n\geq1$, this gives
\begin{align}
 n[\chi^n]\mathcal S
 &=\sum_{\ell=0}^{\lfloor(n-2)/2\rfloor}
 \frac1{\ell+1}\binom{2\ell}{\ell}q^\ell
 \sum_{j=0}^{n-2-2\ell}
 \binom{2\ell+j}{j}(1+\alpha)^j,
 \label{eq:S-Lagrange}\\
 n[\chi^n]\log\frac{z(\chi)}\chi
 &=-\alpha-2\sum_{\ell=1}^{\lfloor n/2\rfloor}
 \frac1\ell\binom{2\ell-2}{\ell-1}q^\ell
 \sum_{j=0}^{n-2\ell}
 \binom{2\ell+j-2}{j}(1+\alpha)^j.
 \label{eq:z-Lagrange}
\end{align}
Here $\mathcal S$ denotes $\mathcal S(z(\chi))$, and empty sums
vanish. The sewing and Ward identities above show that these are the
coefficients of the original matrix chains. To resum them in
$\mathbb Q[\alpha,q][[\chi]]$, set
\[
 D_\chi=1-(1+\alpha)\chi,
 \qquad
 u=\frac{q\chi^2}{D_\chi^2}.
\]

Since \(D_\chi=1+O(\chi)\) is a unit and \(u=O(\chi^2)\), all inverses and
compositions below are well-defined formal series.  Differentiating
Eq.~\eqref{eq:S-Lagrange} and setting
\(k=n-2-2\ell-j\) gives
\begin{align}
 \frac{\dd\mathcal S}{\dd\chi}
 &=\sum_{\ell,j,k\geq0}
 \frac1{\ell+1}\binom{2\ell}{\ell}q^\ell
 \binom{2\ell+j}{j}(1+\alpha)^j
 \chi^{2\ell+j+k+1}
 \nonumber\\
 &=\frac{\chi}{1-\chi}
 \sum_{\ell\geq0}
 \frac1{\ell+1}\binom{2\ell}{\ell}
 q^\ell\chi^{2\ell}
 \sum_{j\geq0}\binom{2\ell+j}{j}
 \bigl((1+\alpha)\chi\bigr)^j
 \nonumber\\
 &=\frac{\chi}{(1-\chi)D_\chi}
 \sum_{\ell\geq0}
 \frac1{\ell+1}\binom{2\ell}{\ell}u^\ell .
 \label{eq:S-resummation-intermediate}
\end{align}
Here, we used
\[
 \sum_{k\geq0}\chi^k=\frac1{1-\chi},
 \qquad
 \sum_{j\geq0}\binom{r+j}{j}t^j
   =\frac1{(1-t)^{r+1}}.
\]
The Catalan generating function can be written in the formally regular
form
\begin{equation}
 \sum_{\ell\geq0}
 \frac1{\ell+1}\binom{2\ell}{\ell}u^\ell
 =\frac{2}{1+\sqrt{1-4u}},
 \label{eq:Catalan-generating-function}
\end{equation}
where the square root denotes the unique formal square root with constant
term one.  Since
\[
 \Delta(\chi)=D_\chi^2-4q\chi^2
             =D_\chi^2(1-4u),
 \qquad
 \sqrt{\Delta(\chi)}=D_\chi\sqrt{1-4u},
\]
Eq.~\eqref{eq:S-resummation-intermediate} becomes
\begin{equation}
 \frac{\dd\mathcal S}{\dd\chi}
 =\frac{2\chi}
 {(1-\chi)\bigl[D_\chi+\sqrt{\Delta(\chi)}\bigr]}
 =\frac{2\chi}{(1-\chi)
 [1-(1+\alpha)\chi+\sqrt{\Delta(\chi)}]} .
 \label{eq:dS-dchi}
\end{equation}

The second sum is treated in the same way.  Differentiating
Eq.~\eqref{eq:z-Lagrange} and now setting \(k=n-2\ell-j\) yields
\begin{align}
 \frac{\dd}{\dd\chi}\log\frac{z(\chi)}{\chi}
 &=-\frac{\alpha}{1-\chi}
 -\frac{2}{1-\chi}
 \sum_{\ell\geq1}\frac1\ell\binom{2\ell-2}{\ell-1}
 q^\ell\chi^{2\ell-1}
 \sum_{j\geq0}\binom{2\ell+j-2}{j}
 \bigl((1+\alpha)\chi\bigr)^j
 \nonumber\\
 &=-\frac{\alpha}{1-\chi}
 -\frac{2D_\chi}{\chi(1-\chi)}
 \sum_{\ell\geq1}
 \frac1\ell\binom{2\ell-2}{\ell-1}u^\ell .
 \label{eq:z-resummation-intermediate}
\end{align}
Since
\[
 \frac1\ell\binom{2\ell-2}{\ell-1}
 =\frac1{(\ell-1)+1}\binom{2(\ell-1)}{\ell-1},
\]
Eq.~\eqref{eq:Catalan-generating-function} gives
\[
 \sum_{\ell\geq1}
 \frac1\ell\binom{2\ell-2}{\ell-1}u^\ell
 =\frac{1-\sqrt{1-4u}}2.
\]
Consequently,
\begin{align}
 \frac{\dd}{\dd\chi}\log\frac{z(\chi)}{\chi}
 &=-\frac{\alpha}{1-\chi}
 -\frac{D_\chi}{\chi(1-\chi)}
   \bigl(1-\sqrt{1-4u}\bigr)
 \nonumber\\
 &=-\frac{\alpha}{1-\chi}
 -\frac{D_\chi-\sqrt{\Delta(\chi)}}{\chi(1-\chi)}
 \nonumber\\
 &=-\frac\alpha{1-\chi}
 -\frac{1-(1+\alpha)\chi-\sqrt{\Delta(\chi)}}
 {\chi(1-\chi)}.
 \label{eq:dlogzchi}
\end{align}
The apparent pole \(1/\chi\) in the last expression cancels, since
\[
 D_\chi-\sqrt{\Delta(\chi)}
 =D_\chi\bigl(1-\sqrt{1-4u}\bigr)=O(\chi^2).
\]
The quotient is therefore a formal power series. Since $u=O(\chi^2)$,
each reindexing above contains only finitely many terms at a given
order.

It follows that
\begin{equation}
 \frac{\dd\log z}{\dd\chi}
 =\frac{\sqrt{\Delta(\chi)}}{\chi(1-\chi)}.
 \label{eq:dlogz}
\end{equation}
Using \(\dd\mathcal A/\dd\chi=(1-\chi)^{-1}\) and
\begin{equation}
 4q\chi^2=[1-(1+\alpha)\chi]^2-\Delta(\chi),
 \label{eq:Delta-relation}
\end{equation}
one obtains
\begin{equation}
 \frac{\dd}{\dd\chi}
 \left(\alpha\mathcal A+2q\mathcal S-\log\frac\chi z\right)=0.
 \label{eq:first-identity-derivative}
\end{equation}
The expression in parentheses vanishes at $\chi=0$, proving
Eq.~\eqref{eq:chi-identity-one} to every order.

Integrating Eq.~\eqref{eq:chain-determinant-ODE} with
$\mathcal C(0)=0$ gives the closed-chain resummation of
\begin{equation}
 \mathcal C=\sum_{r\geq1}\frac1r
 \operatorname{Tr}[Zg(b)Zg(a)]^r
 \label{eq:C-trace-series}
\end{equation}
in the form
\begin{equation}
 12\mathcal C=(1+\alpha)\log(1-\chi)
 -\frac12\log\Delta(\chi)
 -[\alpha(1+\alpha)-2q]\mathcal S.
 \label{eq:C-resummation}
\end{equation}

The open-chain result also implies
\begin{equation}
 \chi'=\frac{\chi(1-\chi)}{z\sqrt{\Delta(\chi)}}.
 \label{eq:chi-prime}
\end{equation}
Finally,
\begin{equation}
 a(1+a)+b(1+b)=\alpha(1+\alpha)-4q,
 \label{eq:alpha-q-identity}
\end{equation}
so Eqs.~\eqref{eq:C-resummation} and \eqref{eq:chi-prime}, together
with the first identity, imply Eq.~\eqref{eq:chi-identity-two}.
This proves the three chain-resummation relations and the two
four-point matrix identities as stabilized formal series, completing
the change of variable in Eq.~\eqref{eq:critical-residue}.
The proof is local to the four-point degeneration. It does not
establish the higher-point moduli-space identities of
section~\ref{sec:berezin-grunsky}.
\section{Transverse-basis independence of the sewing}
\label{app:transverse-basis}

We prove invariance of the sewing under orthogonal rotations of the
transverse bases at propagator endpoints and cubic vertices.
The light-cone directions are fixed, so all bases span the same
transverse space. Independence of the auxiliary null direction follows
separately from physical-state completeness, as discussed in the
main text.

For an internal edge \(e\), let \(\{\varepsilon^i_{e,L}\}\) and
\(\{\varepsilon^j_{e,R}\}\) be orthonormal bases of the same transverse
space, and define
\begin{equation}
 (\Omega_e)_{ij}
 =\langle\varepsilon^i_{e,L},\varepsilon^j_{e,R}\rangle_{\rm BPZ}.
 \label{eq:Omega-def}
\end{equation}
Orthonormality and completeness imply
\begin{equation}
 \Omega_e^T\Omega_e=\Omega_e\Omega_e^T=I_{d_\perp}.
 \label{eq:Omega-orthogonal}
\end{equation}
The one-oscillator BPZ product is therefore
\begin{equation}
 {}_{L(e)}\langle n,i\vert m,j\rangle_{R(e),\rm BPZ}
 =n\delta_{nm}(\Omega_e)_{ij},
 \label{eq:general-bpz-overlap}
\end{equation}
where the half-edge oscillator rephasings of section~\ref{sec:arbitrary-tree-sewing} have
already been applied.  Redefining
the right basis by
\begin{equation}
 \varepsilon_{e,R}^{\prime i}
 =\sum_j\varepsilon_{e,R}^j(\Omega_e^{-1})_{ji}
 \label{eq:right-basis-rotation}
\end{equation}
sets the overlap to \(\delta_{ij}\).  The propagator may consequently always
be written as
\begin{equation}
 \exp\!\left[
 \sum_{n=1}^{L_e}\sum_{i=1}^{d_\perp}\frac{z_e^n}{n}
 \partial_{\lambda_{L(e),e,n}^{i}}
 \partial_{\lambda_{R(e),e,n}^{i}}\right].
 \label{eq:common-frame-propagator}
\end{equation}
Thus the $\delta_{ij}$ contraction follows from the choice of
BPZ-dual orthonormal bases and includes the full transverse state sum.

For the five-point diagram, let the \(P\)- and \(Q\)-channel bases be related
by
\begin{equation}
 \Gamma_{ij}=\varepsilon_P^i\cdot\varepsilon_Q^j,\qquad
 \Gamma^T\Gamma=\Gamma\Gamma^T=I_{d_\perp}.
 \label{eq:Gamma-def}
\end{equation}
The central distinct-leg block is
\begin{equation}
 (G_R)_{(P,n,i),(Q,m,j)}=h_{mn}(c)\Gamma_{ij},
 \label{eq:five-point-Gamma-block}
\end{equation}
so, in composite mode--transverse indices,
\begin{equation}
 G_R(\Gamma)=
 \begin{pmatrix}
 g^{[N]}(c)\otimes I_{d_\perp}&h(c)^T\otimes\Gamma\\
 h(c)\otimes\Gamma^T&g^{[M]}(e)\otimes I_{d_\perp}
 \end{pmatrix}.
 \label{eq:five-point-Gamma-GR}
\end{equation}
Meanwhile,
\begin{equation}
 G_L=\begin{pmatrix}
 g^{[N]}(a)\otimes I_{d_\perp}&0\\
 0&g^{[M]}(b)\otimes I_{d_\perp}
 \end{pmatrix},\qquad
 Z=\diag(Z_x\otimes I_{d_\perp},Z_y\otimes I_{d_\perp}).
 \label{eq:five-point-Gamma-GLZ}
\end{equation}
Apply the same rotation to both half-edges of $Q$:
\begin{equation}
 T=\diag(I_N\otimes I_{d_\perp},I_M\otimes\Gamma),\qquad
 u'=Tu,\qquad w'=Tw.
 \label{eq:five-point-T-rotation}
\end{equation}
Then
\begin{equation}
 TZT^T=Z,\qquad T G_LT^T=G_L,\qquad
 T G_R(\Gamma)T^T=G_R(I_{d_\perp}).
 \label{eq:five-point-congruences}
\end{equation}
For \(K_L(\Gamma)=I-ZG_R(\Gamma)ZG_L\),
\begin{equation}
 K_L(I_{d_\perp})=TK_L(\Gamma)T^T,\qquad
 \det K_L(I_{d_\perp})=\det K_L(\Gamma),
 \label{eq:five-point-K-basis-independence}
\end{equation}
and every source contraction in the Gaussian exponent is invariant.  Hence
\begin{equation}
 \mathcal G_5(\Gamma;u,w)
 =\mathcal G_5(I_{d_\perp};Tu,Tw).
 \label{eq:five-point-generator-basis-independence}
\end{equation}
For example, at \(N=M=1\), where \(h_{11}=1\), direct differentiation gives
\begin{equation}
 R_{1,1}^{(5)}=(u_P^Tw_P)(u_Q^Tw_Q)
 +\sum_{i,j}u_{P,i}\Gamma_{ij}u_{Q,j}.
 \label{eq:five-point-basis-example}
\end{equation}
Defining \(u_Q'=\Gamma u_Q\), \(w_Q'=\Gamma w_Q\), orthogonality gives
\begin{equation}
 (u_Q')^Tw_Q'=u_Q^Tw_Q,\qquad
 \sum_{i,j}u_{P,i}\Gamma_{ij}u_{Q,j}
 =\sum_i u_{P,i}u_{Q,i}'.
 \label{eq:five-point-basis-example-rotation}
\end{equation}
The result is the common-basis expression written in rotated components.
\section{Cubic conventions and coordinate factors}
\label{app:cubic-coordinate-factors}

We first convert the cyclic oscillator coefficients to the comb
convention, then show how the puncture-coordinate powers cancel in
the complete cubic amplitude at fixed levels.

\subsection{From cyclic data to the comb convention}

To recover the comb convention, consider the cyclically ordered
vertex $(P,k,-Q)$, where $Q=P+k$, and define
\[
c=\frac{k^+}{P^+},
\qquad
e=\frac{k^+}{-Q^+}=-\frac{c}{1+c},
\]
\[
D_P=\sqrt{2\alpha'}(k-cP)_\perp,
\qquad
D_Q=\sqrt{2\alpha'}(k+eQ)_\perp.
\]
Suppressing the common transverse index, the cyclic reference
data on its two internal legs are
\[
\rho_P^{\mathrm{cyc}}=c,
\qquad
J_P^{\mathrm{cyc}}=D_P,
\qquad
\rho_{-Q}^{\mathrm{cyc}}=-\frac{1}{1+c}=-1-e,
\qquad
J_{-Q}^{\mathrm{cyc}}=-D_Q.
\]
The finite-product definition gives
\[
f_n(-1-\rho)=(-1)^{n-1}f_n(\rho).
\]
Choosing the half-edge phases
\[
\theta_{P}(n)=-1,
\qquad
\theta_{-Q}(m)=(-1)^m,
\]
gives the oriented sources
\[
\widehat j_{P,n}=-f_n(c)D_P,
\qquad
\widehat j_{-Q,m}=f_m(e)D_Q,
\]
and the two same-leg blocks reduce to $g(c)$ and $g(e)$.
The cyclic orientation of the distinct-leg block is $(-Q,P)$.
Applying the same phases to both oscillator indices gives the mixed-block
identity in Eq.~\eqref{eq:h-from-b}.

\subsection{The fixed-level cubic amplitude}

At fixed levels, the oscillator coordinate powers cancel against
the seed KN factor and ghost correlator. To show this, let
$p_{v,r}$ be the three outgoing physical momenta, with levels
$N_{v,r}$, $r\in\{a,b,c\}$; an external tachyon has $N_{v,r}=0$.
The corresponding seed-tachyon momenta obey
\[
k_{T,v,r}^{+}=p_{v,r}^{+},
\qquad
(k_{T,v,r})_\perp=(p_{v,r})_\perp,
\qquad
k_{T,v,r}^{-}
=
p_{v,r}^{-}
-\frac{N_{v,r}}{2\alpha' p_{v,r}^{+}}.
\]
Hence
\[
2\alpha' k_{T,v,r}\cdot k_{T,v,s}
=
2\alpha'p_{v,r}\cdot p_{v,s}
+
N_{v,r}\frac{\beta_{v,s}}{\beta_{v,r}}
+
N_{v,s}\frac{\beta_{v,r}}{\beta_{v,s}}.
\]
Momentum conservation and the physical mass-shell conditions give
\[
\beta_{v,a}+\beta_{v,b}+\beta_{v,c}=0,
\qquad
2\alpha'p_{v,a}\cdot p_{v,b}
=
-1+N_{v,a}+N_{v,b}-N_{v,c},
\]
and cyclic permutations of the second identity.

With the common ordering and branch convention used to define
$\Xi_v$, the coordinate-dependent seed factor is therefore
\[
\mathcal K_{T,v}
=
\prod_{r<s}
x_{v,rs}^{\,2\alpha'k_{T,v,r}\cdot k_{T,v,s}}
=
\frac{
\Xi_v^{-\sum_r N_{v,r}/\beta_{v,r}}
}{
x_{v,ab}x_{v,ac}x_{v,bc}
}.
\]
The oscillator coefficient at these levels carries
\[
\prod_r \sigma_{v,r}^{N_{v,r}}
=
\left(\prod_r\epsilon_{v,r}^{N_{v,r}}\right)
\Xi_v^{\sum_rN_{v,r}/\beta_{v,r}},
\]
whereas
\[
\left\langle
c(x_{v,a})c(x_{v,b})c(x_{v,c})
\right\rangle
=
x_{v,ab}x_{v,ac}x_{v,bc}.
\]
Thus,
\[
\mathcal K_{T,v}
\left(\prod_r\sigma_{v,r}^{N_{v,r}}\right)
\left\langle
c(x_{v,a})c(x_{v,b})c(x_{v,c})
\right\rangle
=
\prod_r\epsilon_{v,r}^{N_{v,r}}.
\]
All puncture-dependent powers therefore cancel in the fixed-level
cubic amplitude. The remaining constant phases are fixed by the
local-coordinate and BPZ conventions. Before level selection, the
coherent vertex retains its seed KN factor.

\bibliographystyle{JHEP}
\bibliography{DDF_sewing_paper_v2}

\end{document}